\pdfoutput=1
\documentclass[cernpreprint,atlasdraft=false,texlive=2016,UKenglish,texmf]{atlasdoc}
 
\usepackage[subfigure]{atlaspackage}
\usepackage{tabu,multirow}
\usepackage{slashed}
\usepackage{mathrsfs}
\usepackage{xspace} 
\usepackage{units}
\usepackage{rotating}
\usepackage{verbatim}
\usepackage{float}
\usepackage{hyperref}
\usepackage{cleveref}
\usepackage{arydshln}
 
\usepackage{color}

\usepackage{atlaspackage}
\usepackage[backref=false]{atlasbiblatex}
 
\usepackage{atlasphysics}
 
\usepackage{atlasbsm}
 
\graphicspath{{logos/}{figures/}}
 
\usepackage{ANA-SUSY-2018-22-PAPER-defs}

\AtlasTitle{Search for squarks and gluinos in final states with jets
and missing transverse momentum using \\
139~fb$^{-1}$ of $\sqrt{s}$ =13~TeV $pp$ collision data with the ATLAS detector}
 
\AtlasJournal{JHEP}
\AtlasAbstract{
A search for the supersymmetric partners of quarks and gluons (squarks
and gluinos) in final states containing jets and missing
transverse momentum, but no electrons or muons, is presented.
The data used in this search were recorded by the
ATLAS experiment in proton--proton collisions at a centre-of-mass energy of $\sqrt{s}$ = 13~TeV during Run 2 of
the Large Hadron Collider, corresponding to an integrated luminosity of 139~fb$^{-1}$.
The results are interpreted in the context of various $R$-parity-conserving models where
squarks and gluinos are produced in pairs or in association and a neutralino is the lightest supersymmetric particle.
An exclusion limit at the 95\% confidence level on the mass of the
gluino is set at 2.30~TeV for a simplified model containing only
a gluino and the lightest neutralino, assuming the latter
is massless. For a simplified model involving the strong production of
mass-degenerate first- and second-generation squarks, squark masses
below 1.85~TeV are excluded if the lightest neutralino is massless.
These limits extend substantially beyond the region of supersymmetric
parameter space excluded previously by similar searches with the ATLAS
detector.
}
 
\author{The ATLAS Collaboration}
 
\AtlasRefCode{SUSY-2018-022}
 
\PreprintIdNumber{CERN-EP-2020-166}
 
\AtlasJournalRef{JHEP 02 (2021) 143}
\AtlasDOI{10.1007/JHEP02(2021)143}

\hypersetup{pdftitle={ATLAS document},pdfauthor={The ATLAS Collaboration}}
 
\begin{document}
 
\maketitle

\newpage
 
\section{Introduction}
Supersymmetry (SUSY)
\cite{Golfand:1971iw,Volkov:1973ix,Wess:1974tw,Wess:1974jb,Ferrara:1974pu,Salam:1974ig}
is a generalisation of space-time symmetries that
predicts new bosonic partners of the fermions and new fermionic partners of the bosons
of the Standard Model (SM). If $R$-parity is conserved~\cite{Farrar:1978xj},
supersymmetric particles are produced in pairs and the lightest supersymmetric particle (LSP) is stable and represents a possible dark-matter candidate~\cite{Goldberg:1983nd,Ellis:1983ew}.
The scalar partners of the left- and right-handed quarks, the squarks $\squarkL$ and $\squarkR$, mix to form two mass eigenstates  $\tilde{q}_1$ and $\tilde{q}_2$ ordered by increasing mass. Superpartners of the charged and neutral electroweak and Higgs bosons also mix, to form charginos ($\chinopm$) and neutralinos  ($\nino$). Squarks and the fermionic partners of the gluons, the gluinos ($\gluino$), could be produced in strong-interaction processes at the Large Hadron Collider (LHC)~\cite{LHC:2008}  and decay via cascades ending with the stable LSP, which escapes the detector unseen, potentially producing substantial missing transverse momentum (with magnitude denoted $E\mathrm{^{miss}_T}$).
 
The large cross-sections predicted for the strong production of supersymmetric particles make the gluinos and squarks a primary target in searches for SUSY in proton--proton ($pp$) collisions at the LHC.
The large range of possible
parameter values for $R$-parity-conserving models in the Minimal Supersymmetric Standard Model (MSSM)~\cite{Fayet:1976et,Fayet:1977yc} leads to a rich phenomenology. Squarks (including antisquarks) and gluinos can be produced in pairs ($\gluino\gluino$, $\squark\squark$) or in association ($\squark \gluino$) and can decay through $\squark \to q \ninoone$ and $\gluino \to q \bar{q} \ninoone$ to the lightest neutralino, $\ninoone$, assumed to be the LSP.
Additional decay modes can include the production of charginos via $\squark\to q' \chinopm$ (where $\squark$ and $q'$ are of different flavour) and $\gluino\to q\bar{q}'\chinopm$.  Subsequent chargino decays to $W^{\pm}\ninoone$, depending on the decay modes of the $W$ bosons, can increase the jet multiplicity in these events.
 
This paper presents a search for these SUSY particles, using three strategies, in final states containing exclusively hadronic jets and large missing transverse momentum. The first, referred to as the `multi-bin search', extends the previous search from Ref.~\cite{0Lpaper2017} by simultaneously fitting the background expectations to the observed data yields in multiple event selection bins.
The second, referred to as the `BDT search', is a complementary analysis which uses boosted decision trees (BDTs) implemented in the TMVA framework \cite{TMVAref} for the event selection. The BDT search provides improved sensitivity to supersymmetric models in which gluinos decay via an intermediate chargino, by virtue of its highly optimised design and ability to exploit correlations between variables. A final strategy, referred to as the `model-independent search' uses a simple single-bin cut-and-count approach giving sensitivity to generic models characterised by the above final states.
The CMS Collaboration has set limits on similar models in Refs.~\cite{CMS:2015dbr,CMS:2017razor,Khachatryan:2016xvy,Sirunyan:2018vjp,Sirunyan:2017kqq,Sirunyan:2017cwe}.
 
In the search presented here, events with reconstructed high transverse momentum electrons or muons are rejected to reduce the background from events with neutrinos ($W \rightarrow e\nu,\mu\nu$) and to avoid any overlap with a complementary ATLAS search in final states with one lepton, jets and missing transverse momentum  \cite{1leptonPaper-last}. The selection criteria are optimised in the $(m(\gluino), m(\ninoone))$ and $(m(\squark), m(\ninoone))$ planes, (where $m(\gluino)$, $m(\squark)$ and $m(\ninoone)$ are the gluino, squark and the LSP masses, respectively) for simplified models  \cite{Alwall:2008ve,Alwall:2008ag,Alves:2011wf} in which all other supersymmetric particles are assigned masses beyond the reach of the LHC.
Although interpreted in terms of SUSY models, the results of this analysis can also constrain any model of new physics that predicts the production of jets in association with missing transverse momentum.
 
The paper is organised as follows. Section~\ref{sec:detector} describes the  ATLAS experiment and the data sample used for the search, and Section~\ref{sec:montecarlo} the Monte Carlo (MC) simulation samples used for background and signal modelling. The physics object reconstruction and identification are presented in Section \ref{sec:objects}.
The search is performed in signal regions which are defined in Section~\ref{sec:selection}.
Summaries of the background estimation methodology and corresponding systematic uncertainties are presented in Sections~\ref{sec:background} and \ref{sec:systematics}, respectively.
Results obtained by the search are reported in Section~\ref{sec:results}. Section~\ref{sec:conclusions} is devoted to a summary and conclusions.
\section{The ATLAS detector and data samples}
\label{sec:detector}
\interfootnotelinepenalty=10000
The ATLAS detector~\cite{Aad:2008zzm} is a multipurpose detector with a forward--backward symmetric cylindrical
geometry and nearly 4$\pi$ coverage in solid angle.\footnote{
ATLAS uses a right-handed coordinate system with its origin at the nominal
interaction point in the centre of the detector. The positive $x$-axis is defined by the direction from the interaction point to the centre
of the LHC ring, with the positive $y$-axis pointing upwards, while the beam direction defines the $z$-axis. Cylindrical coordinates $(r,\phi)$ are used in the transverse
plane, $\phi$ being the azimuthal angle around the $z$-axis. The transverse momentum $\pt$, the transverse energy $\ET$ and the missing transverse momentum are defined in the $x$--$y$ plane. The pseudorapidity $\eta$ is
defined in terms of the polar angle $\theta$ by $\eta=-\ln\tan(\theta/2)$ and the rapidity is defined as $y = (1/2)\ln[(E+p_z)/(E-p_z)]$ where $E$ is the energy and $p_z$ the longitudinal momentum of the object of interest.
}
The inner detector (ID) tracking system consists of pixel and silicon microstrip detectors
covering the pseudorapidity region $|\eta|<2.5$, surrounded by a transition radiation tracker, which improves electron identification over the region $|\eta|<2.0$.  The innermost
pixel layer, the insertable B-layer \cite{B-layerRef,Abbott:2018ikt}, was added between Run~1 and Run~2 of the LHC, at a radius of 33 mm around a new, narrower and thinner beam pipe.
The ID is surrounded by a thin superconducting solenoid providing an axial 2~T magnetic field and by
a fine-granularity lead/liquid-argon (LAr) electromagnetic calorimeter covering $|\eta|<3.2$.
A steel/scintillator-tile calorimeter provides hadronic coverage in
the central pseudorapidity range ($|\eta|<1.7$).
The endcap and forward calorimeters ($1.5<|\eta|<4.9$) are made of LAr active layers with either copper or tungsten as the absorber material for electromagnetic and hadronic measurements.
A muon spectrometer with an air-core toroid magnet system surrounds the calorimeters.
Three layers of high-precision tracking chambers
provide coverage in the range $|\eta|<2.7$, while dedicated chambers allow triggering in the region $|\eta|<2.4$.
 
The ATLAS trigger system \cite{ATLASTrigger2016Paper} consists of two levels; the first level is a hardware-based system, while the second is a software-based system called the high-level trigger. The events used by the search described in this paper were selected using a trigger logic that accepts events with a missing transverse momentum above $70$--$110 ~\GeV$, depending on the data-taking period.
The trigger is approximately 100\% efficient for the event selections considered in this search.
Auxiliary data samples used to estimate or validate the yields of $Z(\to\nu\bar\nu)$+jets background events were selected using triggers requiring at least one isolated photon ($\pt>120~\GeV$),  electron ($\pt>24~\GeV$) or muon ($\pt>20~\GeV$), for data collected in 2015. For the 2016--2018 data, these events were selected using triggers requiring at least one isolated electron or muon ($\pt>26~\GeV$) or photon ($\pt>140~\GeV$).
 
The data were collected by the ATLAS detector during 2015--2018 with a centre-of-mass energy of 13~\TeV\ and a 25~ns proton bunch crossing interval.
The average number of $pp$ interactions per bunch crossing (pile-up), $\langle\mu\rangle$, ranged from 13 in 2015 to around 38 in 2017--2018.
Application of beam, detector and data-quality criteria \cite{Aad:2700249} resulted in a total integrated luminosity of 139~\ifb. The uncertainty in the combined 2015--2018 integrated luminosity is 1.7\% \cite{ATLAS-CONF-2019-021}, obtained using the LUCID-2 detector \cite{LUCID2} for the primary luminosity measurements.
\section{Simulated event samples}
\label{sec:montecarlo}
 
Monte Carlo (MC) data samples are used by the search presented in this paper to optimise the selections, aid the estimation of backgrounds and assess the sensitivity to specific SUSY signal models.
 
Simplified SUSY model signal samples are used to describe the production of squarks and gluinos. The topologies considered include
squark-pair production, followed by the direct ($\squark \rightarrow q\ninoone$) or one-step ($\squark \rightarrow  q'\chinopm \rightarrow q'W \ninoone$) decays of squarks, shown in Figures~\ref{fig:feynman_directgrids}(a) and \ref{fig:feynman_directgrids}(b), and gluino-pair production, followed by the direct ($\gluino \rightarrow q\bar{q}\ninoone$) or one-step ($\gluino \rightarrow  q\bar{q}'\chinopm \rightarrow q\bar{q}'W \ninoone$) decays of gluinos, as shown in Figures~\ref{fig:feynman_directgrids}(c) and \ref{fig:feynman_directgrids}(d).
`One-step' decays refer to cases where the decays occur via one intermediate on-shell SUSY particle. An additional simplified model scenario in which squark pairs, gluino pairs, and squark--gluino pairs are produced inclusively is also considered. In this scenario, all production processes (gluino--gluino, squark--antisquark, squark--squark, antisquark--antisquark, squark--gluino and antisquark--gluino) are included, and the produced squarks and/or gluinos can follow the direct decays indicated in Figures~\ref{fig:feynman_directgrids}(a), \ref{fig:feynman_directgrids}(c) and \ref{fig:feynman_directgrids}(e), or decays of squarks via gluinos ($\tilde{q}\rightarrow q\tilde{g}$) and decays of gluinos via squarks ($\tilde{g}\rightarrow q\tilde{q}$) if kinematically possible. The branching ratios for these decays are calculated with the SUSY-HIT program~\cite{Djouadi:2006bz}.  The free parameters are $m(\ninoone)$ and $m(\squark)$ ($m(\gluino)$) for squark-pair (gluino-pair) production with direct decays of squark and gluinos. In the case of squark- or  gluino-pair production models with one-step decays, the free parameters are $m(\squark)$ or $m(\gluino)$, and either $m(\chinoonepm)$ (with fixed $m(\ninoone)$ = 60~\GeV) or  $m(\ninoone)$ (with $m(\chinoonepm)$ set equal to $(m(\gluino/\squark)+m(\ninoone))/2$). For models with inclusive production of squarks and gluinos both $m(\squark)$ and $m(\gluino)$ are varied, with $m(\ninoone)$ fixed to 0 \GeV, 995 \GeV~or 1495 \GeV. All other supersymmetric particles, including the squarks of the third generation, have their masses set such that the particles are effectively decoupled. Eightfold degeneracy of first- and second-generation squarks is assumed for the simplified models with direct decays of squarks, while fourfold degeneracy is assumed for the simplified models with one-step decays of squarks. The gluino is allowed to decay into four flavours  ($u$, $d$, $s$, $c$) of quarks in simplified models with gluino-pair production.
 
These samples were generated at tree level with up to two extra partons in the matrix
element (one extra parton for the models with inclusive production of both squarks and gluinos) using the \MGMCatNLOV{2.6.1} or 2.6.2 event generator~\cite{Alwall:2014hca}
interfaced to  \pythia~8.212 and \pythia~8.230~\cite{Sjostrand:2014zea}, respectively.
The CKKW-L merging scheme~\cite{Lonnblad:2011xx} was applied with a scale parameter that was set to a quarter of the mass of the gluino for $\gluino\gluino$ production or a quarter of the mass of the squark for $\squark\squark$ production in simplified models. In models with squark, gluino, and squark--gluino pairs, a quarter of the smaller of the gluino and squark masses was used for the CKKW-L merging scale.
The A14~\cite{A14tune} set of tuned parameters (tune) was used for initial/final-state radiation (ISR/FSR) and underlying-event parameters together with the NNPDF2.3LO~\cite{Ball:2012cx} parton distribution function (PDF) set.
 
Signal cross-sections are calculated to approximate next-to-next-to-leading order in the strong coupling constant, adding the resummation of soft gluon emission at next-to-next-to-leading-logarithm accuracy (approximate NNLO+NNLL)~\cite{Beenakker:2016lwe,Beenakker:2014sma,Beenakker:2013mva,Beenakker:2011sf,Beenakker:2009ha,Kulesza:2009kq,Kulesza:2008jb,Beenakker:1996ch}. The nominal cross-section and its uncertainty are derived using the PDF4LHC15\_mc PDF set, following the recommendations of Ref.~\cite{Butterworth:2015oua}, considering only first- and
second-generation squarks ($\tilde{u}$, $\tilde{d}$, $\tilde{s}$, $\tilde{c}$).
\begin{figure}[t]
\begin{center}
\subfigure[]{\includegraphics[width=0.24\textwidth]{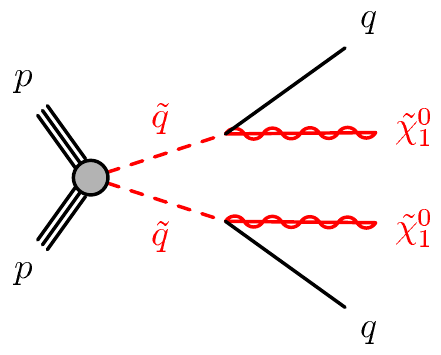}}
\subfigure[]{\includegraphics[width=0.24\textwidth]{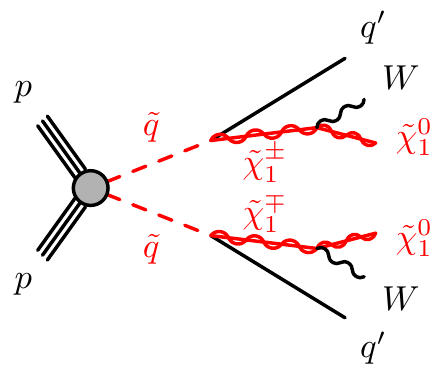}}
\subfigure[]{\includegraphics[width=0.24\textwidth]{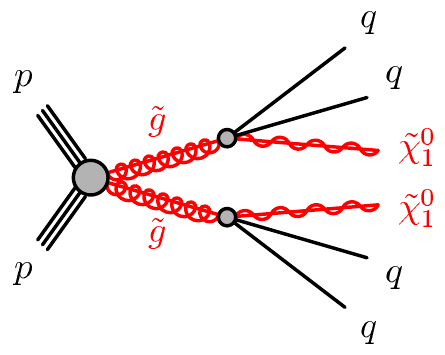}}
\subfigure[]{\includegraphics[width=0.24\textwidth]{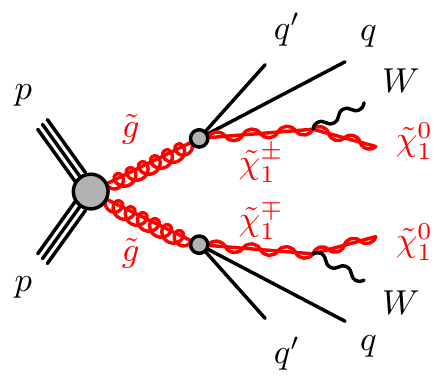}}\\
\subfigure[]{\includegraphics[width=0.24\textwidth]{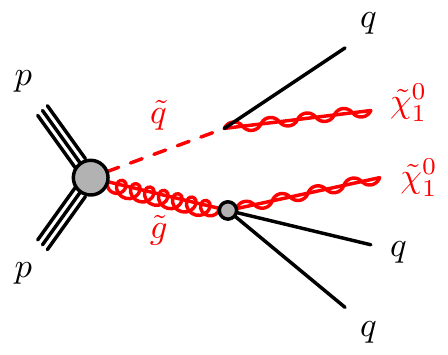}}
\end{center}
\caption{The decay topologies of (a, b) squark-pair production, (c, d) gluino-pair production and (e) squark--gluino production in simplified models with (a, c, e) direct decays of squarks and gluinos or (b, d) one-step decays of squarks and gluinos. }
\label{fig:feynman_directgrids}
\end{figure}
 
A summary of all SM background processes together with the MC event generators, cross-section calculation orders in $\alpha_{\textrm s}$, PDFs, parton shower and tunes used is given in Table~\ref{tab:montecarlo}. Further details of the generator configuration can be found in Ref.~\cite{0Lpaper2017}, with updates for $t\bar{t}$ modelling described in Ref.~\cite{ATL-PHYS-PUB-2018-009}. The most significant change in generator configuration with respect to Ref.~\cite{0Lpaper2017} relates to the simulation of the production of a photon in association with jets ($\gamma$+jets). This process is now simulated with \sherpa~2.2.2 with next-to-leading-order (NLO) cross-sections and the NNPDF3.0NNLO PDF set. Matrix elements are calculated for up to two partons at NLO and three or four additional partons at leading order (LO) using the \textsc{Comix} \cite{comix} and \textsc{Open Loops} \cite{openloops} matrix-element generators, and merged with the \sherpa\ parton shower \cite{sherpashower} using the ME+PS@NLO prescription \cite{mepsnlo}.
 
\begin{table}[t]
\scriptsize
\begin{center}
\renewcommand{\arraystretch}{1.4}
\hspace*{-0.3cm}\begin{tabular}{ l l c c c c }
\toprule
Physics process & Generator& Cross-section & PDF set & Parton shower & Tune \\
&& normalisation & & & \\
\midrule
$W(\rightarrow \ell\nu)$ + jets              & \sherpa~2.2.1~\cite{Gleisberg:2008ta}        & NNLO~\cite{Catani:2009sm}   &  NNPDF3.0NNLO~\cite{Ball:2014uwa}   &  \sherpa~\cite{Hoeche:2009rj}    & \sherpa \\
$Z/\gamma^{*}(\rightarrow \ell \bar \ell)$ + jets & \sherpa~2.2.1         & NNLO  &  NNPDF3.0NNLO   & \sherpa\      & \sherpa\\
$\gamma $ + jets & \sherpa~2.2.2         & NLO  &    NNPDF3.0NNLO   & \sherpa\   & \sherpa\\
$t\bar{t}$              & \textsc{Powheg-Box}~v2~\cite{powheg-box}   & NNLO+NNLL~\cite{Czakon:2013goa,Czakon:2011xx}                   &  NNPDF2.3LO~\cite{Ball:2012cx} &  \pythia~8.230~\cite{Sjostrand:2014zea}  & A14~\cite{ATL-PHYS-PUB-2014-021} \\
Single top ($Wt$-channel) & \textsc{Powheg-Box} v2  &  NNLO+NNLL~\cite{Kidonakis:2010ux,Kidonakis:2011wy}.  &  NNPDF2.3LO &   \pythia~8.230   & A14\\
Single top ($s$-channel)           & \textsc{Powheg-Box} v2  & NLO~\cite{Aliev:2010zk,Kant:2014oha}  &  NNPDF2.3LO &   \pythia~8.230   & A14\\
Single top ($t$-channel)           & \textsc{Powheg-Box} v2  & NLO  &  NNPDF2.3LO &   \pythia~8.230   & A14\\
$t\bar{t}+W/Z/H$       &  MG5\_aMC@NLO~2.2.3~\cite{Alwall:2014hca}  & NLO~\cite{Lazopoulos:2008de,Campbell:2012dh}  & NNPDF2.3LO & \pythia~8.210 & A14    \\
$t\bar{t}+WW$       &  MG5\_aMC@NLO~2.2.2  & NLO  & NNPDF2.3LO & \pythia~8.210 & A14    \\
$WW$, $WZ$, $ZZ$, $W\gamma$, $Z\gamma$    &  \sherpa~2.2.1       & NLO  &  NNPDF3.0NNLO & \sherpa\   & \sherpa \\
\bottomrule
\end{tabular}
\caption{The SM background MC simulation samples used in this paper. The generators, the order in $\alpha_{\textrm s}$ of cross-section calculations used for yield normalisation, PDF sets, parton showers and tunes used for the underlying event are shown. }
\label{tab:montecarlo}
\end{center}
\end{table}

For all SM background samples the response of the detector to particles was modelled with the full ATLAS detector simulation \cite{:2010wqa} based on \textsc{Geant4} \cite{Agostinelli:2002hh}.
Signal samples were prepared using a fast simulation based on a parameterisation of showers in the ATLAS electromagnetic and hadronic calorimeters \cite{ATL-PHYS-PUB-2010-013} coupled to \textsc{Geant4} simulations of particle interactions elsewhere. The {\textsc EvtGen}~v1.2.0 program~\cite{evtgen} was used to describe the properties of the $b$- and $c$-hadron decays in the signal samples, and the background samples except those produced with \sherpa~\cite{Gleisberg:2008ta}.
 
All simulated events were overlaid with multiple $pp$ collisions simulated with \pythia~8.186 using the A3 tune~\cite{A14tune} and the NNPDF2.3LO parton distribution functions~\cite{Ball:2012cx}.
The MC samples were generated with a variable number of additional $pp$ interactions (pile-up), and were reweighted to match the distribution of the mean number of interactions observed in data in 2015--2018.
 
 
\section{Object reconstruction and identification}
\label{sec:objects}
 
The reconstructed primary vertex of the event is required to be consistent with the luminous region and to have at least two associated tracks with $\pt > 500$~\MeV. When more than one such vertex is found, the vertex with the largest  $\sum \pt^2$ of the associated tracks is chosen.
 
Jet candidates are reconstructed using the anti-$k_{t}$ jet clustering algorithm~\cite{Cacciari:2008gp,Cacciari:2011ma} with a jet
radius parameter of $0.4$ starting from clusters of calorimeter cells~\cite{Topocluster_ATLAS}. The jets are corrected for energy from pile-up using the method described in Ref.~\cite{Cacciari:2007fd}: a contribution equal to the product of the jet area and the median energy density of the event is subtracted from the jet energy \cite{ATLAS-CONF-2013-083}.
Further corrections, referred to as the jet energy scale corrections, are derived from MC simulation and data, and are used to calibrate the average energies of jets to the scale of their constituent particles \cite{PERF-2016-04}.
Only corrected jet candidates with $\ourpt > 20~\GeV$ and $|\eta|<2.8$  are considered in this analysis.
An algorithm based on boosted decision trees, `MV2c10'~\cite{ATL-PHYS-PUB-2016-012,Aad:2015ydr}, is used to identify jets containing a $b$-hadron ($b$-jets), with an operating point corresponding to an efficiency of 77\%, and rejection factors of about 130 for jets originating from gluons and light-flavour quarks (light jets) and about 6 for jets induced by charm quarks, determined using MC simulated $t\bar{t}$ events. Candidate $b$-jets are required to possess $\ourpt > 50~\GeV$ and $|\eta|<2.5$. In order to reduce the number of jets generated by pile-up, a significant fraction of the tracks associated with each jet must have an origin compatible with the primary vertex. This is enforced by using the jet vertex tagger (JVT) output using the momentum fraction of such tracks~\cite{JVT_CONF}. The requirement JVT $> 0.59$ is only applied to jets with $\pt < 120~\GeV$ and $|\eta| < 2.5$, while in the region $2.4<|\eta|<2.5$, a looser value, JVT $> 0.11$ is used. No JVT requirement is applied to jets in the region $2.5<|\eta|<2.8$.
Events with jets originating from detector noise and non-collision background are rejected if jets satisfying the jet vertex tagging criteria and passing jet--lepton ambiguity resolution (see below) fail to satisfy the `LooseBad' quality criteria, or if at least one of the two leading jets fails to satisfy the `TightBad' quality criteria, both of which are described in Ref.~\cite{Aad:2013zwa}. The application of these criteria reduces the data sample by $\sim 9\%$ and maintains an efficiency for simulated $Z$+jets events of 99.5\%.
 
Two different classes of reconstructed lepton candidates (electrons or muons) are used in the analyses presented here. When selecting samples for the search, events containing a `baseline' electron or muon are rejected. The selections applied to identify baseline leptons are designed to maximise the efficiency with which $W$+jets and top quark background events are rejected. When selecting events for the purpose of estimating residual $W$+jets and top quark backgrounds, additional requirements are applied to leptons to ensure greater purity of these backgrounds. These leptons are referred to as `high-purity' leptons below and form a subset of the baseline leptons.
 
Baseline muon candidates are formed by combining information from the muon spectrometer and inner detector as described in Ref.~\cite{MuonPerfRun2} and are required to possess $\ourpt > 6~\GeV$ and $|\eta| <2.7$.  Baseline muon candidates must satisfy `Medium' identification criteria~\cite{MuonPerfRun2}. High-purity muon candidates must also have a transverse impact parameter significance of $|d_0^{\mathrm{PV}}|/\sigma(d_0^{\mathrm{PV}}) < 3$ relative to the primary vertex, and a longitudinal impact parameter satisfying $|z_0^{\mathrm{PV}} \mathrm{sin}(\theta)|<$ 0.5~mm. Furthermore, high-purity candidates must satisfy the `FCTight' isolation requirements described in Ref.~\cite{MuonPerfRun2}, which rely on tracking- and calorimeter-based variables and implement a set of $\eta$- and $\pt$-dependent criteria.
 
Baseline electron candidates are reconstructed from an electromagnetic calorimeter energy deposit matched to an ID track \cite{Aad:2019tso} and are required to satisfy $\ourpt > 7~\GeV$, $|\eta| < 2.47$ (including the calorimeter transition region $1.37<|\eta|<1.52$), and the `Loose' likelihood-based identification criteria described in Refs.~\cite{Aad:2019tso,Aaboud:2019ynx}.
High-purity electron candidates must also satisfy `Tight' selection criteria described in Refs.~\cite{Aad:2019tso,Aaboud:2019ynx}. They are also required to satisfy $|d_0^{\mathrm{PV}}|/\sigma(d_0^{\mathrm{PV}}) < 5$, $|z_0^{\mathrm{PV}} \mathrm{sin}(\theta)|<0.5$~mm, and isolation requirements similar to those applied to high-purity muons~\cite{Aaboud:2016vfy}.
 
After the selections described above, ambiguities between electrons and muons are resolved to avoid double counting and/or remove non-isolated leptons: the electron is discarded if a baseline electron and a baseline muon share the same ID track.
Ambiguities between candidate jets with $|\eta|<2.8$ and leptons are resolved as follows: first, any such jet candidate lying within a distance $\Delta
R\equiv\sqrt{(\Delta y)^2+(\Delta\phi)^2}=0.2$ of a baseline electron is discarded.
Additionally, if a baseline electron or muon and a jet are found within $ \Delta R <\textrm{min}(0.4, 0.04 + 10~\GeV/p_{\textrm T}^{e/\mu}$), it is interpreted as a jet and the nearby electron or muon candidate is discarded.
Finally, if a baseline muon and jet are found within $\Delta R < 0.2$, and the jet satisfies $N_\text{trk} < 3$ (where $N_\text{trk}$ refers to the number of tracks with $\pT>500~\MeV$ that are associated with the jet), it is treated as a muon and the overlapping jet is ignored.
This criterion rejects jets consistent with final-state radiation or hard bremsstrahlung. The ambiguity resolution procedure follows that used in previous ATLAS analyses seeking evidence for SUSY particles.
 
Reconstructed photons are used in the measurement of missing transverse momentum
as well as in the control region used to constrain the
$Z$+jets background, as explained in
Section~\ref{sec:background}.
These photon candidates are
required to satisfy $\ourpt > 25~\GeV$ and $|\eta| < 2.37$ (excluding the transition region $1.37 < |\eta| < 1.52$ between the barrel and endcap EM calorimeters), to satisfy photon
shower shape and electron rejection criteria, and to be
isolated~\cite{Aad:2019tso,ATLAS:photon_Run1,PERF-2017-02}. The reduced $\eta$ range for photons
is chosen to avoid a region of coarse granularity at high $\eta$ where discrimination between
photon and $\pi^{0}$ candidates worsens. Ambiguities between candidate jets and photons (when used in the event selection) are resolved by discarding any jet candidates lying within $\Delta R$ = 0.4 of a photon candidate.
Additional selections to remove ambiguities between electrons or muons and photons are applied such that a photon is discarded if it is within $\Delta R$ = 0.4 of a baseline electron or muon.
 
The measurement of the missing transverse momentum vector
$\textbf{\emph{p}}\mathrm{^{miss}_T}$ (and its magnitude $\ourmagptmiss$)
is based on the calibrated transverse momenta of all electron, muon, jet and photon
candidates, and all tracks originating from the primary vertex that are not associated with the preceding reconstructed objects~\cite{PERF-2016-07,ATLAS-CONF-2018-023}.

Corrections derived from data control samples are applied to simulated events to account for differences between data and simulation for the lepton and photon trigger and reconstruction efficiencies, the lepton momentum/energy scale and resolution, the jet vertex tagger, and the efficiency and mis-tag rate of the $b$-tagging algorithm.
 
\section{Event selection and signal regions definitions}
\label{sec:selection}
 
Due to the high mass scale expected for the SUSY models considered in this study, the `effective mass', $\meff$, defined to be the scalar sum of \met{} and the transverse momenta of all jets with $\pt > 50~\GeV$, is a powerful discriminant between the signal and most SM backgrounds. In some regions, when selecting events with at least $N_\mathrm{j}$ jets, $\meff(N_\mathrm{j})$ is calculated using the transverse momenta of the leading $N_\mathrm{j}$ jets with $\pt > 50~\GeV$ and \met{}. Only jets with $\pt > 50~\GeV$ are used directly to select events in the search presented in this paper, although jets with lower \pt{} are taken into account indirectly through their contribution to \met{} and through their use when rejecting noise and non-collision background events, as explained above in Section~\ref{sec:objects}.
 
Following the event reconstruction described in Section~\ref{sec:objects}, a common set of preselection criteria is used in this search. Events are discarded if a baseline electron (muon) with $\pt>7~(6)~\GeV$ remains after resolving the ambiguities between the objects, or if they contain a jet
failing to satisfy quality selection criteria designed to suppress detector noise and non-collision backgrounds (described in Section~\ref{sec:objects}). Events are also rejected if no second jet with $\pt >50~\GeV$ is found, the leading jet $\pt$ is smaller than 200~\GeV, the missing transverse momentum in the event is smaller than 300~\GeV, or the effective mass is smaller than 800~\GeV. In addition, the selection requires the smallest azimuthal separation between the $\textbf{\emph{p}}\mathrm{^{miss}_T}$ and the momenta of the leading two or three jets, $\Delta\phi(j_{1,2,(3)},\textbf{\emph{p}}\mathrm{^{miss}_T})_{\textrm{ min}}$, to be greater than 0.2.  The requirement is applied to the third leading jet whenever such a jet is present in the event. A summary of these preselection criteria is given in Table~\ref{tab:common_preselection}.
The remaining events are then analysed with three complementary search strategies,
which all require the presence of jets and significant missing transverse momentum.
 
\begin{table}[htbp]
\footnotesize
\begin{center}
\renewcommand{\arraystretch}{1.4}
\begin{tabular}{l c}
\toprule
Lepton veto &No baseline electron (muon) with $\pt>$7 (6)~\GeV \\ \midrule
\met [\GeV]  & $>$ 300\\
$\pt(j_1)$ [\GeV]  & $>$ 200 \\
$\pt(j_2)$ [\GeV]  & $>$ 50 \\
$\Delta\phi(j_{1,2,(3)},\textbf{\emph{p}}\mathrm{^{miss}_T})_{\textrm{ min}}$&   $>$ 0.2 \\
$\meff$ [\GeV]  & $>$ 800 \\ \bottomrule
 
\end{tabular}
\end{center}
\caption{Summary of common preselection criteria used for the search presented in this paper. }
\label{tab:common_preselection}
\end{table}
 
To search for a possible signal, selection criteria are defined to
enhance the expected signal yield relative to the SM
backgrounds. Signal regions (SRs) are defined using the MC simulation
of SUSY signals and the SM background processes. The SRs are optimised to
maximise the expected 95\% CL exclusion reach in the signal model parameter spaces considered.
In order to maximise the sensitivity in the $(m(\gluino),m(\squark))$ plane, a variety of signal regions are defined.
Squarks typically produce at least one jet in their decays, for instance through $\squark \to q \ninoone$, while gluinos typically produce at least two jets, for instance through $\gluino\to q q \ninoone$. Processes contributing to $\squark\squark$ and $\gluino\gluino$ final states therefore lead to events containing at least two or four jets, respectively. Decays of heavy SUSY and SM particles (for instance $W$ bosons) produced in longer $\squark$ and $\gluino$ decay cascades tend to further increase the jet multiplicity in the final state.
To target different SUSY particle production scenarios, signal regions with different jet multiplicity requirements and either specific ranges of kinematic variables (in the multi-bin search) or values of the BDT output variable (in the BDT search) are defined.
An additional set of single-bin signal regions used for a model-independent presentation of the results is also defined (in the model-independent search). All signal regions applied in these three search strategies are summarised in the following.

\subsection{The multi-bin search}
 
In this search strategy, three sets of signal regions targeting different scenarios with direct decays of squarks and gluinos are defined: the MB-SSd (`multi-bin squark-squark-direct') and MB-GGd (`multi-bin gluino-gluino-direct') regions target scenarios with large mass difference between the pair-produced squarks or gluinos and the lightest neutralino, respectively,
while the MB-C (`multi-bin compressed') regions target scenarios with small mass difference between the pair-produced squarks or gluinos and the \ninoone.
Events are assigned to three sets of mutually exclusive signal regions based on the jet multiplicity, the effective mass and the missing transverse momentum significance, defined as \met/$\sqrt{\HT}$, where $\HT$ is calculated as a scalar sum of transverse momenta of all jets with  $\pt >50~\GeV$ and $|\eta|<2.8$. This variable is used to suppress backgrounds in which jet energy mismeasurement generates missing transverse momentum, and was found to enhance sensitivity to models characterised by $\squark\squark$ production. The signal regions are mutually exclusive within any given set, but can overlap with signal regions from other sets.
 
After preselecting events as in Table~\ref{tab:common_preselection}, the following selection criteria are applied for the three sets of signal regions, to further suppress the background processes.
At least two jets with $|\eta| <$ 2 are required for MB-SSd regions, where the \pt\ of the sub-leading jet must be greater than 100~\GeV. The MB-C regions rely on the selection of an energetic jet with $\pt > 600~\GeV$, which could be generated by QCD ISR. In the MB-GGd  regions, at least four jets with $\pt > 100~\GeV$, and  $|\eta| < 2$ are required. The smallest azimuthal separation between the $\textbf{\emph{p}}\mathrm{^{miss}_T}$ vector and (i) the momenta of the three leading jets, $\Delta\phi(j_{1,2,(3)},\textbf{\emph{p}}\mathrm{^{miss}_T})_{\textrm{ min}}$, and (ii) the remaining jets with $\pt>50~\GeV$ in the event, $\Delta\phi(j_{i>3},\textbf{\emph{p}}\mathrm{^{miss}_T})_{\textrm{ min}}$, is required to be greater than 0.4 and 0.2, respectively. In MB-SSd, tighter requirements of 0.8 and 0.4, respectively, are applied. These requirements reduce the background from multi-jet processes, where a jet can be mismeasured and generate missing transverse momentum that points along the axis of the jet.
In the regions with at least four jets in the final state, jets from signal processes are distributed isotropically.
The aplanarity variable $A$, defined by $A = 3/2 \lambda_3$, where $\lambda_3$ is the smallest eigenvalue of the normalised momentum tensor of the jets~\cite{PhysRevD.1.1416}, is maximised by such topologies and is therefore used to select events in the MB-GGd regions, where a requirement $A>0.04$ is applied.
 
The missing transverse momentum significance \met/$\sqrt{\HT}$ is required to be greater than $10~\GeV^{1/2}$ and $\meff$~to be greater than $1000~\GeV$ in all signal regions except in MB-C, where a tighter, $\meff >1600~\GeV$, requirement is applied.
An overview of the signal region preselection criteria applied to the MB-SSd, MB-GGd and MB-C regions is presented in Table~\ref{tab:multibin-preselection}.
 
\begin{table}[htbp]
\footnotesize
\begin{center}
\renewcommand{\arraystretch}{1.4}
\begin{tabular}{ l  c  c  c }
\toprule
& MB-SSd &  MB-GGd & MB-C \\ \midrule
$N_\mathrm{j} $               &   $\geq$ 2    &    $\geq$ 4     &   $\geq$ 2   \\
$\pt(j_1)$ [\GeV]        &  $>$  200   &  $>$  200  & $>$  600 \\
$\pt(j_{i=2,...,N_\mathrm{j_{ min}}})$ [\GeV]        &  $>$  100   & $>$   100  & $>$  50  \\
$|\eta(j_{i=1,...,N_\mathrm{j_{ min}}})|$          &    $<$  2.0  &   $<$  2.0  &   $<$  2.8 \\
$\Delta\phi(j_{1,2,(3)},\textbf{\emph{p}}\mathrm{^{miss}_T})_{\textrm{ min}}$         &  $>$  0.8       &  $>$  0.4        & $>$   0.4 \\
$\Delta\phi(j_{i>3},\textbf{\emph{p}}\mathrm{^{miss}_T})_{\textrm{ min}}$        &  $>$   0.4       & $>$    0.4        & $>$    0.2 \\
Aplanarity         &    -    &    $>$    0.04     &  -   \\
\met/$\sqrt{\HT}$ [\GeV$^{1/2}$] &   $>$   10  &   $>$   10 &    $>$  10 \\
\meff\  [\GeV] &   $>$ 1000  &  $>$ 1000  &  $>$ 1600\\  \bottomrule
\end{tabular}
\end{center}
\caption{Summary of preselection criteria used for the multi-bin search.}
\label{tab:multibin-preselection}
\end{table}
 
Following these selections, the three sets of signal regions are defined with selections based upon bins in $\meff$, \met/$\sqrt{\HT}$ and $N_{\mathrm{j}}$, to maximise the sensitivity of the search in the $(m(\squark),m(\ninoone))$ or $(m(\gluino),m(\ninoone))$ planes. The MB-SSd regions are separated into two jet multiplicity bins, up to six bins in \meff\ and up to four bins in \met/$\sqrt{\HT}$, giving a total of 24 signal regions. In the lower jet multiplicity bin ($N_{\mathrm{j}}=[2, 3]$), tighter requirements are applied to the transverse momenta of the leading and sub-leading jets such that $\pt(j_{i=1,2})>250~\GeV$. In order to reduce the total number of signal regions without significant loss of search power, some bins are merged, as represented schematically in Table~\ref{tab:MB-SSd-bins}.
The MB-GGd signal regions are defined by six bins in \meff\
and three bins in \met/$\sqrt{\HT}$,
as shown in Table~\ref{tab:MB-GGd-bins}.
The MB-C signal regions are defined by three bins in jet multiplicity,
three bins in \meff\
and two bins in \met/$\sqrt{\HT}$,
as shown in Table~\ref{tab:MB-C-bins}.
 
\begin{table}[h]
\footnotesize
\begin{center}
\renewcommand{\arraystretch}{1.4}
\begin{tabu}{|c|c|[1pt]c|c|c|c|c|c|}
\toprule
\multicolumn{2}{|c|[1pt]}{\multirow{2}{*}{$N_{\mathrm{j}}=[2, 3]$, $\pt(j_{i=1,2})>250~\GeV$} }& \multicolumn{6}{c|}{\meff\ [\TeV]} \\ \cline{3-8}
\multicolumn{1}{|c}{}  & & [1.0, 1.6) & [1.6, 2.2) & [2.2, 2.8) & [2.8, 3.4) & [3.4, 4.0) &  [4.0, $\infty$) \\ \tabucline[1pt]{1-8}
\multirow{4}{*}{\met/$\sqrt{\HT}$  [\GeV$^{1/2}$]}  & [10, 16) & &  \multicolumn{5}{c|}{  } \\ \cline{2-8}
& [16, 22) & & & & \multicolumn{3}{c|}{  } \\  \cline{2-8}
& [22, 28) & & & & &$N_{\mathrm{j}}=[2, \infty)$ &$N_{\mathrm{j}}=[2, \infty)$ \\ \cline{2-2} \cline{7-8}
& [28, $\infty$) & & & & &$N_{\mathrm{j}}=[2, \infty)$ &$N_{\mathrm{j}}=[2, \infty)$  \\ \toprule
\toprule
\multicolumn{2}{|c|[1pt]}{\multirow{2}{*}{$N_{\mathrm{j}}=[4, \infty)$} }& \multicolumn{6}{c|}{\meff\ [\TeV]}  \\ \cline{3-8}
\multicolumn{1}{|c}{}  & & [1.0, 1.6) & [1.6, 2.2) & [2.2, 2.8)  &  \multicolumn{3}{c|}{[2.8, $\infty$)}  \\  \tabucline[1pt]{1-8}
\multirow{3}{*}{\met/$\sqrt{\HT}$  [\GeV$^{1/2}$]}  & [10, 16) & &  \multicolumn{5}{c|}{  }   \\ \cline{2-8}
& [16, 22) & & & &  \multicolumn{3}{c|}{  }  \\  \cline{2-8}
& [22, $\infty$) & & & &  \multicolumn{3}{c|}{$\meff=[2.8, 3.4)$}   \\  \bottomrule
\end{tabu}
\end{center}
 
\caption{Summary of the bin boundaries for the MB-SSd signal regions. An empty cell indicates that the corresponding bin uses only the selection criteria specified at the top of the column and to the left of the row. A non-empty cell indicates the use of special selection criteria, as specified by the entry. For each jet multiplicity bin (($N_{\mathrm{j}}=[2, 3]$ and $N_{\mathrm{j}}=[4, \infty)$), the highest bins in \meff\ and \met/$\sqrt{\HT}$, respectively, are inclusive in that variable. In order to guarantee sufficient event yields in the highest four \meff\ and \met/$\sqrt{\HT}$ bins of the upper ($N_{\mathrm{j}}=[2, 3]$) table, no upper limits on $N_{\mathrm{j}}$ are imposed, as indicated in the relevant entries. As a result of this, in order to remove overlap with the highest \meff\ and \met/$\sqrt{\HT}$ bin of the lower ($N_{\mathrm{j}}=[4, \infty)$) table, a requirement that $\meff=[2.8, 3.4)$ is imposed, as indicated in the relevant entry.}
\label{tab:MB-SSd-bins}
\end{table}

\begin{table}[h]
\footnotesize
\begin{center}
\renewcommand{\arraystretch}{1.4}
\begin{tabu}{|c|c|[1pt]c|c|c|c|c|c|}
\toprule
\multicolumn{2}{|c|[1pt]}{\multirow{2}{*}{$N_{\mathrm{j}}= [4, \infty)$} }& \multicolumn{6}{c|}{\meff\ [\TeV]} \\ \cline{3-8}
\multicolumn{1}{|c}{}  & & [1.0, 1.6) & [1.6, 2.2) & [2.2, 2.8) & [2.8, 3.4) & [3.4, 4.0) &  [4.0, $\infty$) \\  \tabucline[1pt]{1-8}
\multirow{3}{*}{\met/$\sqrt{\HT}$  [\GeV$^{1/2}$]}  & [10, 16) & & & & & &\\ \cline{2-8}
& [16, 22) & & & & & & \\  \cline{2-8}
& [22, $\infty$) & & & & & & \\ \bottomrule
\end{tabu}
\end{center}
\caption{Summary of the bin boundaries for the MB-GGd signal regions. An empty cell indicates that the corresponding bin uses only the selection criteria specified at the top of the column and to the left of the row. The highest bin for each variable is inclusive in that variable.}
\label{tab:MB-GGd-bins}
\end{table}
 
\begin{table}[h]
\footnotesize
\begin{center}
\renewcommand{\arraystretch}{1.4}
\begin{tabu}{|c|c|[1pt]c|c|c|}
\toprule
\multicolumn{2}{|c|[1pt]}{\multirow{2}{*}{$N_{\mathrm{j}}= [2, 3]$; 4; $[5, \infty)$} }& \multicolumn{3}{c|}{\meff\ [\TeV]} \\ \cline{3-5}
\multicolumn{1}{|c}{}  & & [1.6, 2.2) & [2.2, 2.8) &  [2.8, $\infty$) \\  \tabucline[1pt]{1-5}
\multirow{2}{*}{\met/$\sqrt{\HT}$  [\GeV$^{1/2}$]}  & [16, 22) & & & \\ \cline{2-5}
& [22, $\infty$) & & &  \\ \bottomrule
\end{tabu}
\end{center}
\caption{ Summary of the bin boundaries for the MB-C signal regions. An empty cell indicates that the corresponding bin uses only the selection criteria specified at the top of the column and to the left of the row. The highest bin for each variable is inclusive in that variable.}
\label{tab:MB-C-bins}
\end{table}
 
\clearpage
 
\subsection{The BDT search}
\label{bdtSR}
 
This search strategy is applied separately through two sets of signal regions targeting models with gluino-pair production with direct (BDT-GGd) or one-step (BDT-GGo) $\gluino$ decays. In each set, events are separated into four categories, depending on the mass difference $\Delta m(\gluino, \ninoone)$ in the target model. A dedicated BDT discriminant is used in each signal region, giving eight independently trained BDTs in total, to obtain optimum sensitivity to the models targeted by each SR. The signal regions are listed in Table \ref{tab:BDT-SRs}, with the values of $\Delta m(\gluino, \ninoone)$ targeted by each of the SRs indicated in the last rows of the table.
The signal regions are not mutually exclusive and hence cannot be combined statistically.
 
\begin{table}[htbp]
\footnotesize
\begin{center}
\renewcommand{\arraystretch}{1.4}
\begin{tabular}{|l|c|c|c|c|} \hline
& BDT-GGd1   & BDT-GGd2 & BDT-GGd3 & BDT-GGd4     \\ \hline
$N_{\mathrm{j}}$                & \multicolumn{4}{c|}{$\geq$ 4}     \\ \hline
$\Delta\phi(j_{1,2,(3)},\textbf{\emph{p}}\mathrm{^{miss}_T})_{\textrm{ min}}$  & \multicolumn{4}{c|}{ $>$ 0.4}           \\ \hline
$\Delta\phi(j_{i>3},\textbf{\emph{p}}\mathrm{^{miss}_T})_{\textrm{ min}}$ & \multicolumn{4}{c|}{ $>$  0.4}          \\ \hline
$\met$/$\meff(N_\mathrm{j})$  & \multicolumn{4}{c|}{ $>$ 0.2}       \\ \hline
$\meff $ [\GeV] & \multicolumn{2}{c|}{ $>$  1400}   & \multicolumn{2}{c|}{ $>$  800}       \\ \hline
BDT score                  & $>$   0.97 & $>$   0.94 & $>$   0.94 & $>$   0.87 \\ \hline
$\Delta m(\gluino, \ninoone)$ [\GeV]   & 1600--1900 & 1000--1400 & 600--1000 & 200--600  \\ \hline
\end{tabular}
 
\vspace*{0.5cm}
 
\begin{tabular}{|l|c|c|c|c|c|} \hline
&  BDT-GGo1 & BDT-GGo2 & BDT-GGo3 & BDT-GGo4  \\ \hline
$N_{\mathrm{j}}$                          &  \multicolumn{2}{c}{ $\geq$   6} & \multicolumn{2}{|c|}{ $\geq$   5}\\ \hline
$\Delta\phi(j_{1,2,(3)},\textbf{\emph{p}}\mathrm{^{miss}_T})_{\textrm{ min}}$  & \multicolumn{3}{c|}{ $>$  0.4}         &  $>$  0.2   \\ \hline
$\Delta\phi(j_{i>3},\textbf{\emph{p}}\mathrm{^{miss}_T})_{\textrm{ min}}$  &   \multicolumn{3}{c|}{ $>$  0.4} &  $>$  0.2    \\ \hline
$\met$/$\meff(N_\mathrm{j})$ & \multicolumn{4}{c|}{ $>$ 0.2} \\ \hline
$\meff $ [\GeV]  & \multicolumn{2}{c}{ $>$  1400}   & \multicolumn{2}{|c|}{ $>$  800}      \\ \hline
BDT score                 &  $>$   0.96 & $>$   0.87 & $>$   0.92 & $>$   0.84 \\ \hline
$\Delta m(\gluino, \ninoone)$ [\GeV]   & 1400--2000 &  1200--1400  & 600--1000  & 200--400  \\ \hline
\end{tabular}
\end{center}
\caption{Signal region selections for the BDT search with the benchmark signal model parameters ($\Delta m(\gluino, \ninoone)$) used in the optimisation, for (top) direct and (bottom) one-step gluino decays, respectively. In the BDT-GGo regions the targeted models are characterised by $m(\chinoonepm)=(m(\tilde{g})+m(\ninoone))/2$.}
\label{tab:BDT-SRs}
\end{table}
 
After applying the preselection criteria from Table~\ref{tab:common_preselection}, additional selection criteria are applied to the BDT-GGd and BDT-GGo signal regions to further distinguish between signal and background processes, prior to the final selections based on the BDT discriminants. All BDT-GGd regions require the presence of at least four jets, with  $\Delta\phi(j_{1,2,(3)},\textbf{\emph{p}}\mathrm{^{miss}_T})_{\textrm{ min}} > 0.4$, $\Delta\phi(j_{i>3},\textbf{\emph{p}}\mathrm{^{miss}_T})_{\textrm{ min}} > 0.4$ and $\met / \meff(4j) > 0.2$ to further suppress the multi-jet background. Additionally, $\met / \meff(N_\mathrm{j}) > 0.2$ is required in all regions. The BDT-GGo regions require the presence of at least six (BDT-GGo1 and BDT-GGo2) or five (BDT-GGo3 and BDT-GGo4) jets, with $\Delta\phi(j_{1,2,(3)},\textbf{\emph{p}}\mathrm{^{miss}_T})_{\textrm{ min}} > 0.4$ and $\Delta\phi(j_{i>3},\textbf{\emph{p}}\mathrm{^{miss}_T})_{\textrm{ min}} > 0.4$ in all regions except in BDT-GGo4, where looser requirements of $\Delta\phi(j_{1,2,(3)},\textbf{\emph{p}}\mathrm{^{miss}_T})_{\textrm{ min}} > 0.2$ and $\Delta\phi(j_{i>3},\textbf{\emph{p}}\mathrm{^{miss}_T})_{\textrm{ min}} > 0.2$ are applied.
To select events close to the kinematic regions of interest, $\meff >1400~\GeV$ is required in the BDT-GGd1, BDT-GGd2, BDT-GGo1 and BDT-GGo2 regions, and  $\meff >800~\GeV$ in the BDT-GGd3, BDT-GGd4, BDT-GGo3 and BDT-GGo4 regions.
 
For the final selection in each of the eight signal regions, a dedicated BDT is trained for events satisfying the dedicated selection criteria for the signal region, listed above. In order to increase the size of the signal MC samples used for BDT training, and at the same time keep the output performance stable, signal MC events with similar mass differences between \gluino\ and \ninoone (leading to similar event kinematics), normalised to their corresponding cross-sections, are combined into a single sample for training. All MC samples for the SM background processes listed in Table~\ref{tab:montecarlo} are taken into account. The multi-jet background events are not used in the BDT training since the contribution from these processes is expected to be negligible. All MC events used in the BDT training are randomly divided into two sets. In order to avoid a decrease of the total MC sample size to a half of the full dataset, the BDT training is performed on both sets of events, following the procedure described in Refs.~\cite{HIGG-2016-29,HIGG-2018-04}. The BDT score calculated using one set of events is applied to the other set, which is then used as input for the signal and background evaluation. The data events used for the evaluation are also randomly divided into two sets. Up to 12 variables are selected among  \met, \meff, aplanarity $A$, and the $\pt$ and $\eta$ of selected jets, and are then used in the training for the eight signal regions.  The selections based on the BDT scores providing the maximal expected sensitivity for a benchmark signal model are then used to define the signal regions. The aplanarity is particularly important for enabling the BDT discriminants to separate signal and background for models with large $\Delta m(\gluino, \ninoone)$, because in such models signal events are more spherical than the background.

\subsection{Model-independent search}
\label{single-binSR}
 
In addition to the multi-bin and BDT searches described above, several signal regions, optimised to maximise sensitivity to generic SUSY models with specific jet multiplicities in the final state, are defined. These comprise the model-independent search. These signal regions rely on the single-bin approach described in Ref.~\cite{0Lpaper2017}.
After applying the preselection criteria of Table~\ref{tab:common_preselection}, ten inclusive SRs characterised by increasing minimum jet multiplicity are defined, listed in Tables~\ref{tab:SB1def} and \ref{tab:SB2def}.  The signal region definitions follow those used for the multi-bin search, but with the requirements on $\meff$, $N_\mathrm{j}$ and \met/$\sqrt{\HT}$ made inclusive. Some of these SRs require the same jet multiplicity, but are distinguished by requiring higher \meff\, values. These regions overlap, and therefore cannot be combined statistically.

\begin{table}[htbp]
\footnotesize
\begin{center}
\renewcommand{\arraystretch}{1.4}
\hspace*{-0.5cm}\begin{tabular}{| l | c | c | c | c |c | c |}
\hline
& SR2j-1600& SR2j-2200 & SR2j-2800 & SR4j-1000 & SR4j-2200 & SR4j-3400 \\ \hline
$N_{\mathrm{j}} $         &  \multicolumn{3}{c|}{$\geq$ 2}             & \multicolumn{3}{c|}{$\geq$ 4}         \\  \hline
$\pt(j_1)$ [\GeV] & $>$ 250 & $>$ 600 & $>$ 250           &  \multicolumn{3}{c|}{ $>$ 200}        \\  \hline
$\pt(j_{i=2,...,N_\mathrm{j_{ min}}})$ [\GeV] & $>$ 250 & $>$ 50  & $>$ 250           &   \multicolumn{3}{c|}{ $>$ 100}        \\  \hline
$|\eta(j_{i=1,...,N_\mathrm{j_{ min}}})|$   &   $<$  2.0 & $<$  2.8 & $<$  1.2          &  \multicolumn{3}{c|}{ $<$  2.0}      \\  \hline
$\Delta\phi(j_{1,2,(3)},\textbf{\emph{p}}\mathrm{^{miss}_T})_{\textrm{ min}}$         & $>$  0.8 & $>$  0.4 & $>$  0.8 & \multicolumn{3}{c|}{ $>$  0.4}   \\  \hline
$\Delta\phi(j_{i>3},\textbf{\emph{p}}\mathrm{^{miss}_T})_{\textrm{ min}}$        &  $>$  0.4 &  $>$  0.2 &  $>$  0.4 &   \multicolumn{3}{c|}{  $>$  0.2} \\  \hline
Aplanarity           &     \multicolumn{3}{c|}{-}    &  \multicolumn{3}{c|}{$>$  0.04} \\  \hline
\met/$\sqrt{\HT}$ [\GeV$^{1/2}$]  &   \multicolumn{5}{c|}{$ >$ 16} &  $ >$  10 \\  \hline
\meff\ [\GeV]&   $>$ 1600  &  $>$ 2200  &  $>$ 2800 &  $>$ 1000 &  $>$ 2200  &  $>$ 3400 \\  \hline
 
\end{tabular}
\end{center}
\caption{Selection criteria used for model-independent search signal regions with minimum jet
multiplicities up to four.}
\label{tab:SB1def}
\end{table}

\begin{table}[htbp]
\footnotesize
\begin{center}
\renewcommand{\arraystretch}{1.4}
\begin{tabular}{| l | c | c | c | c |}
\hline
& SR5j-1600 &  SR6j-1000 & SR6j-2200 & SR6j-3400 \\ \hline
$N_{\mathrm{j}}$         &  $\geq$ 5             & \multicolumn{3}{c|}{ $\geq$ 6}         \\  \hline
$\pt(j_1)$ [\GeV] &  $>$ 600           &  \multicolumn{3}{c|}{ $>$ 200}        \\  \hline
$\pt(j_{i=2,...,N_\mathrm{j_{ min}}})$ [\GeV] &  $>$ 50            &   \multicolumn{3}{c|}{$>$ 75}        \\  \hline
$|\eta(j_i=1,...,N_\mathrm{j_{ min}})|$ &     $<$ 2.8          &  \multicolumn{3}{c|}{  $<$  2.0}      \\  \hline
$\Delta\phi(j_{1,2,(3)},\textbf{\emph{p}}\mathrm{^{miss}_T})_{\textrm{ min}}$        &  \multicolumn{4}{c|}{$>$   0.4}   \\  \hline
$\Delta\phi(j_{i>3},\textbf{\emph{p}}\mathrm{^{miss}_T})_{\textrm{ min}}$        &   \multicolumn{4}{c|}{ $>$  0.2} \\  \hline
Aplanarity            &    -    &  \multicolumn{3}{c|}{ $>$0.08} \\  \hline
\met /$\sqrt{\HT}$ [\GeV$^{1/2}$]   &   \multicolumn{3}{c|}{ $>$ 16} &   $>$ 10 \\  \hline
\meff\ [\GeV] &  $>$ 1600  & $>$ 1000  & $>$ 2200 &  $>$ 3400 \\  \hline
 
\end{tabular}
\end{center}
\caption{Selection criteria used for model-independent search signal regions with high
jet multiplicities.}
\label{tab:SB2def}
\end{table}

\section{Background estimation}
\label{sec:background}
 
Standard Model background processes contribute to the event counts in
the signal regions. The most important backgrounds in the search are: $Z+$jets, $W+$jets, top quark
pair, single top quark, diboson and multi-jet production. Non-collision backgrounds were found to be negligible.
 
Generally, the $Z$+jets background events originate from an irreducible component in which $Z\to\nu\bar\nu$ decays generate large $\ourmagptmiss$. The $W$+jets background is mostly composed of $W\to \tau\nu$ events in which the $\tau$-lepton decays to hadrons, with additional contributions from $W\to e\nu, \mu\nu$ events in which no baseline electron or muon is reconstructed, with $\ourmagptmiss$ due to neutrinos. Top quark pair production, followed by semileptonic decays, in particular $t \bar t \to b \bar b \tau \nu q q' $ (with the $\tau$-lepton decaying to hadrons), as well as single-top-quark events,
can also generate large $\ourmagptmiss$ and satisfy the jet
and lepton veto requirements. Each of these primary backgrounds is estimated using dedicated control regions, as described in the following subsection, while diboson production is estimated with MC simulation normalised using NLO cross-section predictions, as described in Section~\ref{sec:montecarlo}.
 
The multi-jet background in the signal regions is due to missing transverse momentum from misreconstruction of jet energies in the calorimeters, jets lost due to the JVT requirement,
as well as neutrinos from semileptonic decays of heavy-flavour hadrons. It is estimated in a data-driven way described below.
 
\subsection{Control regions}
 
To estimate the SM backgrounds in an accurate and robust fashion, control regions (CRs) are defined for each of the signal regions. For the BDT and model-independent searches, a dedicated unique set of CRs is defined for each SR such that the shapes of the background distributions of SR events cannot bias the analysis. For the multi-bin search, CR bins are defined with similar kinematics to the SR bins to account for potential mismodelling of the shapes of background distributions, as shall be described below.
The CRs are chosen to be exclusive with respect to the SR selections in order to provide independent data samples enriched in particular backgrounds and are used to normalise the background MC simulation used to estimate SR event yields. Equivalently, the MC simulation can be considered to provide multiplicative extrapolation factors for the contributing background processes, relating the observed CR event yields to the expected yield in the SR.  The CR selections are designed to have negligible expected SUSY signal contamination for the models near the exclusion boundary established by previous searches. Cross-checks of the background estimates are performed with data in several validation regions (VRs, described in Section~\ref{sec:valreg}) selected with requirements such that these regions do not overlap with the CR and SR selections and also have a low expected signal contamination.
 
Four control regions are defined for each signal region used in the search.
The CR selections maintain adequate statistical precision while minimising the systematic uncertainties arising from the extrapolation of the CR event yield to estimate the background in the SR. This latter requirement is addressed through the use of CR jet $\pt$ thresholds and $\meff$ and BDT score (where appropriate) selections which match those used in the SR. In some cases, in order to increase the number of CR data events without significantly increasing the theoretical uncertainties associated with the background estimation procedure, some SR selection requirements are omitted or loosened, as indicated in the text below. The CR definitions for the multi-bin (MB) and BDT search strategies are listed in Table~\ref{tab:crdefs}.
For the multi-bin search, only the preselection requirement on \met/$\sqrt{\HT}$, indicated in Table~\ref{tab:multibin-preselection}, is used, rather than the final SR selection on this variable, in order to increase the number of CR data events without significantly increasing the theoretical uncertainties associated with the background estimation procedure. Multi-bin regions selected with the same $\meff$ and $N_\mathrm{j}$ bin but different \met/$\sqrt{\HT}$ bin share the same control region.
The signal region definitions for the model-independent search closely follow those used for the multi-bin search, as discussed in Section~\ref{single-binSR}. For this reason the CR definitions for the model-independent search also closely follow those used for the multi-bin search, adjusted in a similar way.
 
\begin{table}[H]
\footnotesize
\begin{center}
\renewcommand\arraystretch{1.4}
\hspace*{-0.7cm}\begin{tabular}{ l  c  c  c }
\toprule
CR & SR background &  CR process & CR selection \\ \midrule
MB/BDT-CR$\gamma$ & $Z(\to\nu\bar\nu)$+jets & $\gamma$+jets & Isolated photon \\
MB/BDT-CRQ & Multi-jet & Multi-jet & reversed requirements on (i) $\Delta\phi($j$,\textbf{\emph{p}}\mathrm{^{miss}_T})$  \\
& & & and (ii) $\met/\meff(N_\mathrm{j})$ or $\met/\sqrt{\HT}$\\
MB/BDT-CRW & $W(\to\ell\nu)$+($b$)jets & $W(\to\ell\nu)$+jets & one lepton, 30~\GeV $<m_\mathrm{T}(\ell,\met) < 100$~\GeV, $b$-veto\\
MB/BDT-CRT & $t\bar{t}$(+EW) and single top & $t\bar{t}\to b\bar{b}q\bar{q}'\ell\nu$ & one lepton, 30~\GeV $<m_\mathrm{T}(\ell,\met) < 100$~\GeV, $b$-tag\phantom{o}\\
\bottomrule
\end{tabular}
\end{center}
\caption{Control regions used in the analysis. Also listed are the main targeted background in the SR in each case, the process used to model the background, and the main CR requirement(s) used to select this process. The jet $\pt$ thresholds and $\meff$ and BDT score (where appropriate) selections match those used in the corresponding SRs.}
\label{tab:crdefs}
\end{table}

The $\gamma$+jets region in both the multi-bin and BDT search strategies (labelled MB/BDT-CR$\gamma$ in Table~\ref{tab:crdefs}) is used to estimate the contribution of $Z(\to\nu\bar\nu)$+jets background events to each SR by selecting a sample of $\gamma$+jets events with $\pt(\gamma)>150~\GeV$ and then treating the reconstructed photon as contributing to $\met$. For $\pt(\gamma)$ significantly larger than $m_Z$ the kinematic properties of such events strongly resemble those of $Z$+jets events \cite{SUSY-2011-20}.  In order to correct for differences in the $Z$+jets to $\gamma$+jets ratio between data and MC simulation, likely arising from mismodelling of the $\gamma$+jets process, a correction factor ($\kappa$) is applied to simulated $\gamma$+jets events in the CR$\gamma$ regions. This correction factor is determined by comparing CR$\gamma$ observations in data and MC simulation with those in similar regions defined by selecting events with two electrons or muons for which the invariant mass lies within 25~\GeV\ of the mass of the $Z$ boson, satisfying $\met/\sqrt{\HT} > 10~\GeV^{1/2}$ and $\meff >1000~\GeV$. This selection corresponds to the kinematically lowest bins of the multi-bin analysis MB-SSd with $N_\mathrm{j}=[2, 3]$ and $N_\mathrm{j}=[4, \infty]$. The correction factor is obtained from the double ratio
 
$$\kappa=\frac{N_{\gamma}^\mathrm{data}}{N_{\gamma}^\mathrm{MC}}  \bigg/ \frac{N_{Z}^\mathrm{data}}{N_{Z}^\mathrm{MC}},$$
 
where $N_{\gamma}^\mathrm{data}$ and $N_{Z}^\mathrm{data}$ are the data observations in the $\gamma$ and $Z$ control regions, respectively, following subtraction of the respective non-$\gamma$+jet and non-$Z$+jet background expectations obtained from MC simulation. $N_{\gamma}^\mathrm{MC}$ and $N_{Z}^\mathrm{MC}$ are the equivalent $\gamma$+jet and $Z$+jet yields obtained from MC simulation.
The value of $\kappa$ is found to depend on jet multiplicity, but is independent of $\meff$ and $\met/\sqrt{\HT}$. Consequently, $\kappa$ is calculated separately for regions with up to three and at least four jets, and is found to take values $\kappa =0.77 \pm 0.04$ and $\kappa =0.85 \pm 0.05$, respectively. The quoted uncertainty in $\kappa$ is statistical only -- systematic uncertainties in the yields cancel by construction in the ratio and the resulting uncertainties in $\kappa$ are found to be negligible. In both search strategies, the CR$\gamma$ selections omit the SR requirement on the aplanarity variable $A$. Additionally, for the BDT-GGo1 and BDT-GGo2 SRs, the $\Delta\phi($j$,\textbf{\emph{p}}\mathrm{^{miss}_T})$, and $\met/\meff(N_\mathrm{j})$ selections are removed for the corresponding CR selections.

The $W$+jets and top quark background control regions in both the multi-bin and BDT search strategies (labelled MB/BDT-CRW and MB/BDT-CRT in Table~\ref{tab:crdefs}) select samples rich in $W(\to \ell \nu)$+jets events and in semileptonic $t\bar{t}$ and single-top events (referred to generically as `top quark background'), respectively. They use events with one high-purity lepton and differ in the number of $b$-jets required (zero or $\geq 1$, respectively). In both of these search strategies, a requirement on
the transverse mass $m_\mathrm{T}$ computed with $E_{\textrm  T}^{\mathrm{miss}}$ and the selected lepton\footnote{$m_{\mathrm T} =
\sqrt{2\pt^{\ell}\met  (1 - \mathrm{cos}[\Delta\phi(\ell, \textbf{ \emph{p}}\mathrm{^{miss}_T} )] )} $.} is applied, as indicated in
Table~\ref{tab:crdefs}.
Events are selected using a trigger based on the missing transverse momentum, as described in Section~\ref{sec:detector}. This approach allows the use of leptons with transverse momenta as low as 6~\GeV\ (muons) or 7~\GeV\ (electrons), which maximises the proximity of the CRs closer to the SRs in the event selection parameter space. The selected lepton is treated as a jet with the same
momentum to model background events in which a hadronically decaying
$\tau$-lepton is produced \cite{SUSY-2013-02}. The application of this procedure to the offline CRW and CRT selections but not in the trigger introduces an additional inefficiency with respect to the offline and online SR selections of less than 0.1\%.
The CRW and CRT selections omit the SR selection requirements on $\Delta\phi($j$,\textbf{\emph{p}}\mathrm{^{miss}_T})$ in both search strategies.
 
The multi-jet background is estimated using a data-driven technique \cite{SUSY-2011-20}, which applies a jet resolution function to well-measured multi-jet events in order to estimate the impact of jet energy mismeasurement and heavy-flavour semileptonic decays on \met{} and other variables. The resolution function of jets is initially estimated from MC simulation by matching jets reconstructed from generator-level particles including muons and neutrinos to detector-level jets in multi-jet samples, and then is modified to agree with data in dedicated samples used to
measure the resolution function.
The multi-jet region (labelled as MB/BDT-CRQ in Table~\ref{tab:crdefs}) uses reversed selection requirements on $\Delta\phi($j$,\textbf{\emph{p}}\mathrm{^{miss}_T})$ and on $\met/\sqrt{H_{\textrm  T}}$  in the multi-bin search, or on $\met/\meff(N_{\textrm  j})$ in the case of the BDT search, to produce samples enriched in multi-jet background events. For the two signal regions targeting the lowest mass splittings $\Delta m(\gluino, \ninoone)$ in the BDT search, BDT-GGd4 and BDT-GGo4, the BDT score selections are slightly loosened from 0.87 to 0.70 and from 0.84 to 0.60, respectively. The MB/BDT-CRQ regions are used to normalise the shape of the distributions obtained with the data-driven technique.
 
Example $ \meff$ distributions in control regions based on the MB-GGd preselection requirements listed in Table~\ref{tab:multibin-preselection}  are shown in Figure~\ref{fig:MB_GGd_CR}. Figure~\ref{fig:BDT_GGo1_CR} shows the BDT score discriminating variable distributions in control regions corresponding to the BDT-GGo1 signal region selections. Discrepancies between data and MC simulation in these figures (evident particularly for the top quark processes dominating Figure~\ref{fig:MB_GGd_CR}(d)) replicate those observed in the signal regions. The background estimation procedure uses the CR observations to compensate for these discrepancies, as shall now be described. As a result of this procedure these discrepancies do not affect the analysis.
\begin{figure}[H]
\begin{center}
\subfigure[]{\includegraphics[width=0.45\textwidth]{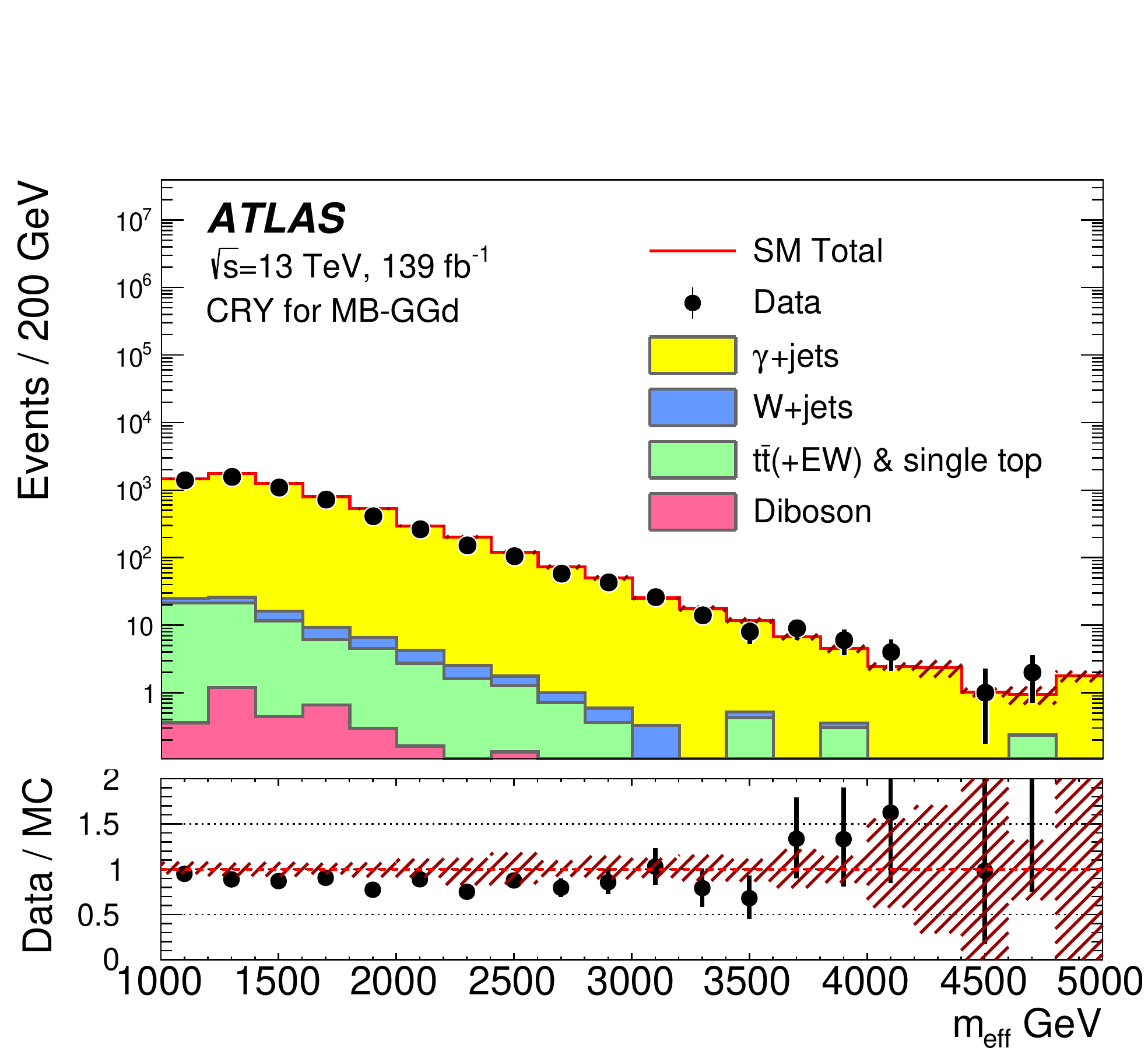}}
\subfigure[]{\includegraphics[width=0.45\textwidth]{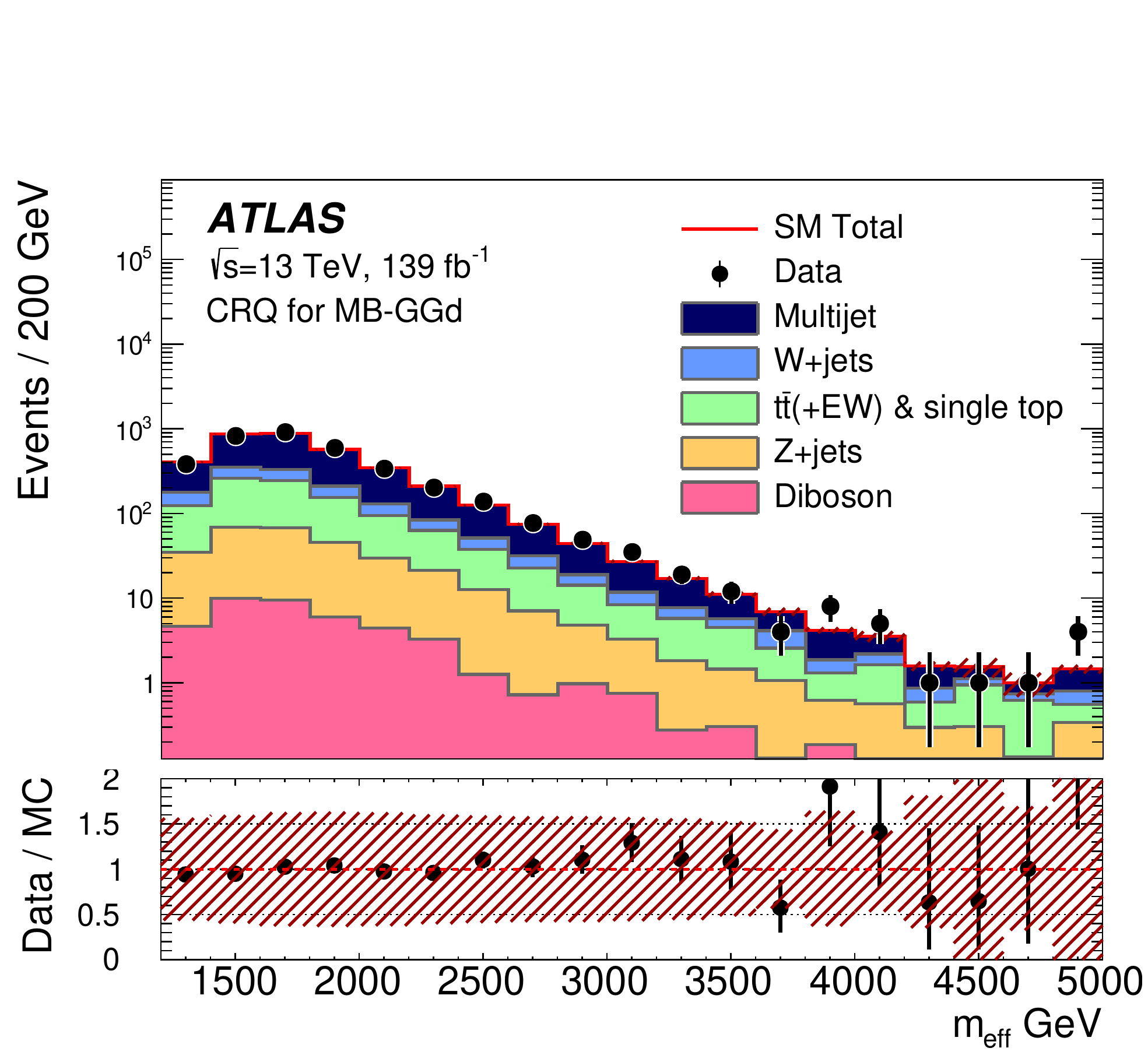}}
\subfigure[]{\includegraphics[width=0.45\textwidth]{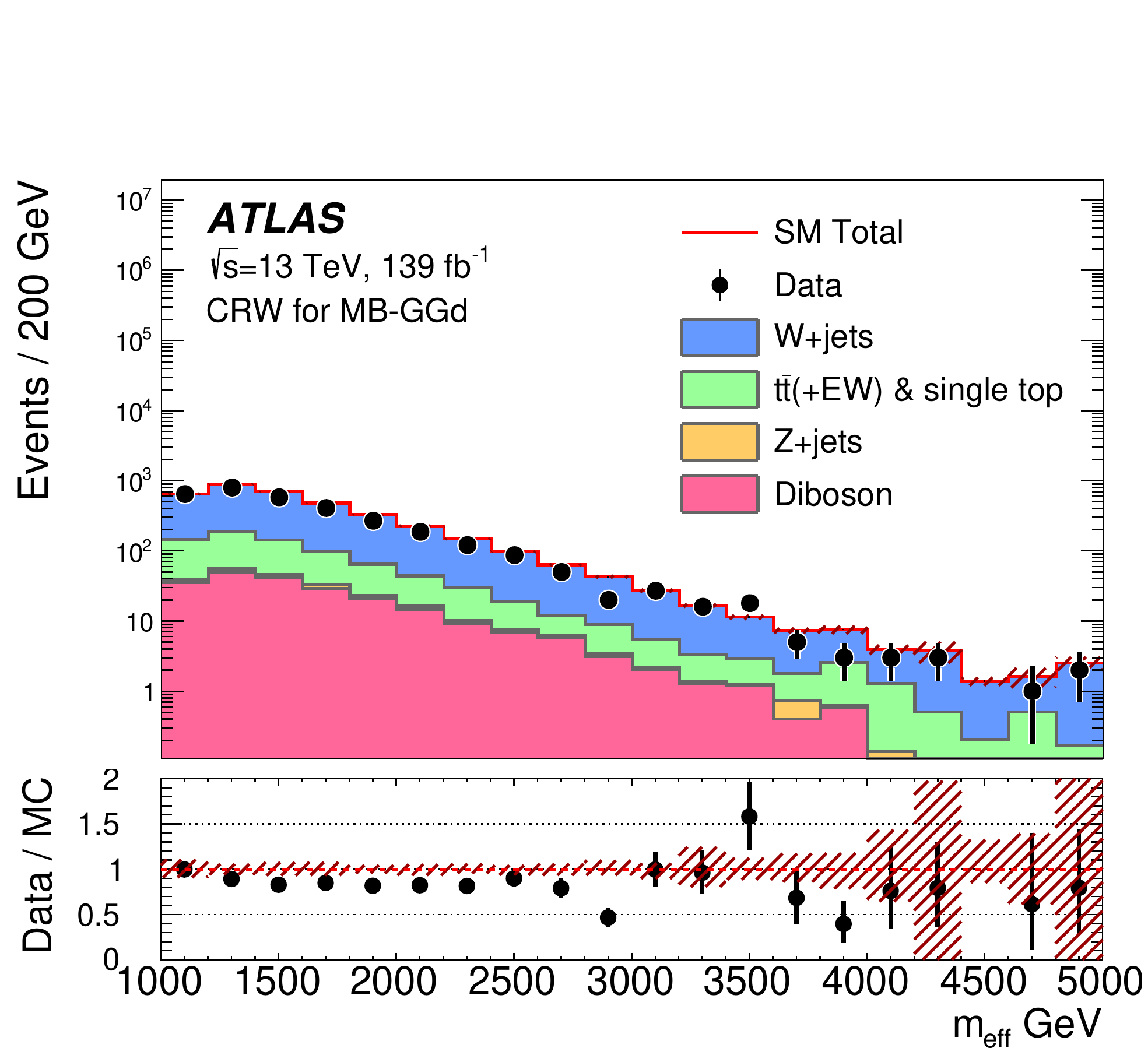}}
\subfigure[]{\includegraphics[width=0.45\textwidth]{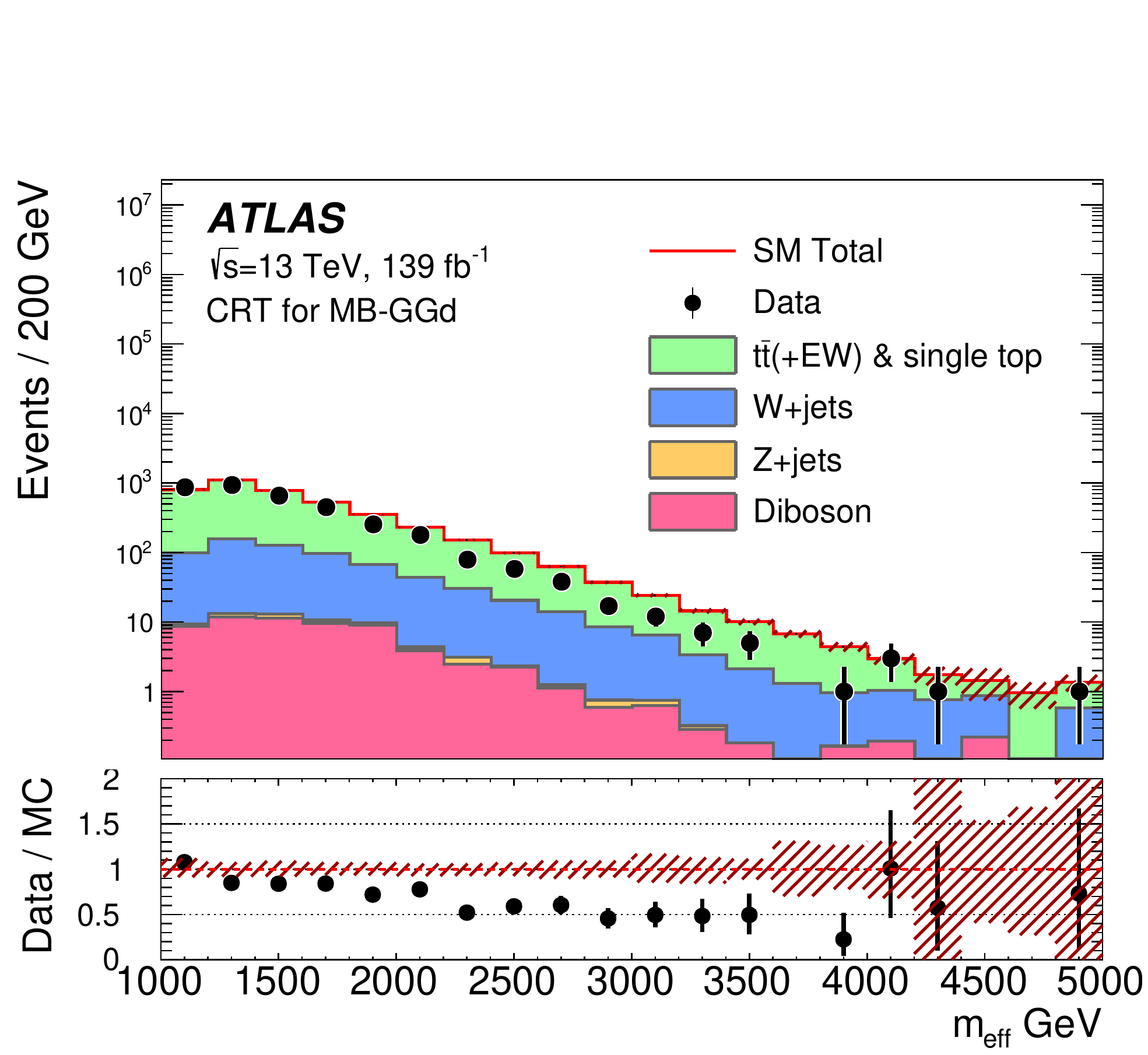}}
\end{center}
\caption{Observed $\meff$ distributions in control regions (a) MB-CR$\gamma$, (b) MB-CRQ, (c) MB-CRW and (d) MB-CRT after applying the MB-GGd preselection requirements listed in Table~\ref{tab:multibin-preselection}.
The histograms show the MC background predictions normalised using cross-section times integrated luminosity, with the exception of multi-jet background which is normalised using data. In the case of the $\gamma$+jets background, a $\kappa$ factor described in the text is applied. The last bin includes overflow events. The lower panels show the ratio of data to the background prediction. The hatched (red) error bands indicate the combined experimental and MC statistical uncertainties on these background predictions.
}
\label{fig:MB_GGd_CR}
\end{figure}
 
\begin{figure}[H]
\begin{center}
\subfigure[]{\includegraphics[width=0.45\textwidth]{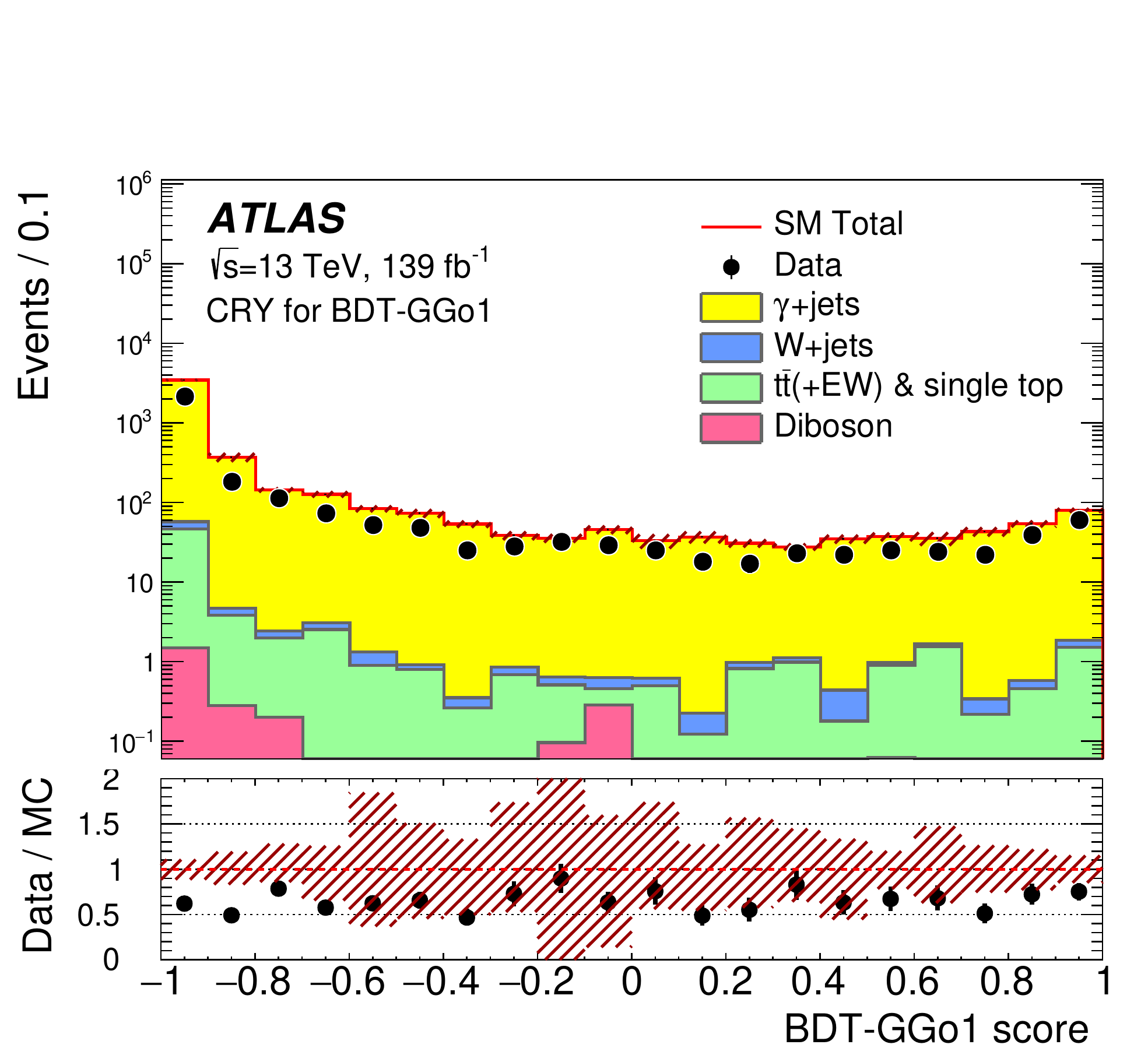}}
\subfigure[]{\includegraphics[width=0.45\textwidth]{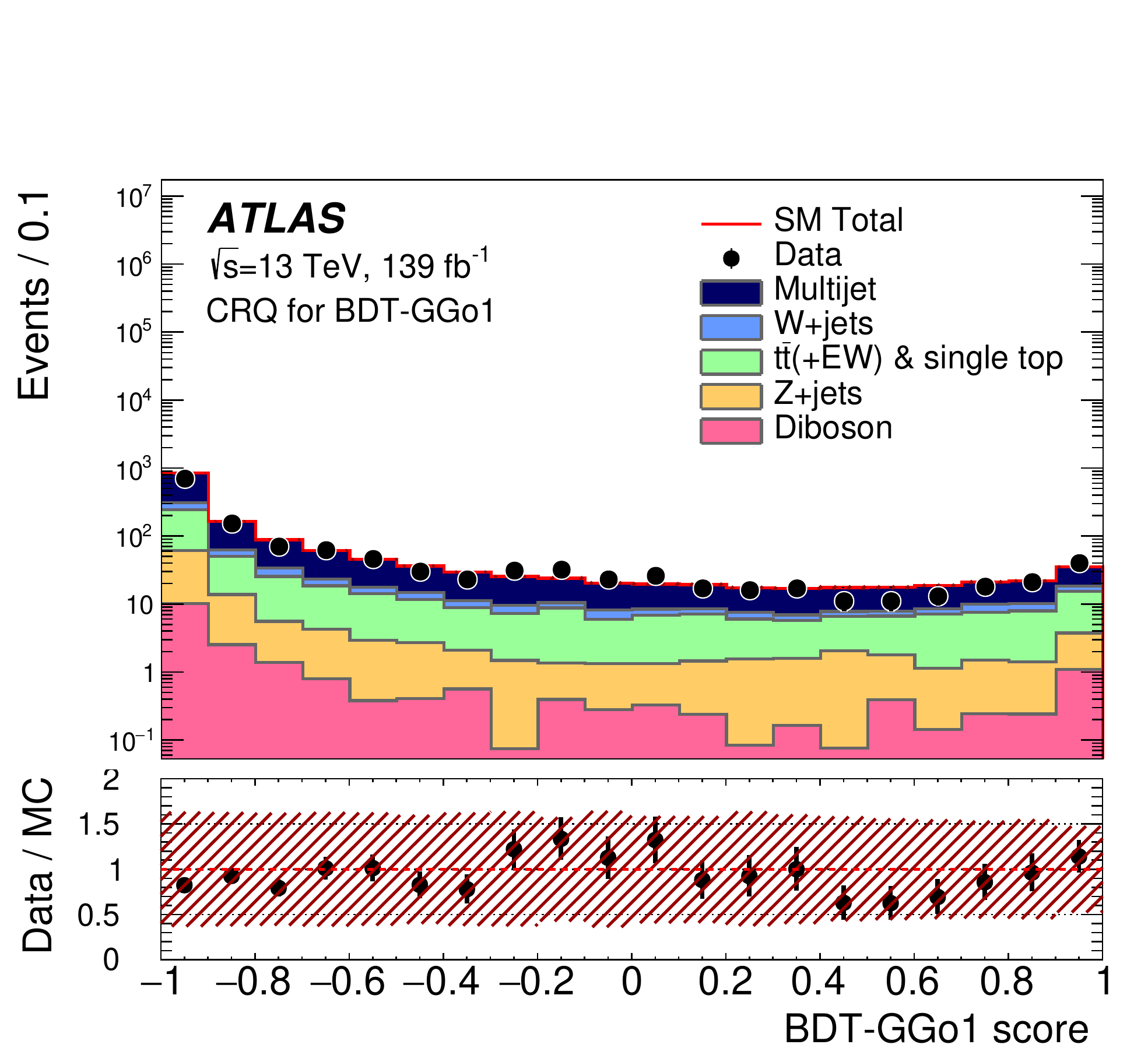}}
\subfigure[]{\includegraphics[width=0.45\textwidth]{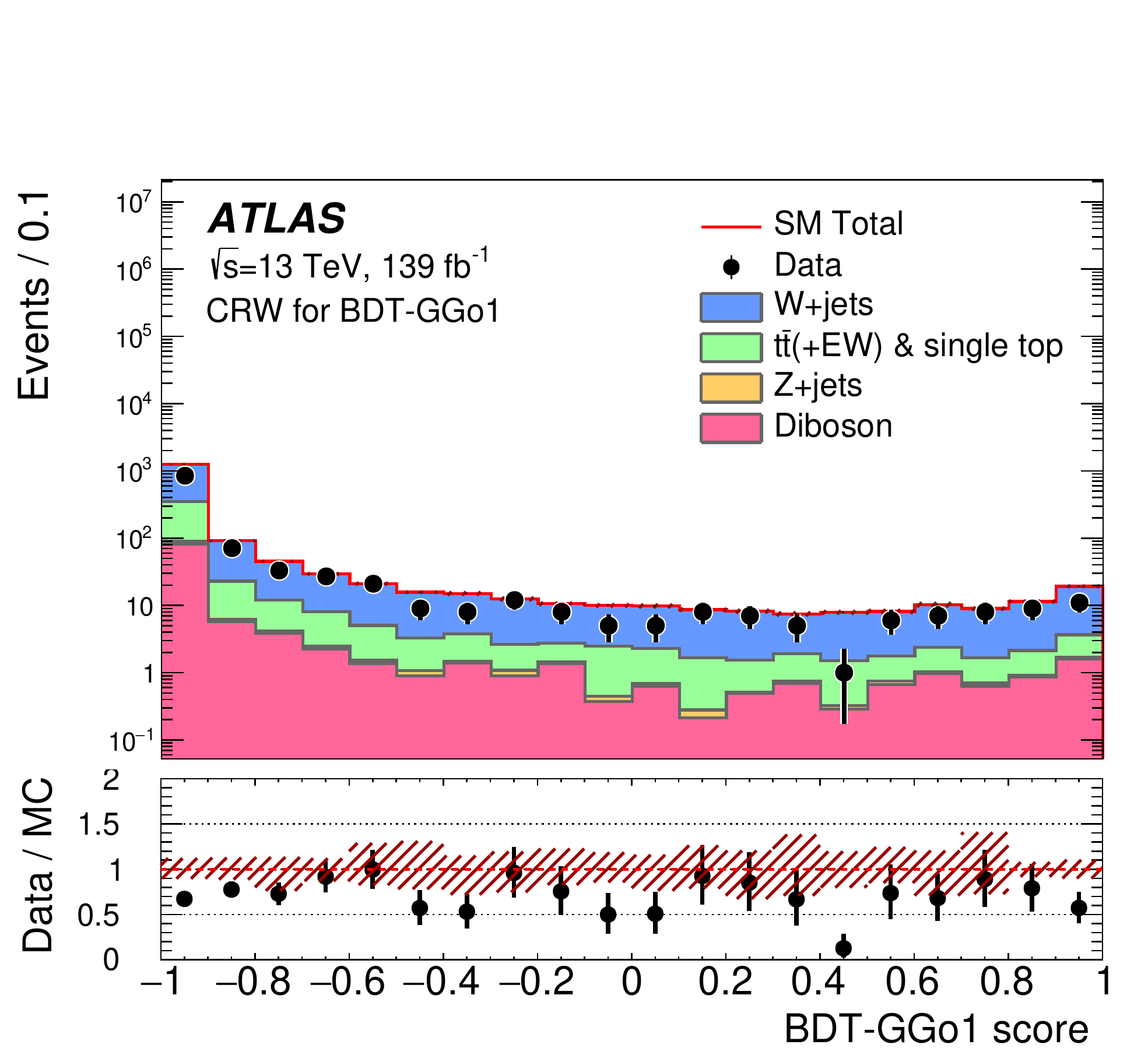}}
\subfigure[]{\includegraphics[width=0.45\textwidth]{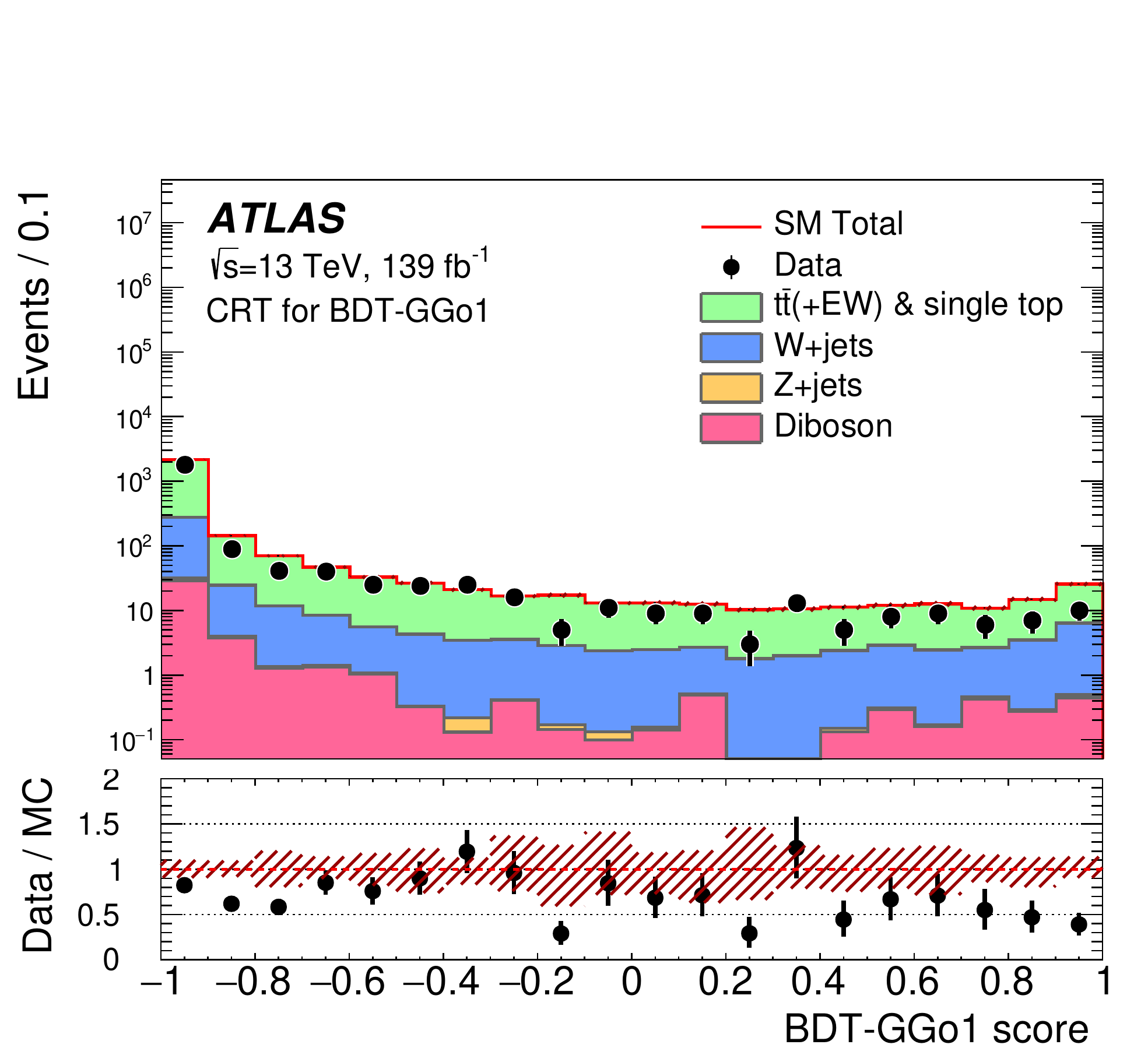}}
\end{center}
\caption{Observed BDT score distributions in control regions (a) BDT-CR$\gamma$, (b) BDT-CRQ, (c) BDT-CRW and (d) BDT-CRT after applying the BDT-GGo1 selection requirements described in Section~\ref{bdtSR}, excluding the BDT score cut.
The histograms show the MC background predictions normalised using cross-section times integrated luminosity, with the exception of multi-jet background which is normalised using data. In the case of the $\gamma$+jets background, a $\kappa$ factor described in the text is applied. The lower panels show the ratio of data to the background prediction. The hatched (red) error bands indicate the combined experimental and MC statistical uncertainties on these background predictions.
}
\label{fig:BDT_GGo1_CR}
\end{figure}
 
In order to estimate the background yields, a background-only fit is used \cite{HFpaper}. The fit is performed using the observed event yields in the CRs associated with the SRs as the only constraints, so that the fit is not constrained by the yields in the SRs. It is assumed that signal events from beyond the Standard Model (BSM) processes do not contribute to the CR yields. Scale factors denoted by $\mu(W\textrm{+jets})$, $\mu(Z\textrm{+jets})$ and $\mu(\textrm{Top})$ represent the normalisation of background components relative to MC predictions, and are simultaneously determined in the fit to all the CRs associated with a SR. The expected background in the SR is based on the yields predicted by simulation for $W/Z$+jets and background processes containing top quarks, corrected by the scale factors derived from the fit.
The systematic and MC statistical uncertainties of the expected values are included in the fit as nuisance parameters that are constrained by Gaussian terms. The means of the Gaussian terms are defined by the nominal predictions, while the standard deviations are determined by the sizes of the systematic uncertainties considered (see Section~\ref{sec:systematics}). Poisson distributions are used for the statistical uncertainties arising from the limited number of data events in the estimation of the background sources, or the limited number of simulated events. The background-only fit is also used to estimate the background event yields in the validation regions.
 
\begin{figure}[H]
\begin{center}
\subfigure[]{\includegraphics[width=0.49\textwidth]{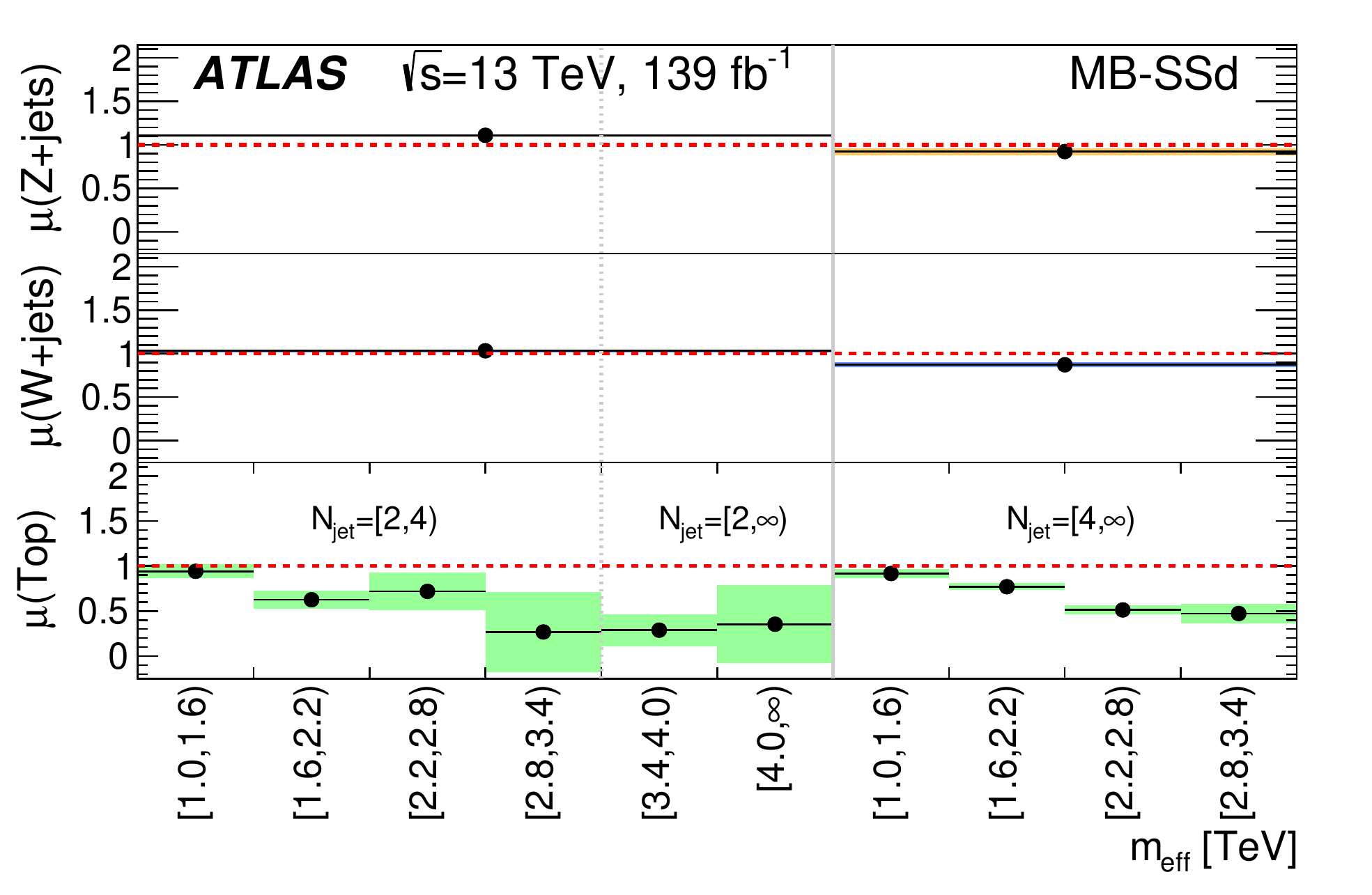}}
\subfigure[]{\includegraphics[width=0.49\textwidth]{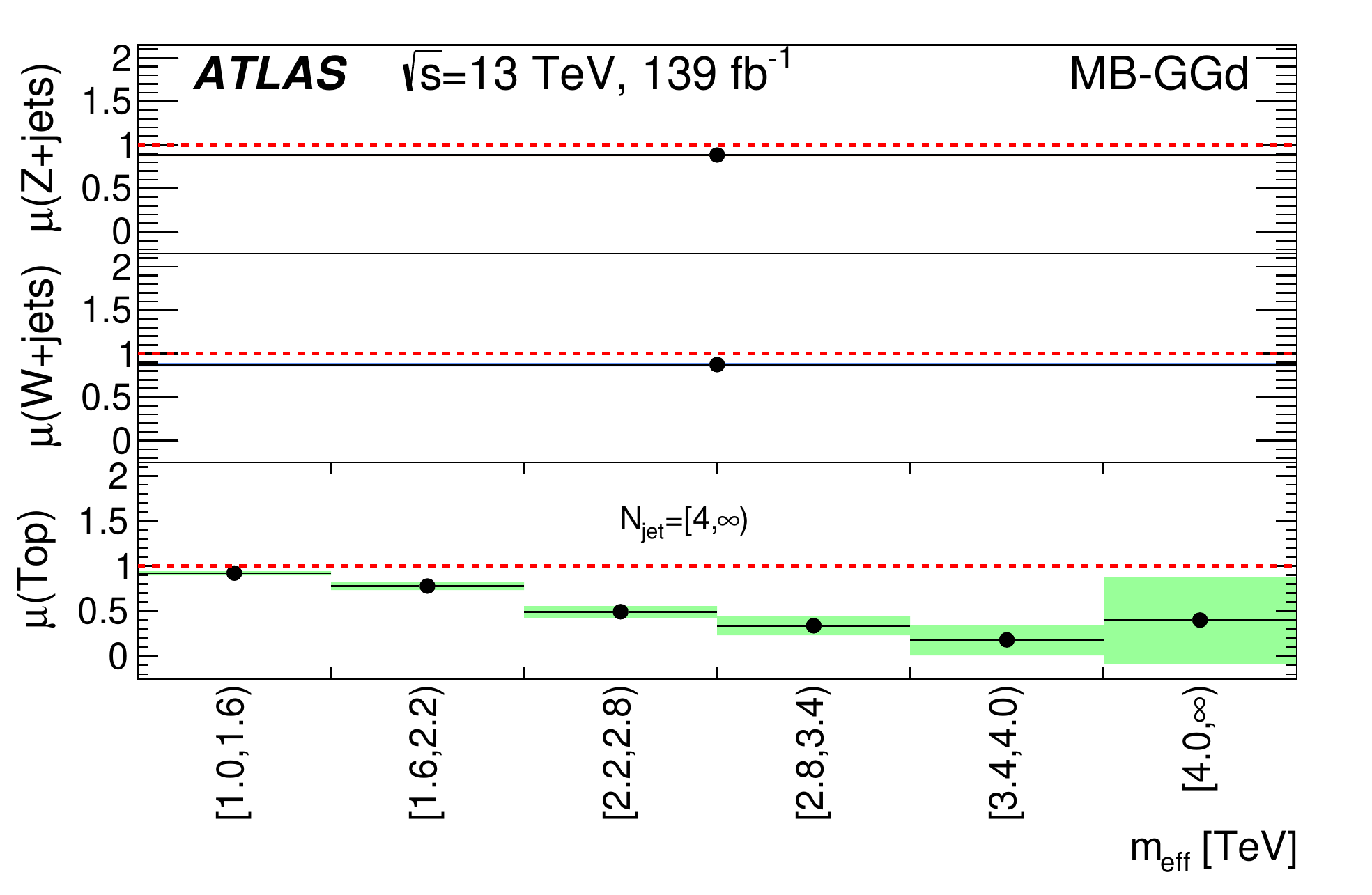}}\\
\subfigure[]{\includegraphics[width=0.49\textwidth]{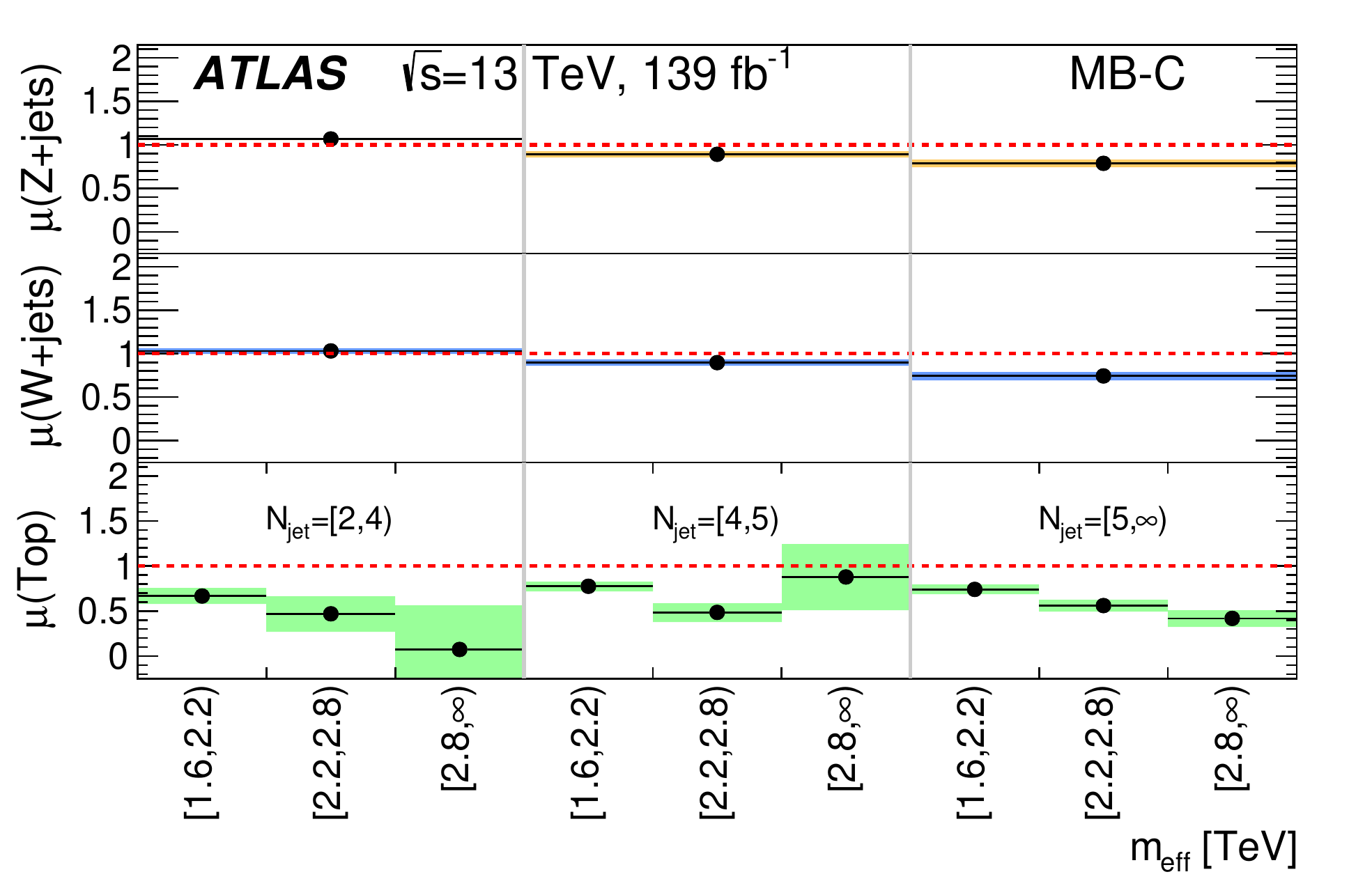}}
\subfigure[]{\includegraphics[width=0.49\textwidth]{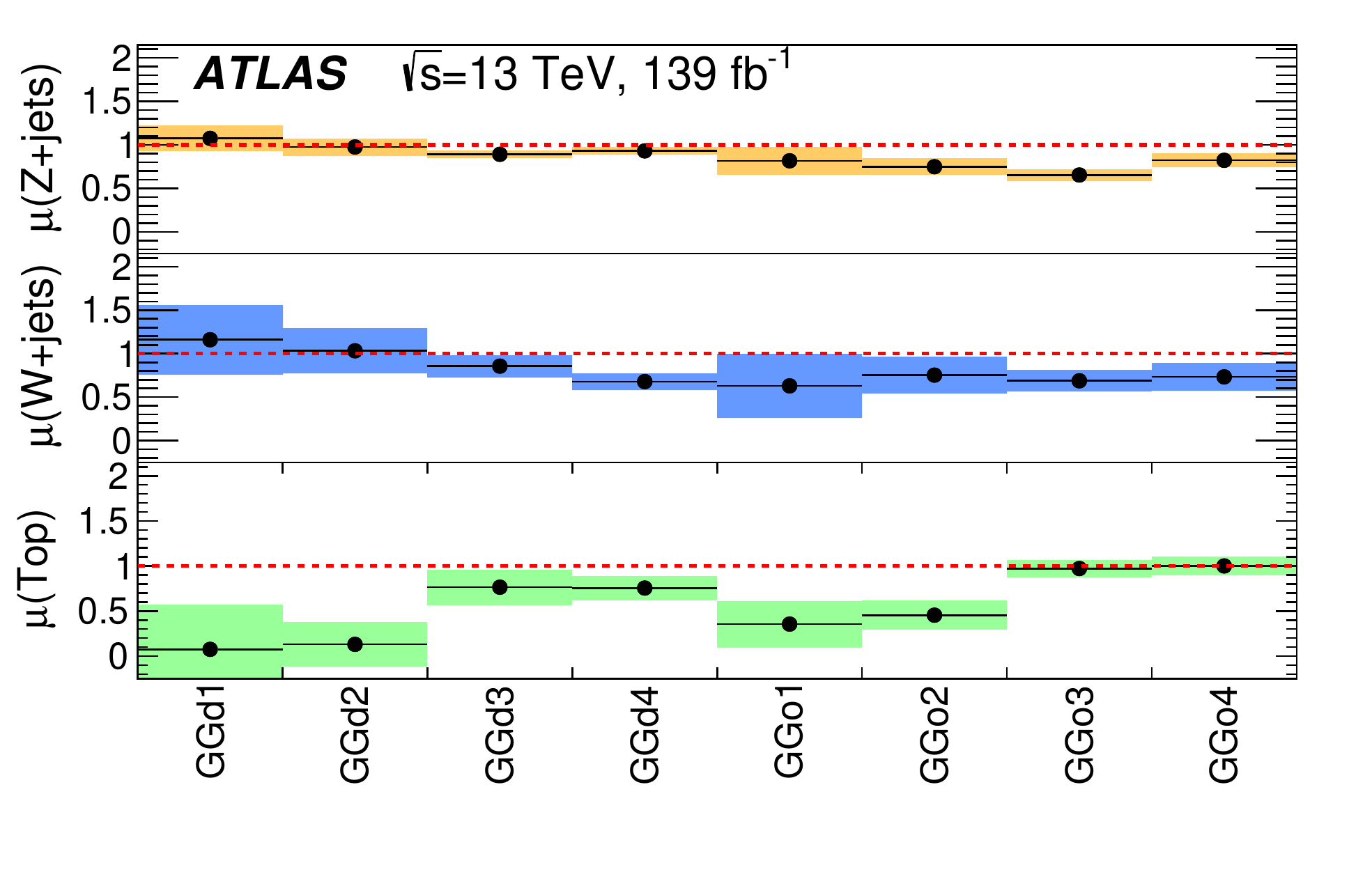}}
\end{center}
\caption{Fitted normalisation factors per process as a function of the signal region considered in the (a) MB-SSd, (b) MB-GGd, (c) MB-C regions from the multi-bin search, and (d) regions from the BDT search.
The dashed horizontal lines at 1.0 correspond to pure MC estimates. The coloured bands correspond to the uncertainties in the normalisation factors for the different background processes.}
\label{fig:summaryMu}
\end{figure}
 
The MC normalisation factors determined from the background-only fits in each CR for each background process are shown in Figure~\ref{fig:summaryMu}.
For the BDT and model-independent searches, three such factors are extracted for each signal region, corresponding to the $W$+jets, $Z$+jets and top quark backgrounds. For the multi-bin search a single normalisation factor is applied to each of the $W$+jets and $Z$+jets processes in all regions associated with each jet multiplicity bin, while a dedicated normalisation factor is applied to the top quark process in each bin. Some trends in these normalisation factors are observed, with those for the top quark background becoming smaller with increasingly tight selection requirements for the multi-bin search signal regions. Similarly, the measured top quark background normalisation factors decrease with increasingly tight BDT score requirements in the BDT search. This behaviour follows from the simulated top quark MC samples exhibiting generally harder kinematics than observed in data~\cite{ATL-PHYS-PUB-2018-009}. Before the top quark background normalisation factors are applied, the contribution of the top quark background is expected to be less than 10\% (typically 1--2\%) in most of the signal regions, with the exception of signal regions requiring large jet multiplicities, where the contribution of the top quark background can reach 50\% of the total background yield. The normalisation factors for the $W$+jets and $Z$+jets processes are generally stable with changing kinematic selections, with the exception of a slight decrease with increasing jet multiplicity.

\subsection{Validation regions}
\label{sec:valreg}
 
The background estimation procedure is validated by comparing the numbers of events observed in the VRs with the corresponding SM background predictions obtained from the background-only fits. Several VRs are defined for all three search strategies, with requirements distinct from those used in the CRs but that maintain low expected signal contamination. The VRs for the model-independent search closely follow those used for the multi-bin search, similarly to the CR definitions discussed previously, and so are not described separately below. As is the case with the CRs, the majority of the VRs are defined using final states with leptons and photons, allowing the different expected background contributions to the SRs to be validated with high-purity selections. The VR event selections are not defined exclusively and hence the observed event yields can be correlated between regions.
 
The MB/BDT-CR$\gamma$ estimates of the $Z(\to \nu\bar{\nu})$+jets background are validated using samples of $Z(\to\ell\bar\ell)$+jets events selected by requiring high-purity lepton pairs of opposite sign and identical flavour for which the dilepton invariant mass lies within 25~\GeV\ of the $Z$ boson mass. The MB/BDT-CRW and MB/BDT-CRT estimates of the $W$+jets and top quark backgrounds are potentially subject to systematic uncertainties arising from extrapolating over $\Delta\phi($j$,\textbf{\emph{p}}\mathrm{^{miss}_T})$, $\met/\meff(N_{\textrm  j})$ or $\met/\sqrt{\HT}$, and aplanarity $A$ from the CRs to the SRs. This extrapolation procedure is checked with validation regions based upon the CR event selection requirements, modified to more closely resemble those used in the equivalent SR.
 
The MB/BDT-CRQ estimates of the multi-jet background are validated with VRs for which the MB/BDT-CRQ selection is applied, but with the SR $\met/\sqrt{\HT}$ (MB-VR0LMETsig) or $\met/\meff(N_\mathrm{j})$ (BDT-VR0LMETMeff) requirements reinstated,  or with a requirement on $\Delta\phi($j$,\textbf{\emph{p}}\mathrm{^{miss}_T})$ applied (MB/BDT-VR0LdPhi).  These VRs, which are independent of all CRs by construction, test not only the multi-jet background estimates, but also the estimates of all backgrounds in cases where the multi-jet background does not dominate. Some representative results are shown in Figures~\ref{fig:VR0LmetSigmefmeff-pulls} and \ref{fig:VR0LdPhi-pulls}, illustrating the level of agreement typically observed between data and the background estimates.
 
For the BDT search, the event yields in the validation regions are often very small. For this reason, additional validation regions with lower BDT score requirements are defined, for which a minimum of 10 background events is expected in each case.
 
No significant systematic biases are observed among all the 542 VRs used by the three search strategies. The largest discrepancy is $2.6\sigma$ in the MB-VR0LMETsig region associated with the MB-SSd signal region that selects events with two or three jets in the \meff\ bin range 2800~\GeV~to infinity, with the $\met/\sqrt{\HT}$ bin requirement 16--22~\GeV$^{1/2}$ reinstated (see Figure~\ref{fig:VR0LmetSigmefmeff-pullsa}).
The $2.6\sigma$ significance is computed following the profile likelihood method of Ref.~\cite{Cousins:2007bmb} including the systematic uncertainties described in Section~\ref{sec:systematics}.

\begin{figure}[H]
\begin{center}
\subfigure[]{\includegraphics[width=0.49\textwidth]{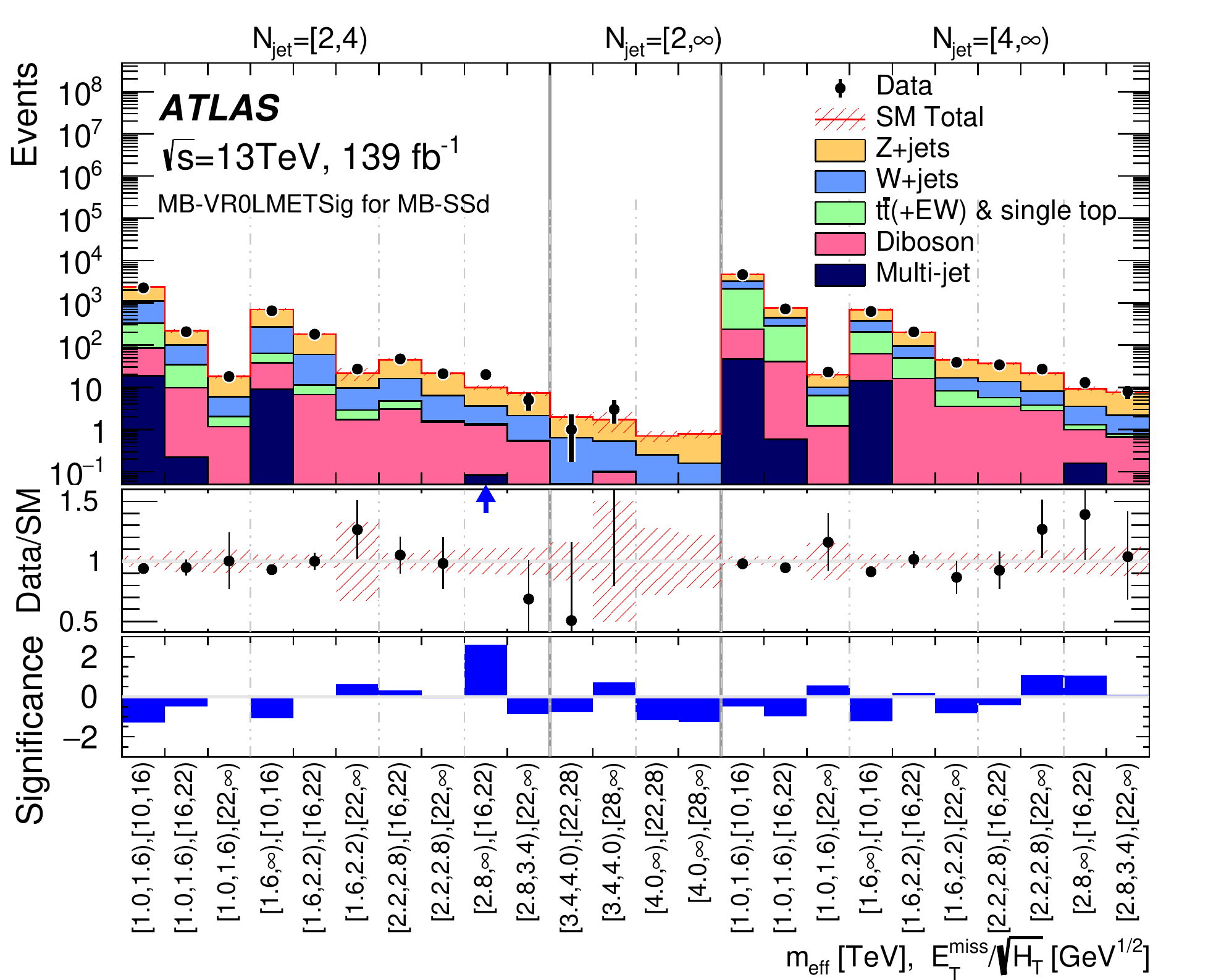}\label{fig:VR0LmetSigmefmeff-pullsa}}
\subfigure[]{\includegraphics[width=0.49\textwidth]{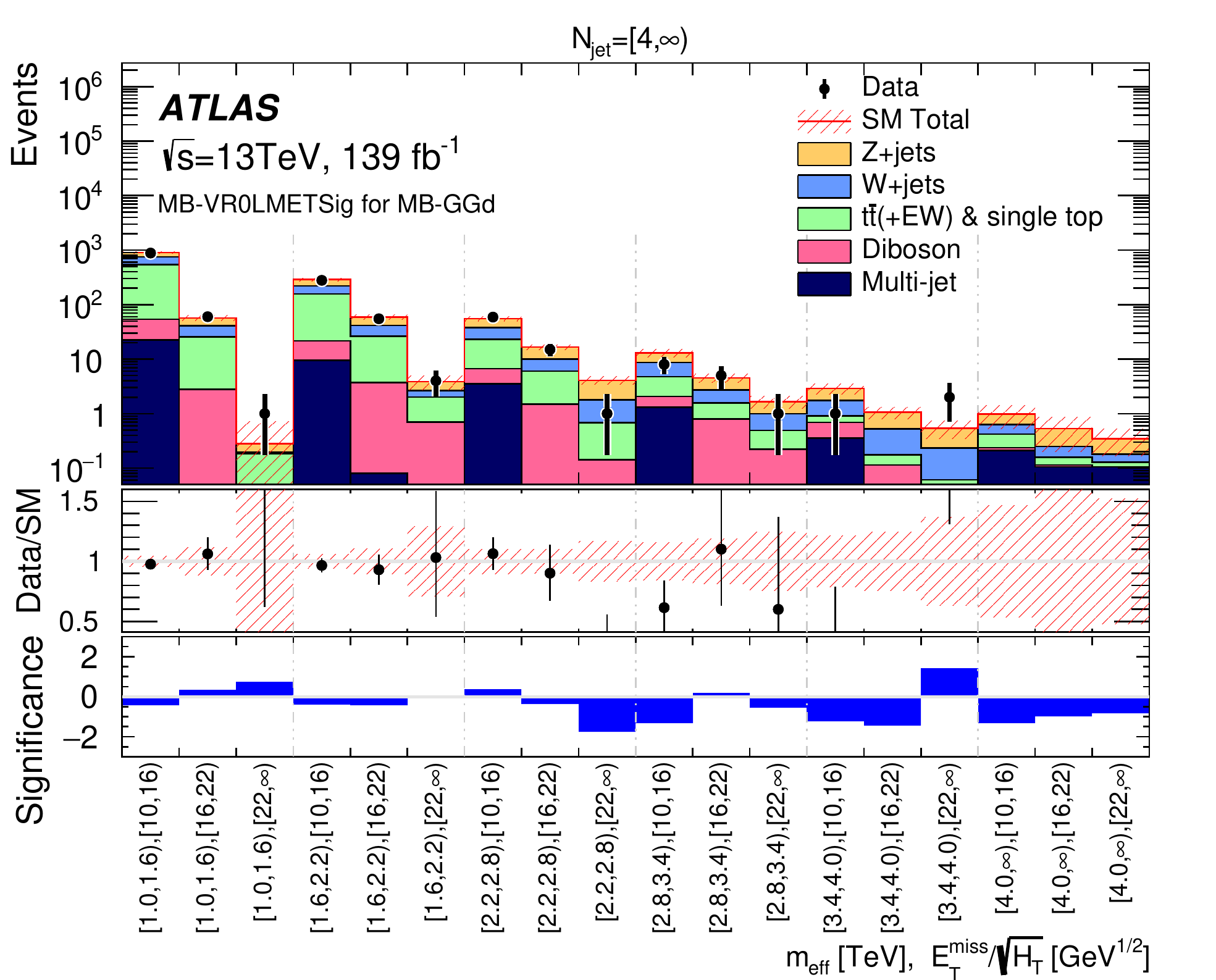}}\\
\subfigure[]{\includegraphics[width=0.49\textwidth]{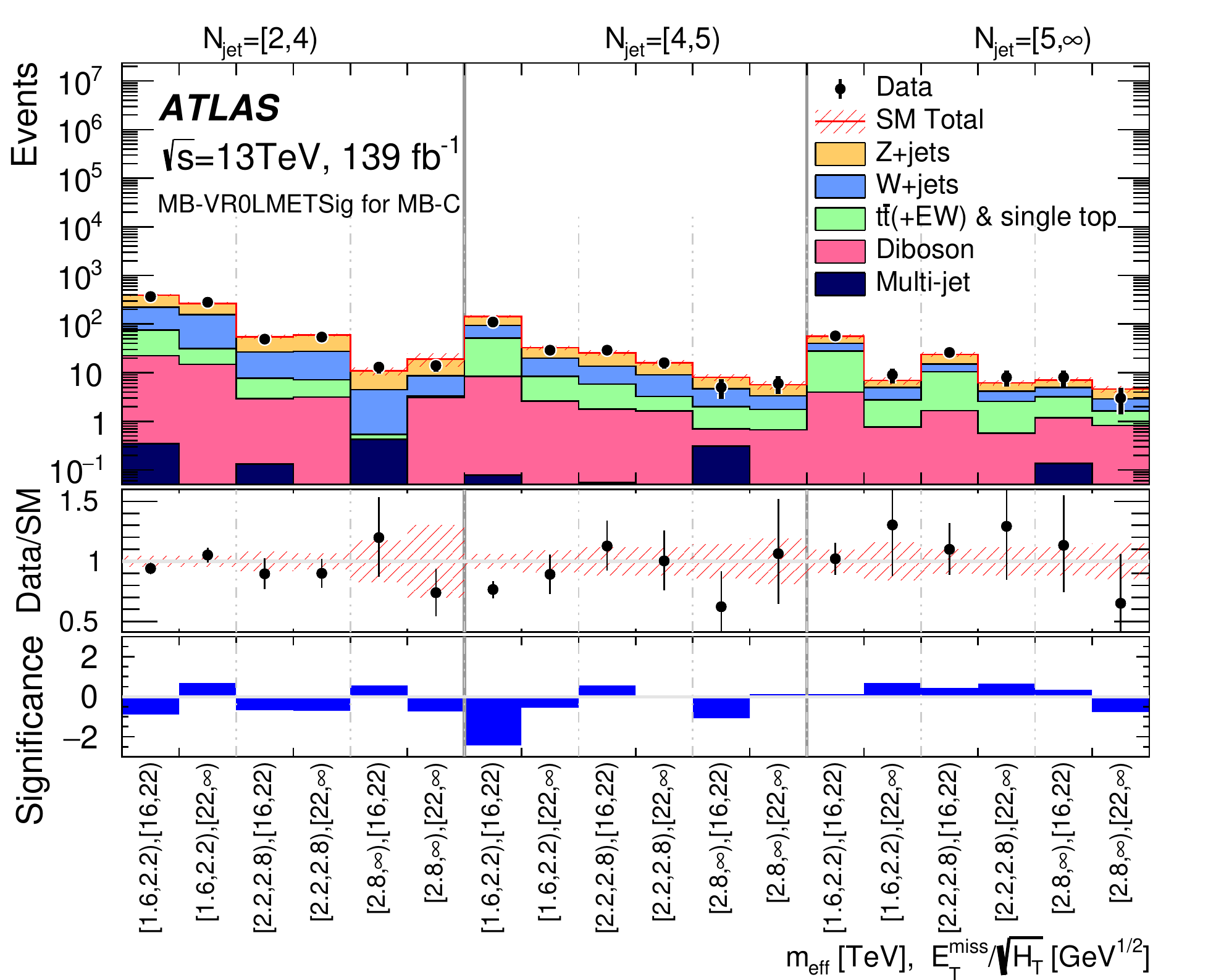}}
\subfigure[]{\includegraphics[width=0.49\textwidth]{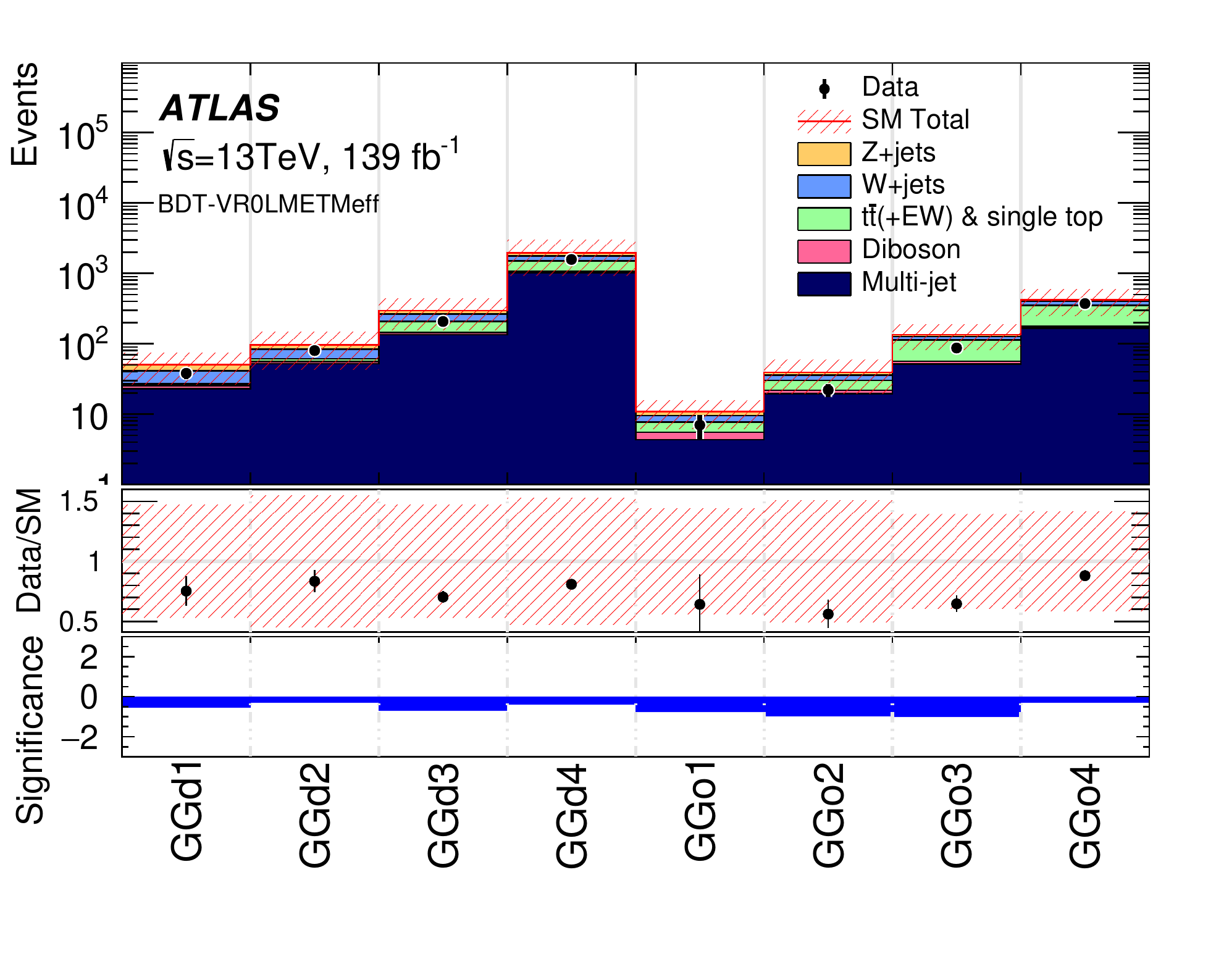}}\\
\end{center}
\caption{Observed numbers of events in data and SM background predictions for the VR0LMETSig regions corresponding to the (a) MB-SSd, (b) MB-GGd, (c) MB-C signal regions from the multi-bin search, and (d) the BDT-VR0LMETMeff regions corresponding to the BDT search signal regions. The lower panels in each case show the ratio of observed data yields to the total predicted background and the observed significance of the data relative to the background-only hypothesis. The significance is computed following the profile likelihood method of Ref.~\cite{Cousins:2007bmb} in the case where the observed yield exceeds the prediction, and using the same expression with an overall minus sign if the yield is below the prediction. The hatched (red) error bands indicate the combined experimental, theoretical and MC statistical uncertainties.}
\label{fig:VR0LmetSigmefmeff-pulls}
\end{figure}
 
\begin{figure}[H]
\begin{center}
\subfigure[]{\includegraphics[width=0.49\textwidth]{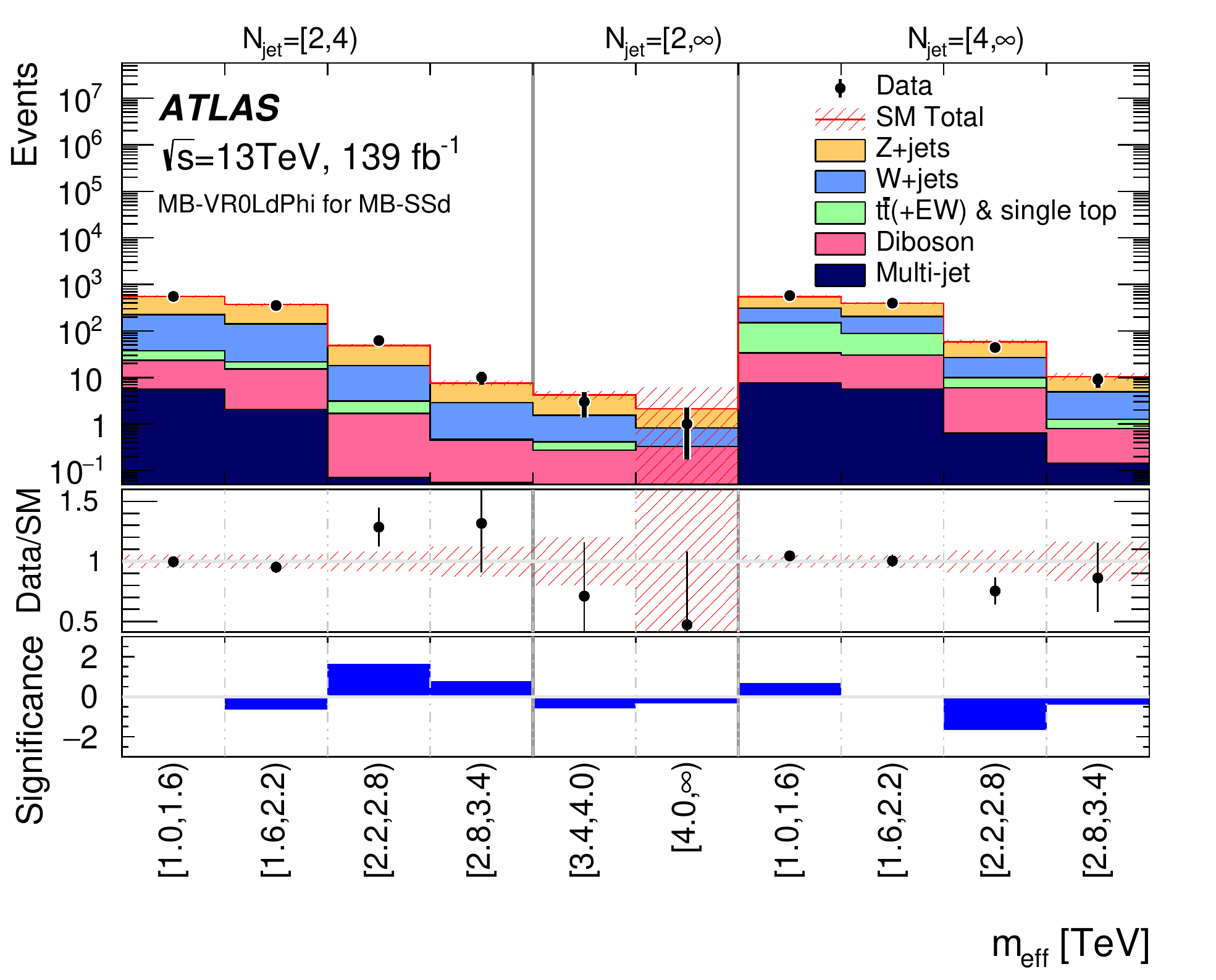}}
\subfigure[]{\includegraphics[width=0.49\textwidth]{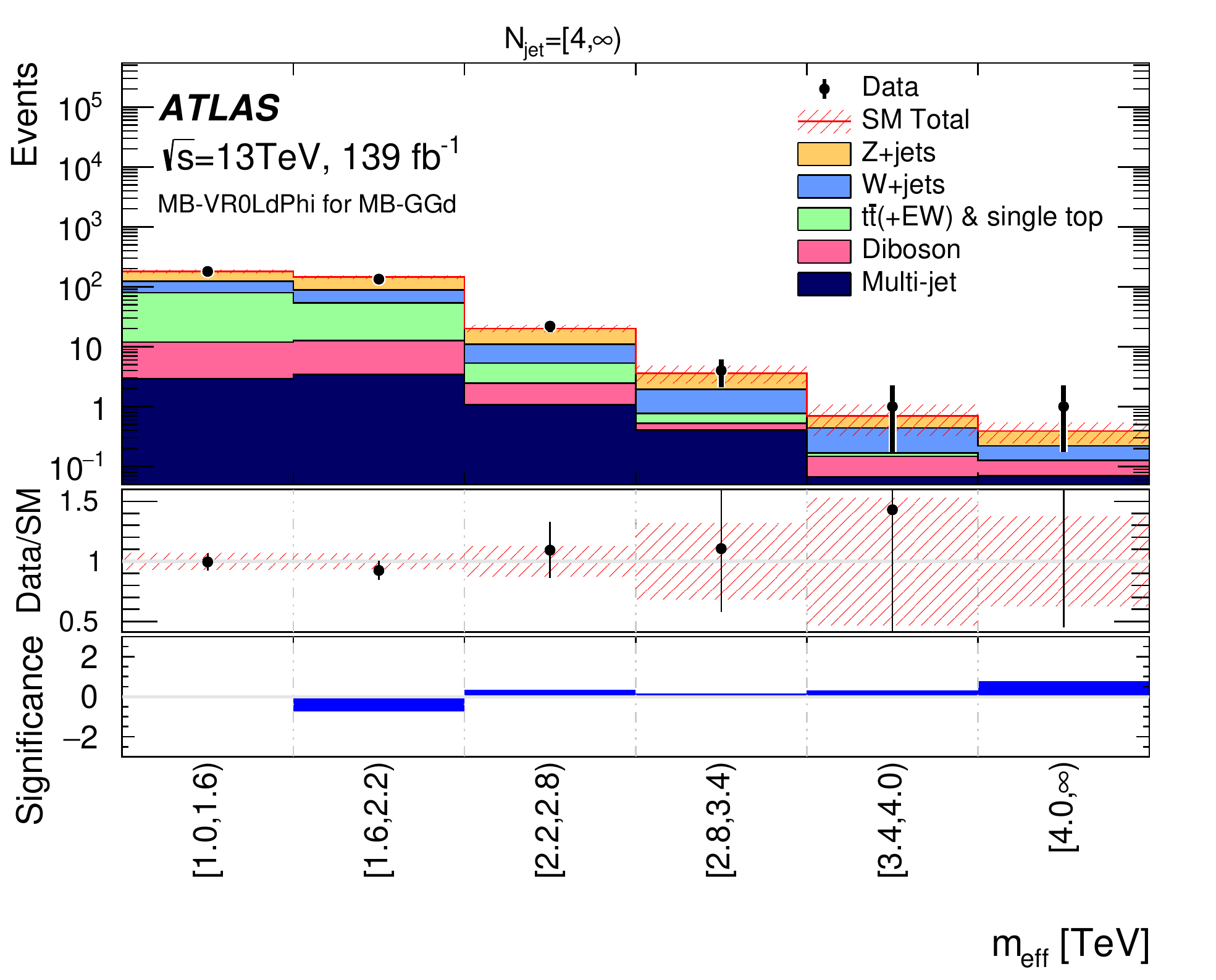}}\\
\subfigure[]{\includegraphics[width=0.49\textwidth]{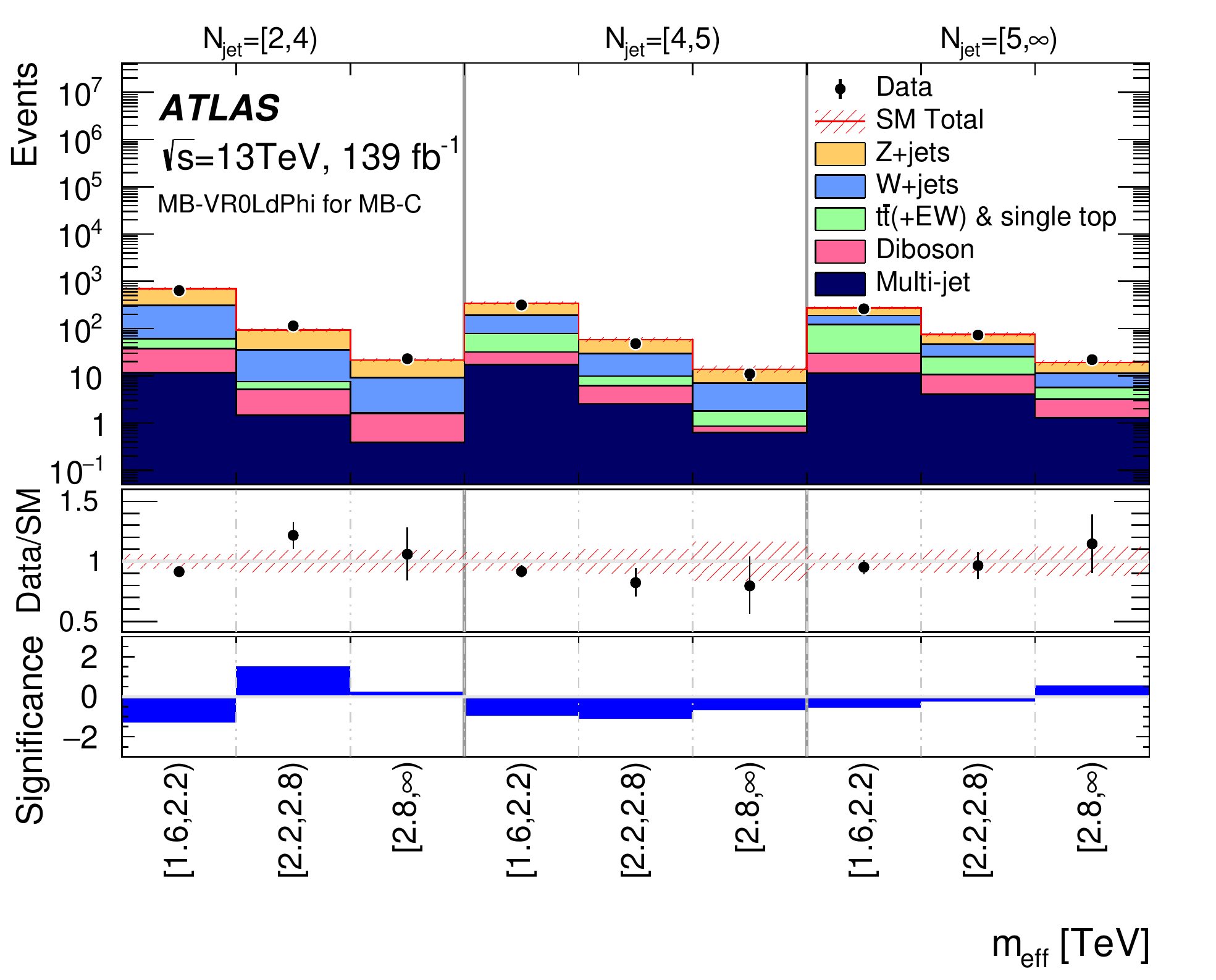}}
\subfigure[]{\includegraphics[width=0.49\textwidth]{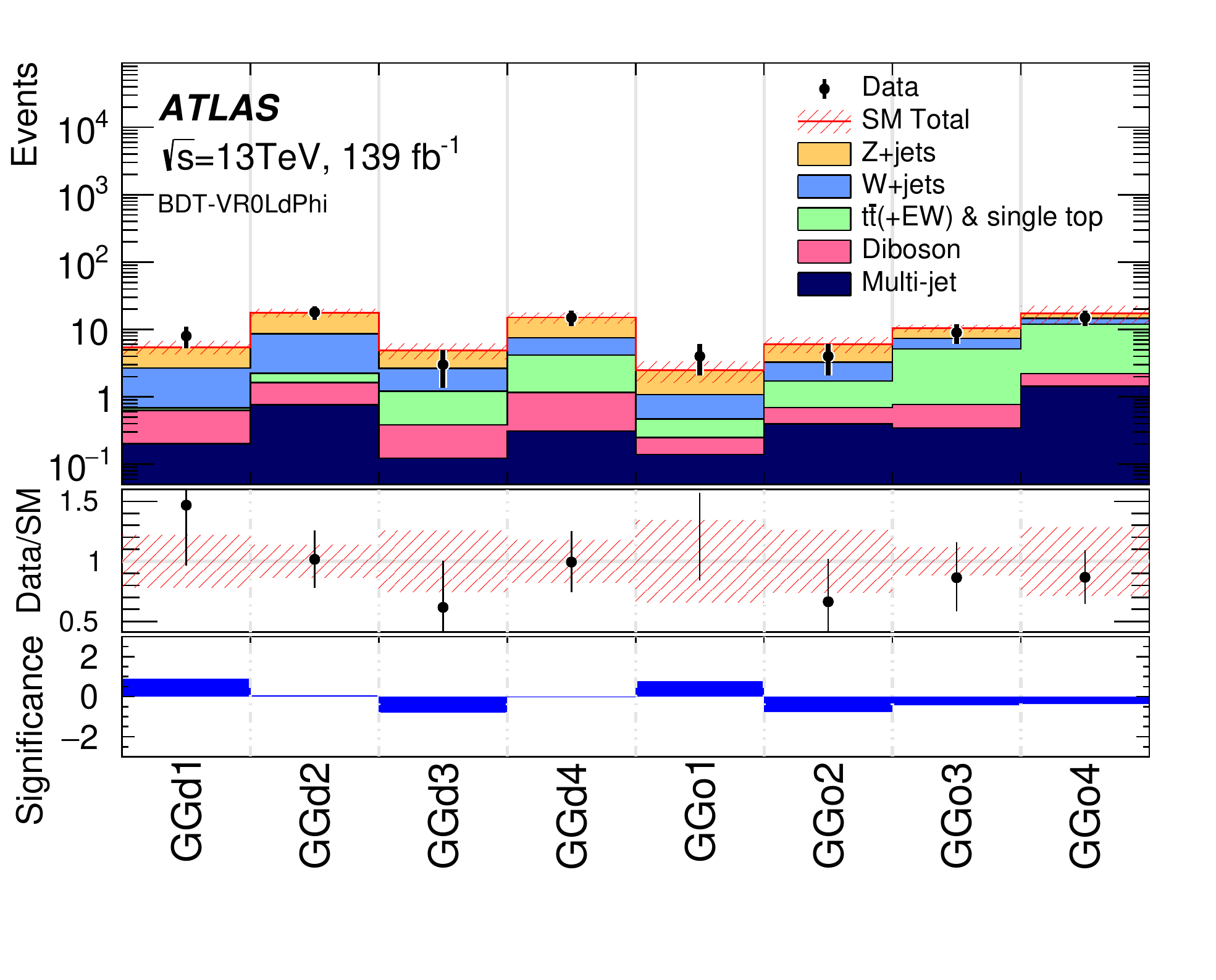}}\\
\end{center}
\caption{Observed numbers of events in data and SM background predictions for the MB-VR0LdPhi regions corresponding to the (a) MB-SSd, (b) MB-GGd, (c) MB-C signal regions from the multi-bin search, and (d) the BDT-VR0LdPhi regions corresponding to the BDT search signal regions. The lower panels in each case show the ratio of observed data yields to the total predicted background and the observed significance of the data relative to the background-only hypothesis. The significance is computed following the profile likelihood method of Ref.~\cite{Cousins:2007bmb} in the case where the observed yield exceeds the prediction, and using the same expression with an overall minus sign if the yield is below the prediction. The hatched (red) error bands indicate the combined experimental, theoretical and MC statistical uncertainties.}
\label{fig:VR0LdPhi-pulls}
\end{figure}
 
\section{Systematic uncertainties}
\label{sec:systematics}
 
The systematic uncertainties (experimental and theoretical) in the background estimates feed into the analysis via the extrapolation factors that relate observations in the control regions to background predictions in the signal regions, and via the MC modelling of minor backgrounds.
The overall post-fit background uncertainties for the multi-bin signal regions, detailed in Figure~\ref{fig:BreakdownSysSRCompressed}, range from 5\% in most of the MB-SSd regions to 60\% in one MB-GGd region. The uncertainty in this last region is dominated by a statistical fluctuation in the MC samples used to evaluate the experimental JER uncertainty, which arises from tight requirements placed on $\meff$ and $\met/\sqrt{\HT}$. This fluctuation has a negligible impact on the results presented later in this paper. In the BDT signal regions, the post-fit background uncertainties range from 8\% in BDT-GGd3 to 28\% in BDT-GGo1, as shown in Figure~\ref{fig:BreakdownSysSRCompressed}(d).
 
\begin{figure}[h]
\begin{center}
\subfigure[]{\includegraphics[width=0.49\textwidth]{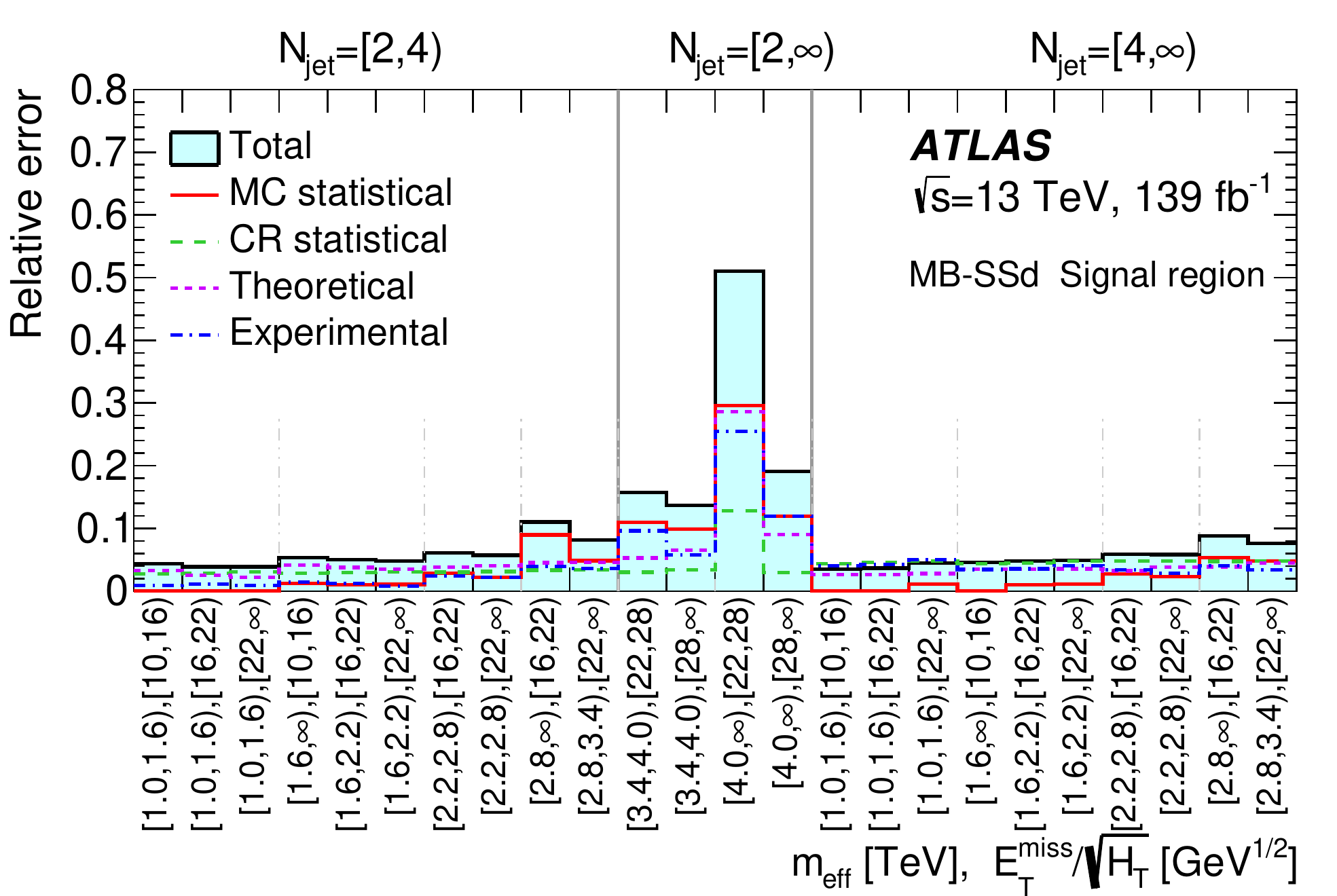}}
\subfigure[]{\includegraphics[width=0.49\textwidth]{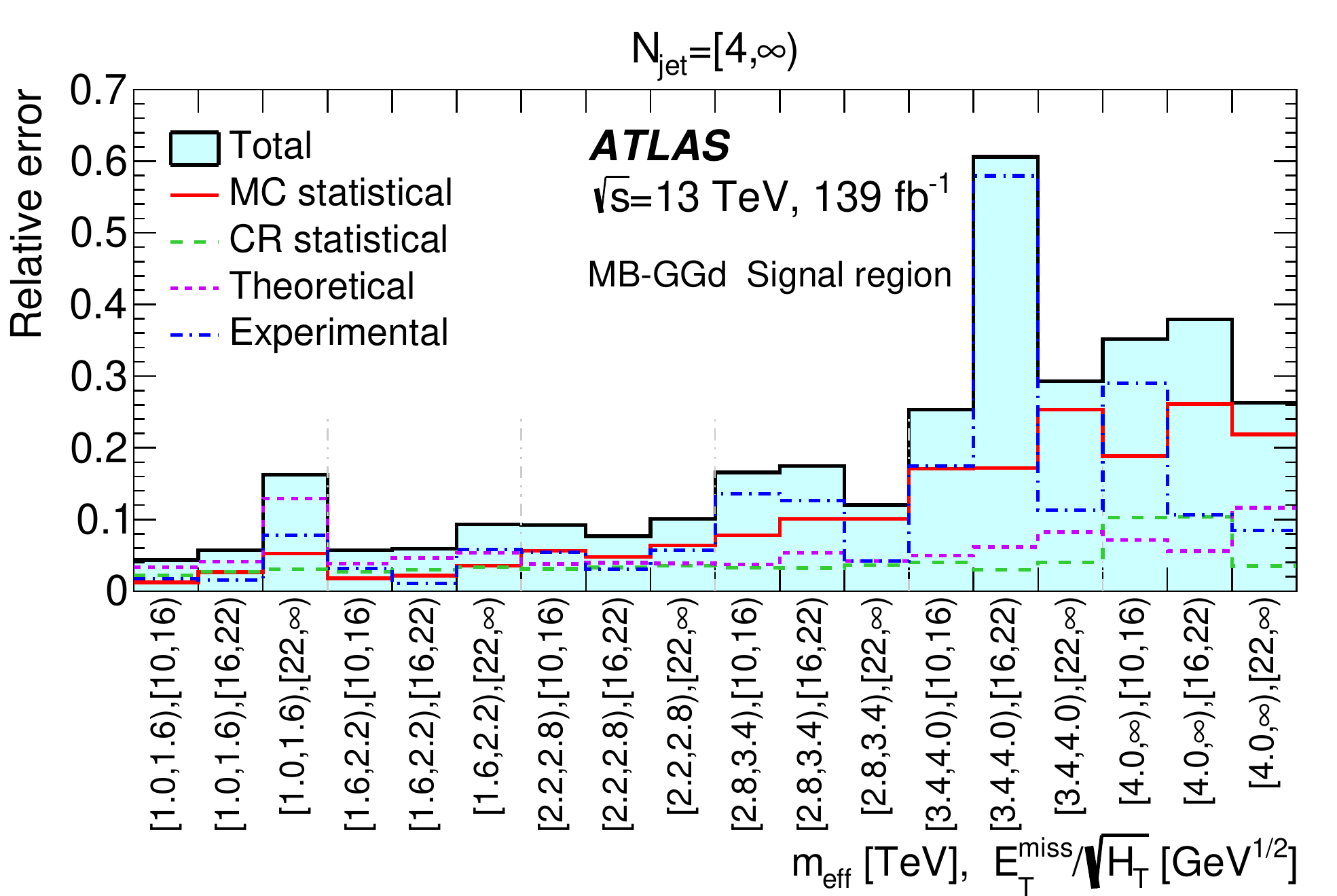}}\\
\subfigure[]{\includegraphics[width=0.49\textwidth]{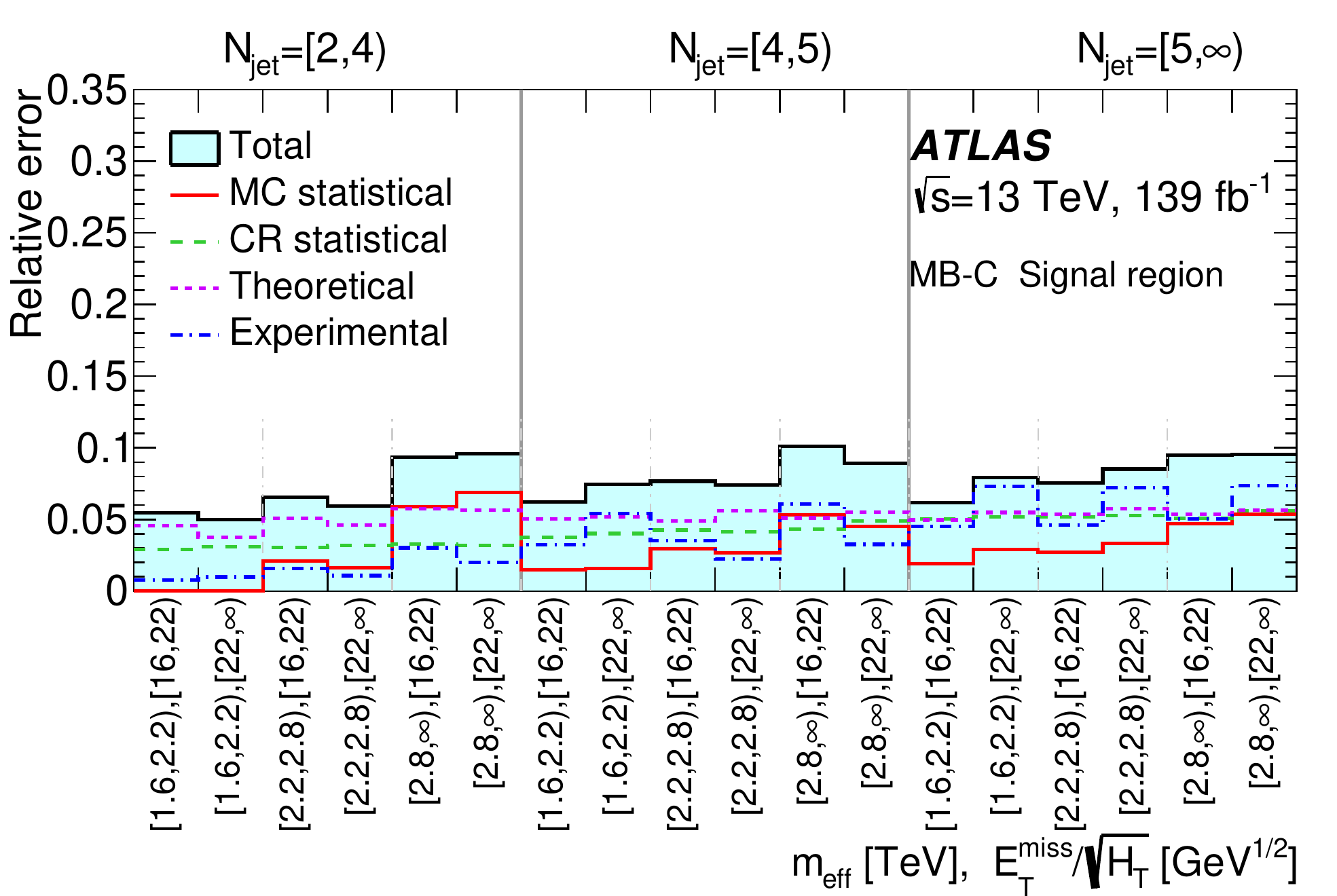}}
\subfigure[]{\includegraphics[width=0.49\textwidth]{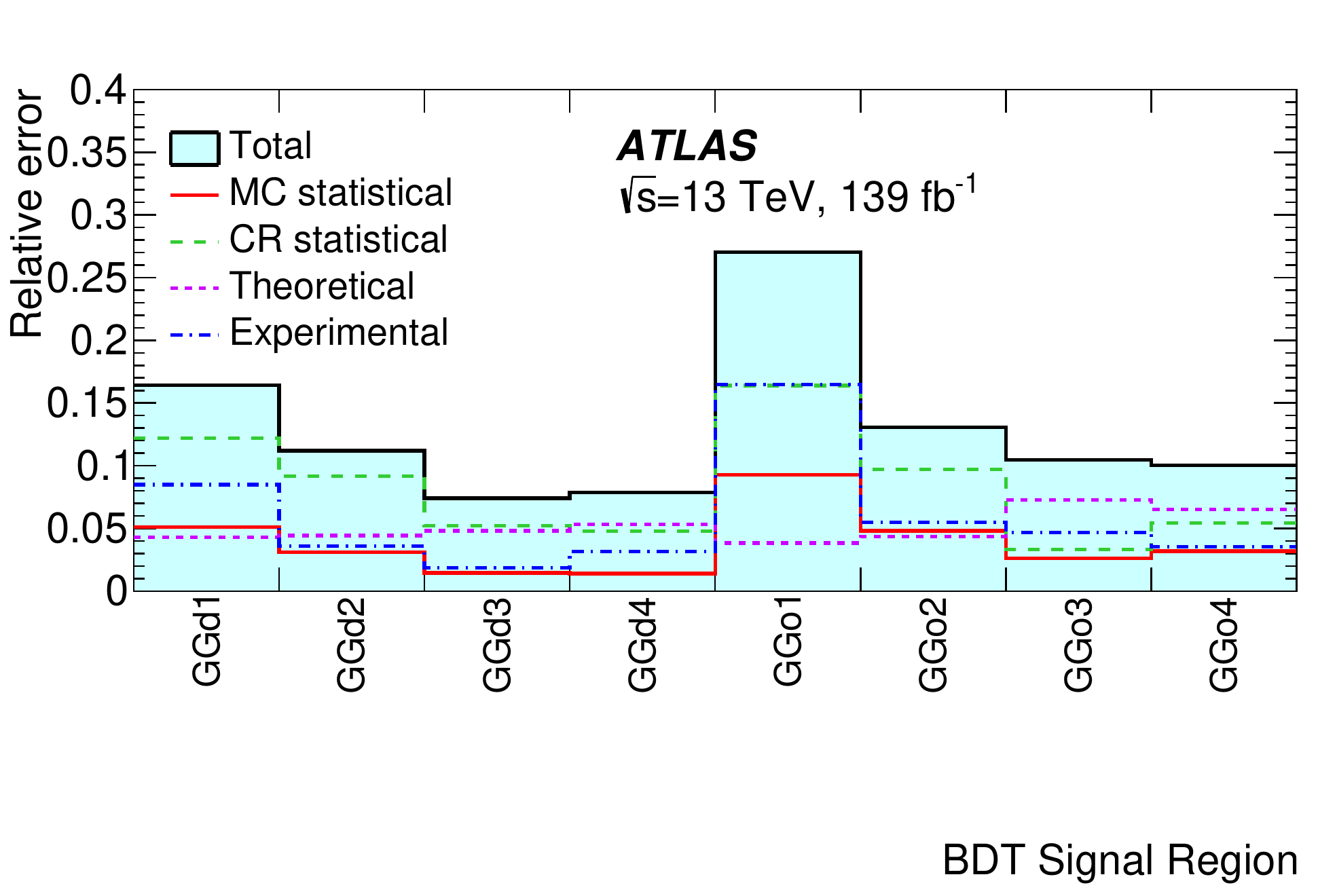}}
\end{center}
\caption{
Breakdown of the largest systematic uncertainties in the background estimates, obtained following the fits described in the text, for the (a) MB-SSd, (b) MB-GGd, (c) MB-C regions from the multi-bin search, and (d) all regions from the BDT search.
The individual uncertainties can be correlated, such that the total background uncertainty is not necessarily their sum in quadrature. }
\label{fig:BreakdownSysSRCompressed}
\end{figure}
 
For the backgrounds estimated with extrapolation factors derived from MC simulation, the primary common sources of systematic uncertainty are the jet energy scale (JES) calibration, jet energy resolution (JER), theoretical uncertainties in the modelling of these backgrounds, and limited event yields in the MC samples and data CRs. Correlations between uncertainties (for instance between JES or JER uncertainties in CRs and SRs) are taken into account where appropriate.
 
The JES and JER uncertainties are estimated using the methods discussed in Refs.~\cite{PERF-2016-04, ATL-PHYS-PUB-2015-015}. Variations according to the scale and resolution of the missing transverse momentum are also considered~\cite{PERF-2016-07}. The combined JES, JER and $\met$ uncertainty (the last arising from soft tracks not associated with other identified objects) ranges from 1\% of the expected background in multi-bin signal regions which select events with two jets, to $\sim$ 60\% in the MB-GGd region with tight requirements on $\meff$ and $\met/\sqrt{\HT}$. In the BDT search, the same uncertainties range from 1\% in BDT-GGd3 to 15\% in BDT-GGo1.
 
Uncertainties arising from theoretical modelling of background processes are estimated by comparing samples produced with different MC generators or by varying the renormalisation and factorisation scales. Uncertainties in each background from scale variations are fully correlated across regions and bins, and uncorrelated between processes. In some cases this may result in uncertainties cancelling out, while the higher-order corrections may not cancel out. Different fits with scale variations uncorrelated in all bins and regions result in limits on the excluded cross-section near the edge of the exclusion region that are weaker by up to 30\% for gluino pair production models with direct decays and moderate neutralino mass. For similar models with lower neutralino mass the degradation of the limits is a few percent.
 
The $W/Z$+jets modelling uncertainties are estimated by considering different merging (CKKW-L) and resummation scales using alternative samples, PDF and strong coupling constant ($\alpha_{\textrm s}$)  variations from the NNPDF3.0NNLO replicas~\cite{Ball:2014uwa}, and variations of factorisation and renormalisation scales in the matrix-element calculations. The last are evaluated using seven point variations, changing the renormalisation and factorisation scales by factors of 0.5 and 2.
Uncertainties in the modelling of top quark pair production are
estimated by comparing the nominal sample listed in Table~\ref{tab:montecarlo} with alternative samples.
The systematic uncertainty due to the hard-scattering process is evaluated using a comparison of the nominal sample with the sample generated with \MGMCatNLO interfaced to \pythia~8. Fragmentation and hadronisation uncertainties are assessed using a comparison of the nominal sample with a sample generated with  \textsc{Powheg-Box} interfaced to the \textsc{Herwig}~7 package~\cite{Bellm:2015jjp} for parton showering. Initial-state radiation uncertainties, as well as uncertainties arising from factorisation and renormalisation scale assumptions, and uncertainties from the \pythia~8 parton shower settings, are calculated by varying the relevant parameters described in Ref.~\cite{ATL-PHYS-PUB-2016-020} and encapsulated in dedicated event weights in the nominal sample.
Uncertainties in diboson production due to PDF, strong coupling constant, and renormalisation and factorisation scale uncertainties are estimated in a way similar to that for the $W/Z$+jets modelling uncertainties.
The combined theoretical uncertainty ranges from 3\% to 13\% in the multi-bin signal regions, except in a single bin of the MB-SSd region with tight kinematic requirements, where it rises to 30\%. The combined theoretical uncertainty lies in the range 3\% to 8\% in the BDT search regions.
 
The impact of lepton reconstruction uncertainties, and of the uncertainties related to the $b$-tag/$b$-veto efficiency, on the overall background uncertainty is found to be negligible for all SRs.
 
The uncertainties arising from the data-driven correction procedure applied to events selected in the CR$\gamma$ region, described in Section~\ref{sec:background}, are included in Figure~\ref{fig:BreakdownSysSRCompressed}
under `CR statistical uncertainty'. The total uncertainties due to CR
data sample sizes range from 3\% to 14\% for multi-bin SRs and from 5\% to
16\% for BDT SRs. The statistical uncertainty arising from the use of
MC samples is largest in the MB-SSd and MB-GGd SRs (up to 30\%) and the BDT-GGo1 SR
(8\%). A uniform uncertainty of 100\% related to the multi-jet background estimates
is applied to the multi-jet yield in all SRs, motivated by studies carried out in a previous iteration of this analysis \cite{SUSY-2011-20}. In most of the SRs the impact of these uncertainties is negligible, and the maximum resulting contribution to the overall background uncertainty is less than 1\%.
Uncertainties in background estimates arising from the reweighting of MC samples to match the distribution of the mean number of pile-up interactions observed in the dataset are found to be negligible.

Experimental uncertainties (JES, JER and $\met$) and MC statistical uncertainty in the SUSY signal samples are estimated in the same way as for the background
and are less than a few percent for most models. The signal cross-section uncertainty is estimated
by computing the changes when the renormalisation and factorisation scales, PDF and the strong coupling constant are varied.
The uncertainties in the generation of ISR and FSR in SUSY signal events are estimated by varying generator tunes in the simulation as well as scales
used in the matrix-element generator as a function of the mass difference, $\Delta m$, between the gluino (or squark) and the $\ninoone$.
When $\Delta m = 25$~\GeV, this uncertainty ranges from $\sim$10\% for low jet multiplicities to 25--30\% for large jet multiplicities. At higher values of $\Delta m$ the uncertainty falls steeply and is negligible for $\Delta m > 400$~\GeV.

\section{Results, interpretation and limits}
\label{sec:results}

Distributions of $\meff$ and $\met/\sqrt{\HT}$ for events satisfying the selection criteria for any of the bins in the (a) MB-SSd, (b) MB-GGd or (c) MB-C signal regions
are shown in Figures~\ref{fig:MB-SR_meff} and~\ref{fig:MB-SR_metSig} for data and for MC samples normalised using the background-only fit described in Section~\ref{sec:background}. Similarly, distributions of the final discriminating variable used in the BDT search obtained after applying the SR selection criteria but before the final selection on the variable is applied, are shown in Figure~\ref{fig:BDT_score} for selected signal regions.
Examples of SUSY signals are also shown for illustration. These signals correspond to the processes to which each SR is primarily sensitive: $\squark\squark$ production for the lower jet multiplicity SRs and $\gluino\gluino$ production for the higher jet multiplicity SRs.
In these figures, data and background distributions largely agree within uncertainties.
\begin{figure}[H]
\begin{center}
\subfigure[]{\includegraphics[width=0.49\textwidth]{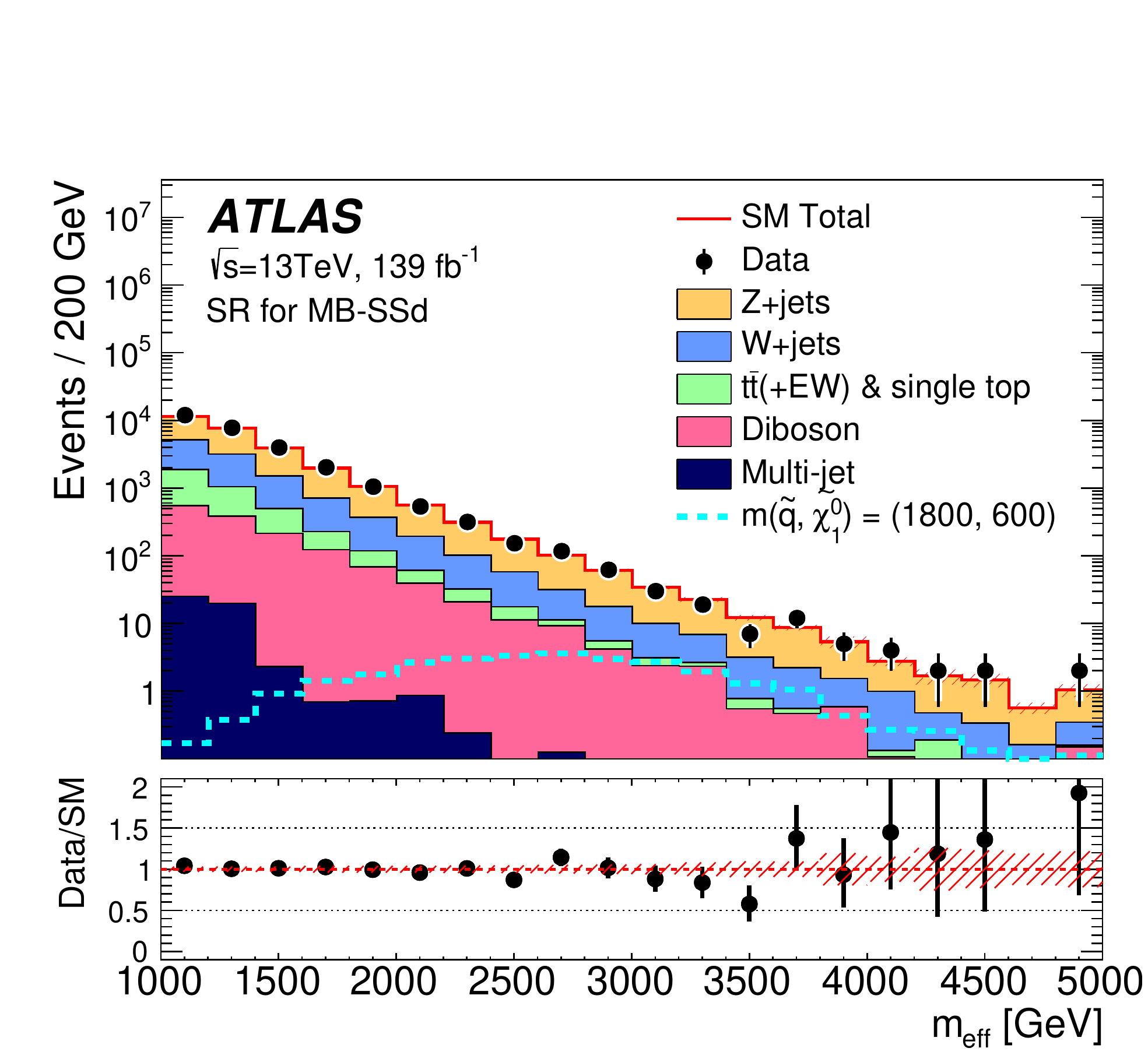}}
\subfigure[]{\includegraphics[width=0.49\textwidth]{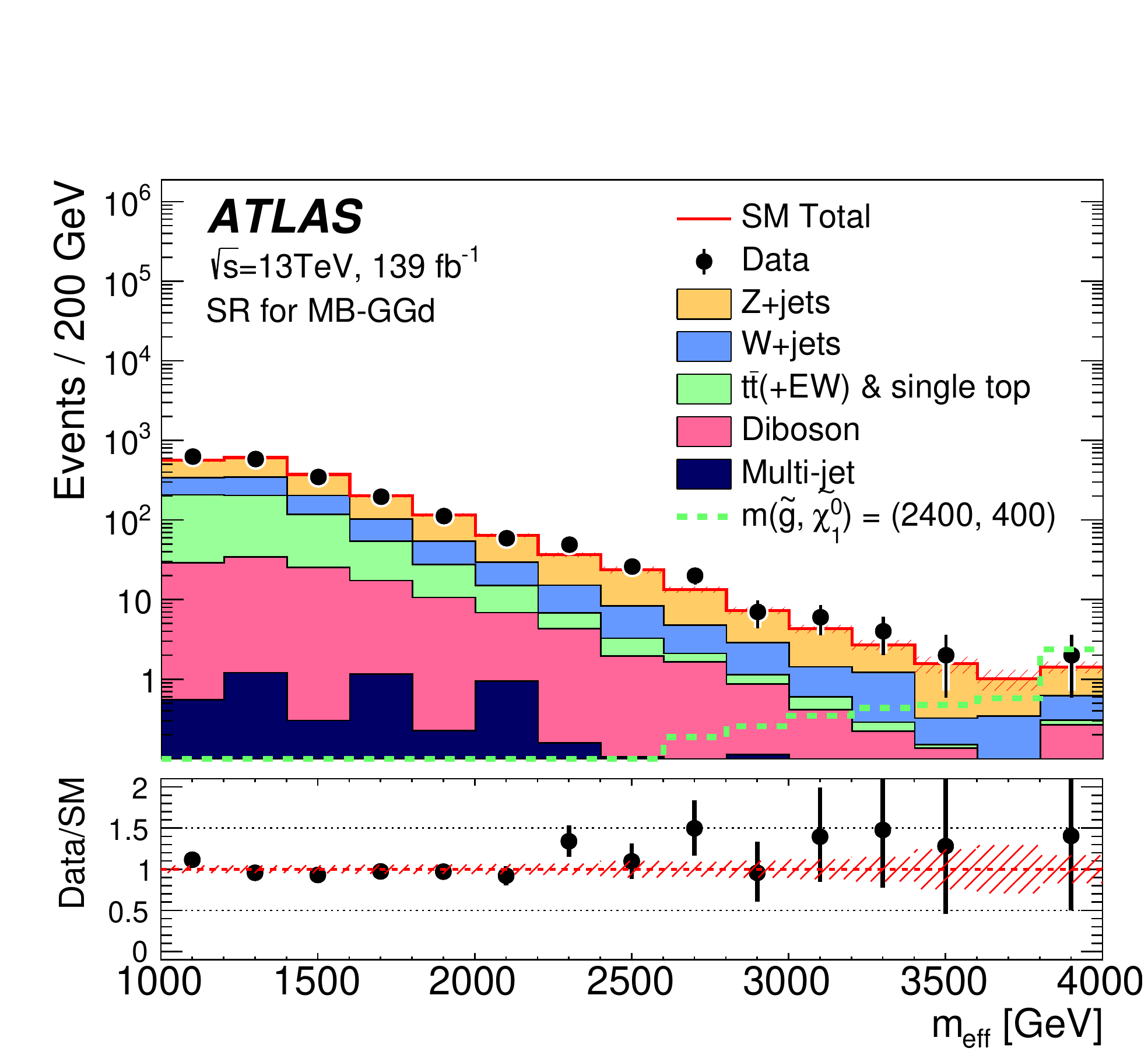}}
\subfigure[]{\includegraphics[width=0.49\textwidth]{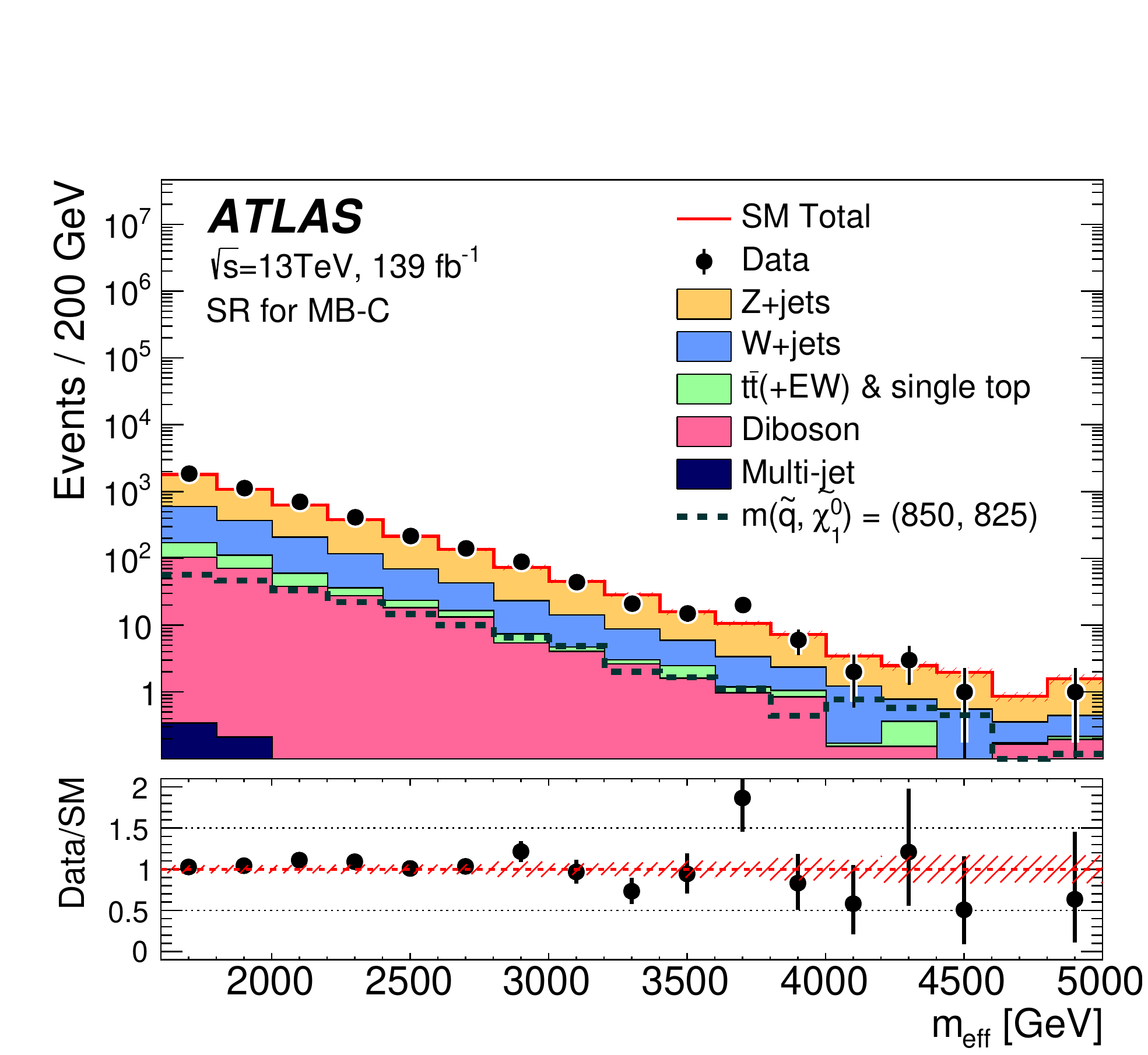}}
\end{center}
\caption{Observed $\meff$ distributions of events satisfying the selection criteria for any of the bins in the (a) MB-SSd,
(b) MB-GGd or (c) MB-C signal regions. The histograms show the MC background predictions normalised by the background-only fit described in the text. The lower panels show the ratio of data to the background prediction. The hatched (red) error bands indicate combined post-fit experimental, theoretical and MC statistical uncertainties, with the experimental and theoretical uncertainties calculated using the coarser SR binning used in the fit rather than the finer binning used in the histograms.
Expected distributions for benchmark signal model points, normalised using the approximate NNLO+NNLL cross-section (Section~\ref{sec:montecarlo}) times integrated luminosity, are also shown for comparison (masses in \GeV). In each case the overflow is included in the final bin.
}
\label{fig:MB-SR_meff}
\end{figure}
 
\begin{figure}[H]
\begin{center}
\subfigure[]{\includegraphics[width=0.49\textwidth]{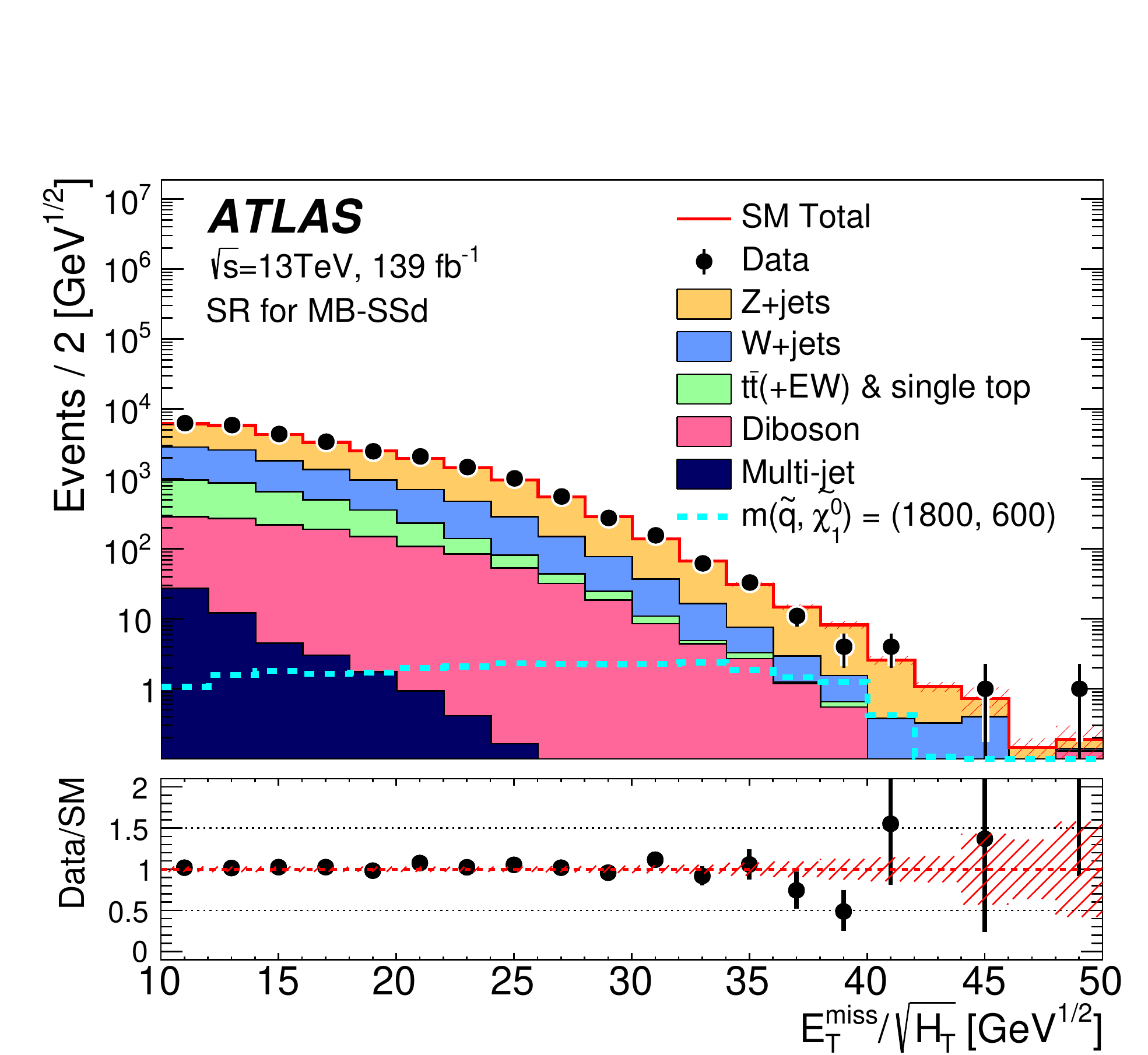}}
\subfigure[]{\includegraphics[width=0.49\textwidth]{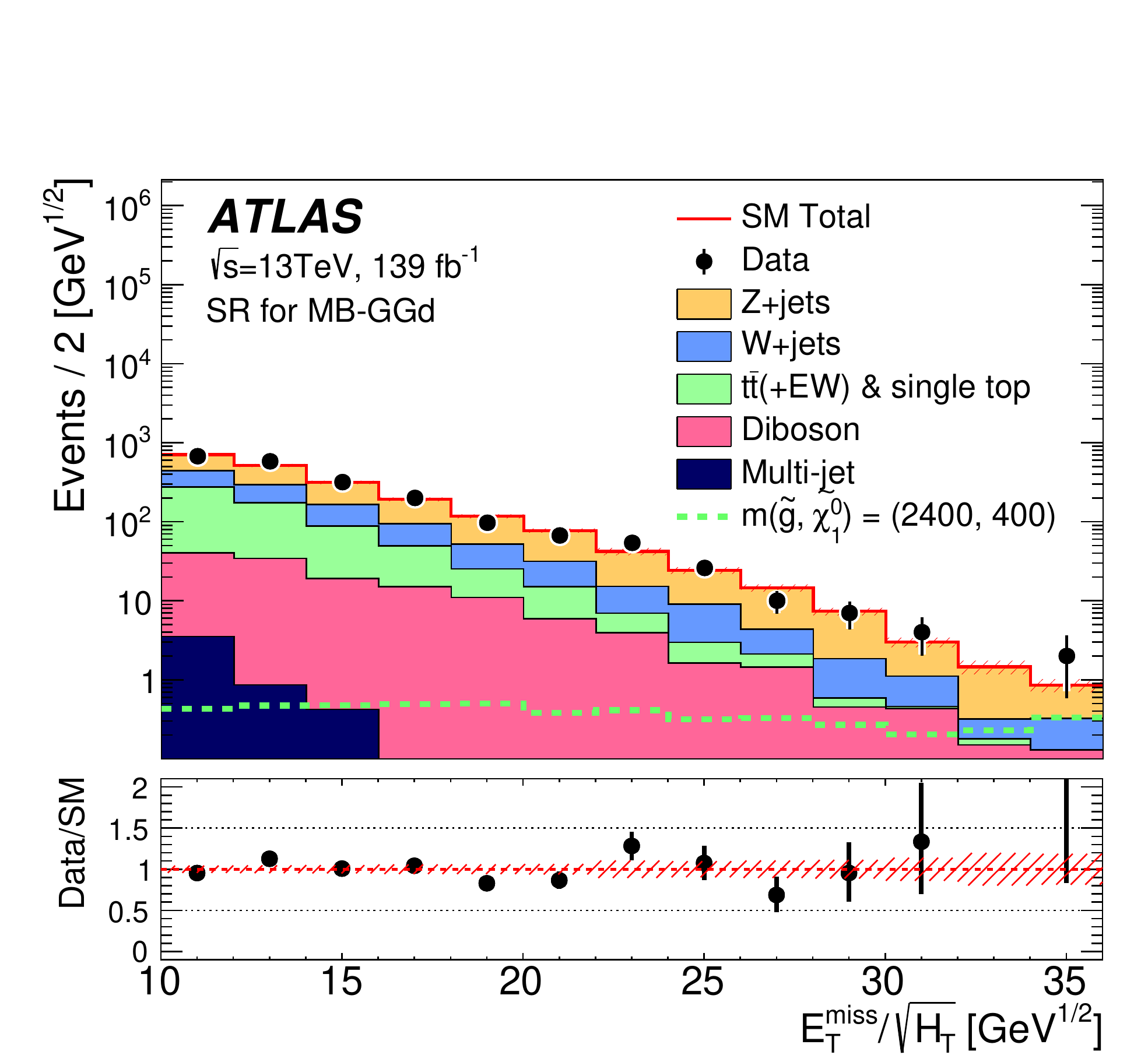}}
\subfigure[]{\includegraphics[width=0.49\textwidth]{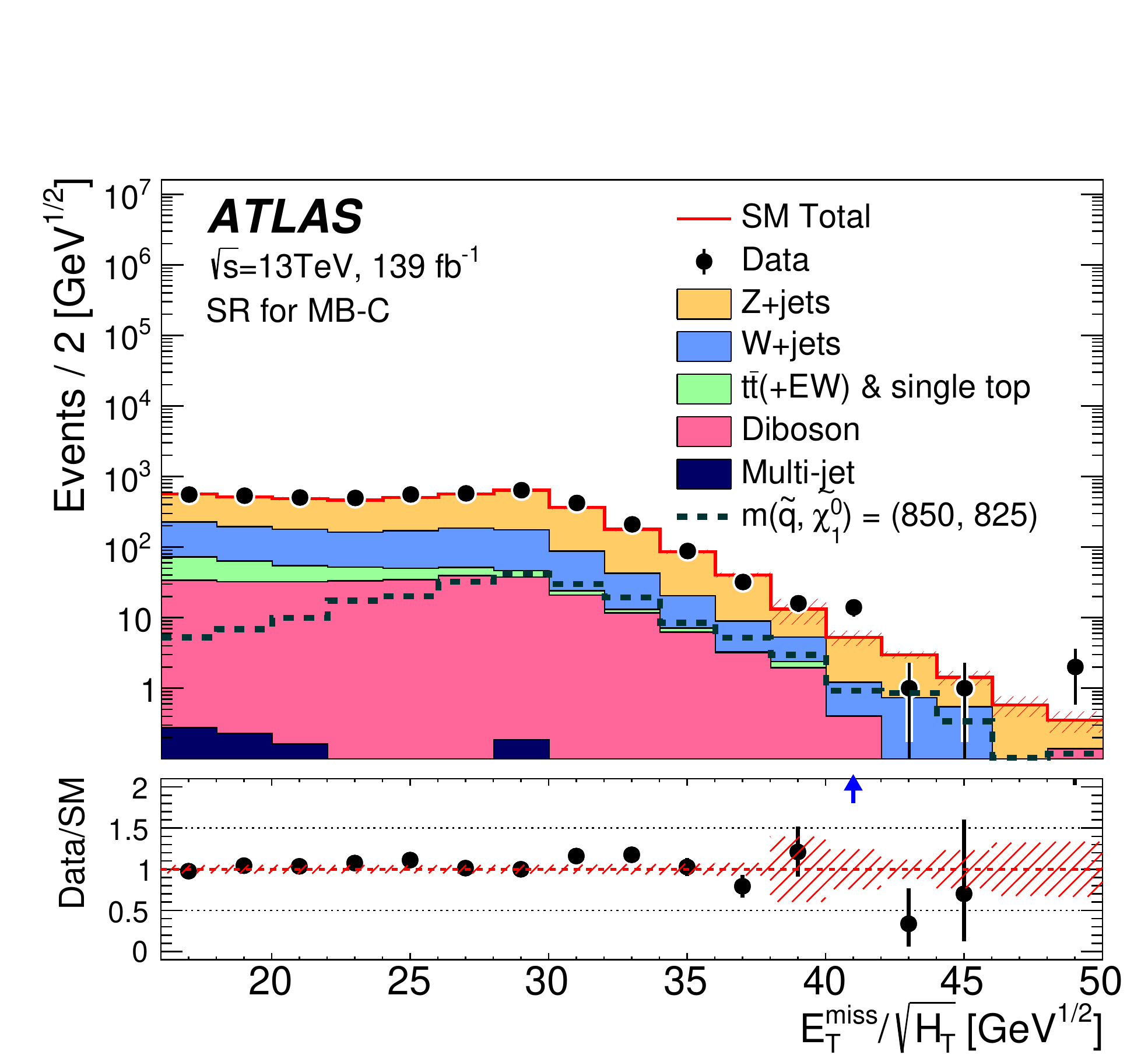}}
\end{center}
\caption{Observed $\met/\sqrt{\HT}$ distributions of events satisfying the selection criteria for any of the bins in the (a) MB-SSd,
(b) MB-GGd or (c) MB-C signal regions. The histograms show the MC background predictions normalised by the background-only fit described in the text. The lower panels show the ratio of data to the background prediction. The hatched (red) error bands indicate combined post-fit experimental, theoretical and MC statistical uncertainties, with the experimental and theoretical uncertainties calculated using the coarser SR binning used in the fit rather than the finer binning used in the histograms.
Expected distributions for benchmark signal model points, normalised using the approximate NNLO+NNLL cross-section (Section~\ref{sec:montecarlo}) times integrated luminosity, are also shown for comparison (masses in \GeV). In each case the overflow is included in the final bin.
}
\label{fig:MB-SR_metSig}
\end{figure}
 
\begin{figure}[H]
\begin{center}
\subfigure[]{\includegraphics[width=0.49\textwidth]{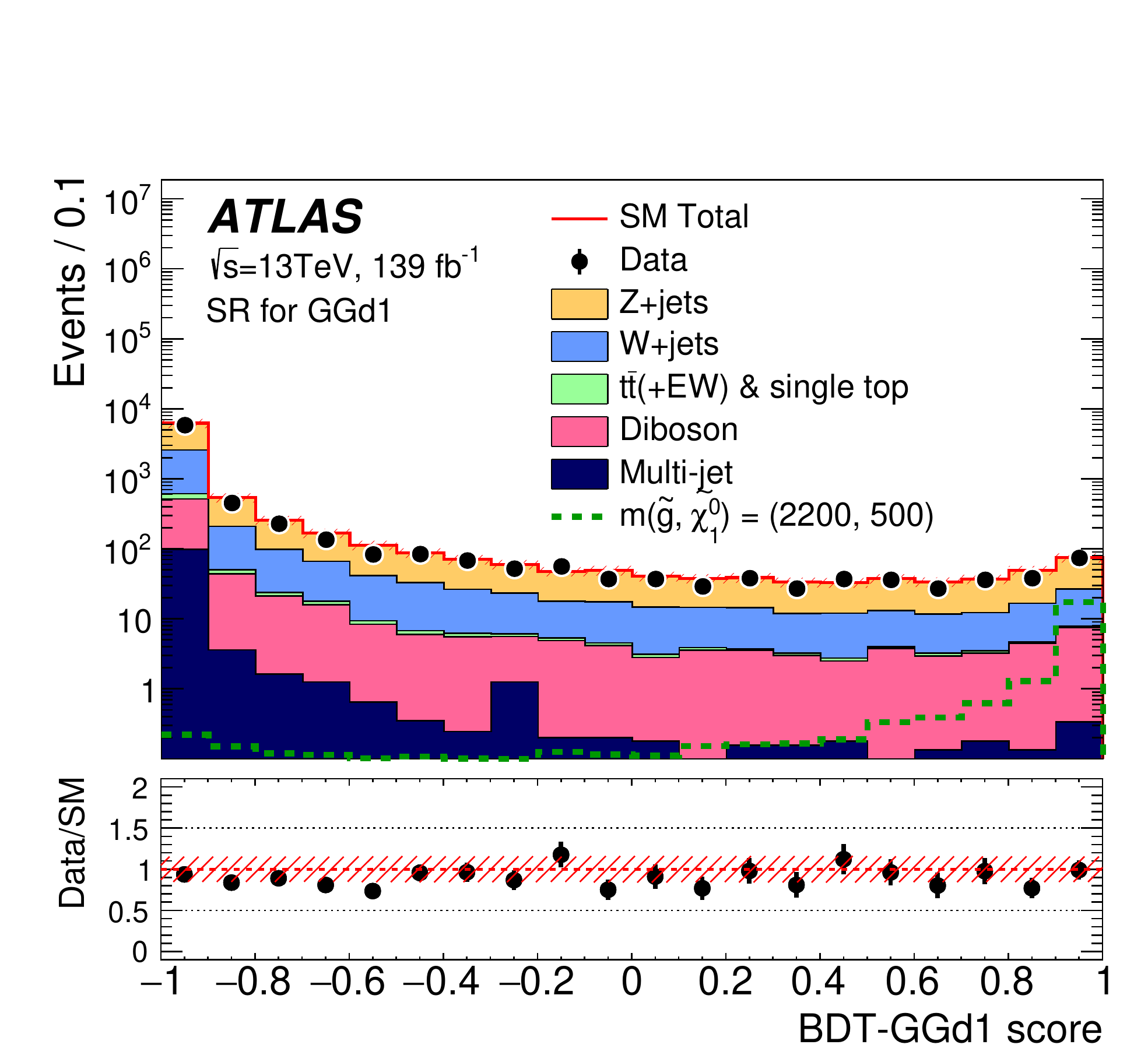}}
\subfigure[]{\includegraphics[width=0.49\textwidth]{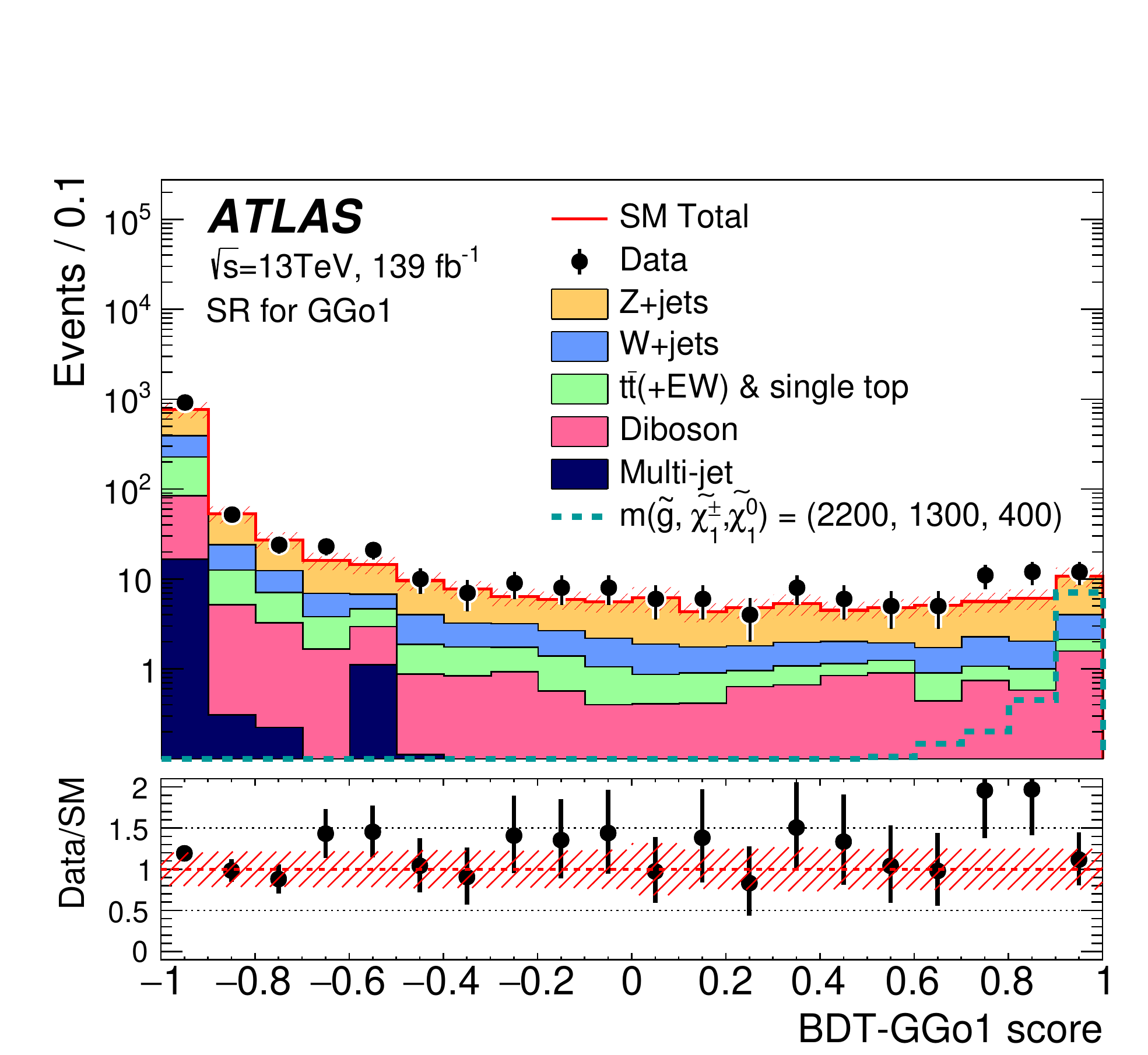}}
\end{center}
\caption{Observed BDT score distributions for the (a) BDT-GGd1 and
(b) BDT-GGo1 regions obtained after applying the SR selection criteria but before the final selection on this quantity is applied.
The histograms show the MC background predictions normalised by the background-only fit described in the text. The lower panels show the ratio of data to the background prediction. The hatched (red) error bands indicate the combined post-fit experimental, theoretical and MC statistical uncertainties, with the experimental and theoretical uncertainties calculated using the SR selection used in the fit rather than on a histogram bin-by-bin basis.
Expected distributions for benchmark signal model points, normalised using the approximate NNLO+NNLL cross-section (Section~\ref{sec:montecarlo}) times integrated luminosity, are also shown for comparison (masses in \GeV).
}
\label{fig:BDT_score}
\end{figure}

The number of events observed in the data and the number of SM events expected to enter each of the signal regions determined using the background-only fit are shown in Table~\ref{tab:results_BDT} for the BDT search and in Table~\ref{tab:p0_UL} for the model-independent search.  The results of all searches presented in this paper are also summarised in Figures~\ref{fig:PlotSR} and~\ref{fig:PlotSRb}.
To quantify the level of agreement
between background predictions and observed yields and to set upper
limits on the number of BSM signal events in each SR, a model-independent fit is used \cite{HFpaper}. This fit
proceeds in the same way as the background-only fit, where yields in
the CRs are used to constrain the predictions of backgrounds in each
SR, while the SR yield is also used in the likelihood function with an
additional parameter-of-interest describing potential signal
contributions. The observed and expected upper limits at 95\%
confidence level (CL) on the number of events from BSM phenomena for
each signal region ($S_{\textrm{obs}}^{95}$ and $S_{\textrm{exp}}^{95}$) are
derived using the CL$_{\textrm s}$ prescription \cite{Read:2002hq}, neglecting any possible signal contamination in the CRs. These limits, when normalised by the integrated luminosity of the data sample, may be interpreted as upper limits on the visible cross-section of BSM physics ($\langle\epsilon\sigma\rangle_{\textrm{obs}}^{95}$), where the visible cross-section is defined as the product of production cross-section, acceptance and efficiency. The model-independent fit is also used to compute the one-sided $p$-value ($p_0$) of the background-only hypothesis, which quantifies the statistical significance of an excess. The fit results are evaluated using asymptotic formulae~\cite{statforumlimits} except in SRs where less than 10 events are observed, where pseudo-experiments are used.
No statistically significant deviation from the background expectation is found for any of the presented search strategies.
 
 
\begin{table}
\scriptsize
\begin{center}
\renewcommand{\arraystretch}{1.4}
\vspace*{-0.035\textwidth}
\begin{tabular}{lrrrr}
\toprule
\multicolumn{5}{c}{\textbf{BDT regions}} \\ 
Signal Region & \textbf{GGd1} & \textbf{GGd2} & \textbf{GGd3} & \textbf{GGd4} \\
\midrule
 
Total bkg pre-fit &  $29$               &  $56$               &  $253$               &  $348$               \\
\midrule
\multicolumn{5}{c}{Fitted background events} \\ \midrule
Diboson & $3.0 \pm 0.9$ & $4.9 \pm 1.4$ & $21 \pm 5$ & $26 \pm 7$ \\
$Z/\gamma^{*}$+jets & $20 \pm 4$ & $33 \pm 5$ & $139 \pm 14$ & $180 \pm 18$ \\
$W$+jets & $7.1 \pm 2.6$ & $13 \pm 4$ & $48 \pm 8$ & $52 \pm 9$ \\
$\ttbar$(+EW) + single top &  $0.1_{-0.1}^{+0.3}$               &  $0.6_{-0.6}^{+0.8}$               & $16 \pm 5$ & $39 \pm 11$ \\
Multi-jet &  $0.1_{-0.1}^{+0.1}$               &  --               &  $0.1_{-0.1}^{+0.1}$               &  $0.1_{-0.1}^{+0.1}$               \\
 
Total bkg post-fit & $30 \pm 5$ & $52 \pm 6$ & $223 \pm 17$ & $298 \pm 23$ \\
 
Observed &  $34$                     &  $68$                     &  $227$                     &  $291$                     \\
\midrule

$\langle\epsilon{\mathrm \sigma}\rangle_{\textrm{obs}}^{95}$ [fb]   & $0.13$ & $0.24$ &  $0.33$ & $0.36$\\
$S_{\textrm{obs}}^{95}$   &  $18$ &  $33$ &  $46$ & $50$ \\
$S_{\textrm{exp}}^{95}$   &  $ { 15 }^{ +6 }_{ -4 }$   & $ { 20 }^{ +8 }_{ -6 }$   &   $ { 44 }^{ +17 }_{ -12 }$  &  $ { 54 }^{ +21 }_{ -15 }$   \\
$p_{0}$ ($Z$)     & $ 0.30$~$(0.51)$ & $ 0.05$~$(1.60)$ & $ 0.44$~$(0.15)$ & $ 0.50$~$(0.00)$\\
 
\bottomrule
\end{tabular}
 
\begin{tabular}{lrrrr}
\toprule
 
Signal Region & \textbf{GGo1} & \textbf{GGo2} & \textbf{GGo3} & \textbf{GGo4} \\
\midrule

Total bkg pre-fit &  $7$               &  $25$               &  $111$               &  $177$               \\
\midrule
\multicolumn{5}{c}{Fitted background events} \\ \midrule
Diboson & $0.6 \pm 0.2$ & $2.2 \pm 0.6$ & $6.6 \pm 2.2$ & $6.8 \pm 2.1$ \\
$Z/\gamma^{*}$+jets & $3.8 \pm 1.3$ & $10.9 \pm 1.9$ & $35 \pm 6$ & $39 \pm 7$ \\
$W$+jets & $0.9 \pm 0.5$ & $3.9 \pm 1.3$ & $16 \pm 4$ & $27 \pm 6$ \\
$\ttbar$(+EW) + single top & $0.2 \pm 0.2$ & $1.3 \pm 0.8$ & $28 \pm 6$ & $85 \pm 14$ \\
Multi-jet &  --               &  --               &  $0.1_{-0.1}^{+0.1}$               &  $0.7_{-0.7}^{+0.7}$               \\
 
Total bkg post-fit & $5.5 \pm 1.5$ & $18 \pm 2.4$ & $85 \pm 9$ & $159 \pm 16$ \\
 
Observed &  $6$                     &  $25$                     &  $80$                     &  $135$                     \\
\midrule

$\langle\epsilon{\mathrm \sigma}\rangle_{\textrm{obs}}^{95}$ [fb]   & $0.05$ & $0.12$ &  $0.16$ & $0.18$ \\
$S_{\textrm{obs}}^{95}$   &  $7.1$ & $17$  & $22$ & $25$ \\
$S_{\textrm{exp}}^{95}$   & $ { 6.9 }^{ +2.3 }_{ -1.6 }$ & $ { 11 }^{ +5 }_{ -2 }$ & $ { 25 }^{ +10 }_{ -7 }$ & $ { 37 }^{ +14 }_{ -10 }$\\
$p_{0}$ ($Z$)     & $ 0.49$~$(0.01)$ & $ 0.10$~$(1.28)$ & $ 0.50$~$(0.00)$  & $ 0.50$~$(0.00)$\\
\bottomrule
\end{tabular}
\end{center}
\vspace*{-0.01\textheight}\caption[p0 and UL]{Numbers of events observed in the signal regions used in the BDT search compared with background expectations obtained from the fits described in the text. Empty cells (indicated by a `--') correspond to estimates lower than $0.01$. The $p$-values ($p_{0}$) give the probabilities of the observations being consistent with the estimated backgrounds. For an observed number of events lower than expected, the $p$-value is capped at 0.5. Between parentheses, $p$-values are also presented in terms of the number of equivalent Gaussian standard deviations ($Z$).
Also shown are 95\% CL upper limits on the visible cross-section ($\langle\epsilon\sigma\rangle_{\textrm{obs}}^{95}$),
the visible number of signal events ($S_{\textrm{obs}}^{95}$ ) and the number of signal events ($S_{\textrm{exp}}^{95}$)
given the expected number of background events (and $\pm 1\sigma$ excursions of the expectation).}
\label{tab:results_BDT}
\end{table}
 
 
\begin{table}[h]
\scriptsize
\begin{center}
\renewcommand{\arraystretch}{1.4}
\vspace*{-0.035\textwidth}
\begin{tabular}{lrrrrr}
\toprule
\multicolumn{6}{c}{\textbf{Model independent regions}} \\
Signal Region & \textbf{SR2j-1600} & \textbf{SR2j-2200} & \textbf{SR2j-2800} & \textbf{SR4j-1000} & \textbf{SR4j-2200} \\
\midrule

Total bkg pre-fit &  $2120$               &  $979$               &  $82$               &  $610$               &  $71$               \\
\midrule
\multicolumn{6}{c}{Fitted background events} \\ \midrule
Diboson & $130 \pm 29$ & $74 \pm 17$ & $5.8 \pm 1.7$ & $44 \pm 12$ & $6.3 \pm 1.7$ \\
$Z/\gamma^{*}$+jets & $1510 \pm 120$ & $670 \pm 50$ & $64 \pm 7$ & $281 \pm 23$ & $35 \pm 4$ \\
$W$+jets & $500 \pm 50$ & $225 \pm 16$ & $15.5 \pm 2.4$ & $144 \pm 12$ & $15.4 \pm 1.9$ \\
$\ttbar$(+EW) + single top & $44 \pm 9$ & $14 \pm 5$ & $1.4 \pm 0.8$ & $67 \pm 14$ & $2.4 \pm 0.9$ \\
Multi-jet & $0.2 \pm 0.2$ & $0.3 \pm 0.3$ & -- & $0.2 \pm 0.2$ &  --               \\
 
Total bkg post-fit & $2190 \pm 130$ & $980 \pm 50$ & $87 \pm 8$ & $536 \pm 32$ & $60 \pm 5$ \\
 
Observed &  $2111$                     &  $971$                     &  $78$                     &  $535$                     &  $60$                     \\
\midrule

$\langle\epsilon{\mathrm \sigma}\rangle_{\textrm{obs}}^{95}$ [fb] & 1.47  & 0.78 & 0.14 & 0.52 & 0.14 \\
$S_{\textrm{obs}}^{95}$   &  204 &108 &19 &72 &19\\
$S_{\textrm{exp}}^{95}$   &$ { 247 }^{ +90 }_{ -67 }$ &$ { 114 }^{ +43 }_{ -31 }$ &  $ { 24 }^{ +9 }_{ -7 }$  &  $ { 73 }^{ +27 }_{ -20 }$  &  $ { 19 }^{ +8 }_{ -5 }$ \\
$p_{0}$ ($Z$)     &0.50 (0.00)&0.50 (0.00) &0.50 (0.00) &0.50 (0.00) &0.48 (0.05)  \\
 
\bottomrule
\end{tabular}

\begin{tabular}{lrrrrr}
\toprule
Signal Region & \textbf{SR4j-3400} & \textbf{SR5j-1600} & \textbf{SR6j-1000} & \textbf{SR6j-2200} & \textbf{SR6j-3400} \\
\midrule

Total bkg pre-fit &  $7$               &  $427$               &  $29$               &  $7$               &  $1.1$               \\
\midrule
\multicolumn{6}{c}{Fitted background events} \\ \midrule
Diboson & $0.7 \pm 0.2$ & $36 \pm 10$ & $1.8 \pm 0.6$ &  $0.3_{-0.3}^{+0.8}$               & $0.1 \pm 0.0$ \\
$Z/\gamma^{*}$+jets & $3.3 \pm 0.8$ & $170 \pm 16$ & $9.3 \pm 1.8$ & $2.4 \pm 0.6$ & $0.3 \pm 0.2$ \\
$W$+jets & $1.6 \pm 0.4$ & $80 \pm 7$ & $7.2 \pm 1.6$ & $1.5 \pm 0.5$ & $0.4 \pm 0.3$ \\
$\ttbar$(+EW) + single top &  $0.1_{-0.1}^{+0.1}$               & $33 \pm 6$ & $2.7 \pm 1.5$ & $0.4 \pm 0.3$ &  --               \\
Multi-jet &  $0.1_{-0.1}^{+0.1}$               & $0.2 \pm 0.2$ & -- & -- & -- \\
 
Total bkg post-fit & $5.7 \pm 1.0$ & $319 \pm 20$ & $21 \pm 3$ & $4.6 \pm 1.0$ & $0.8 \pm 0.4$ \\
 
Observed &  $4$                     &  $320$                     &  $25$                     &  $5$                     &  $0$                     \\
\midrule

$\langle\epsilon{\mathrm \sigma}\rangle_{\textrm{obs}}^{95}$ [fb]  &0.04 &0.37 &0.11 &0.04 &0.02 \\
$S_{\textrm{obs}}^{95}$  &5.0 &51 &16 &6.2 &3.1  \\
$S_{\textrm{exp}}^{95}$  &  $ { 6 .2}^{ +2.6 }_{ -1.7 }$  & $ { 51 }^{ +20 }_{ -14 }$  & $ { 12 }^{ +5 }_{ -3 }$  &  $ { 6.1 }^{ +2.1 }_{ -1.4 }$  &  $ { 3.1 }^{ +1.2 }_{ -0.0 }$   \\
$p_{0}$ ($Z$)  &0.50 (0.00) & $ 0.48$~$(0.06)$ &$ 0.24$~$(0.71)$ &0.47 (0.06) &0.50 (0.00)\\
 
\bottomrule
\end{tabular}
\end{center}
\vspace*{-0.01\textheight}\caption[p0 and UL]{Numbers of events observed in the signal regions used in the model-independent search, compared with background expectations obtained from the fits described in the text. Empty cells (indicated by a `--') correspond to estimates lower than $0.01$.
The $p$-values ($p_{0}$) give the probabilities of the observations being consistent with the estimated backgrounds. For an observed number of events lower than expected, the $p$-value is capped at 0.5. Between parentheses, $p$-values are also presented in terms of the number of equivalent Gaussian standard deviations ($Z$).
Also shown are 95\% CL upper limits on the visible cross-section ($\langle\epsilon\sigma\rangle_{\textrm{obs}}^{95}$),
the visible number of signal events ($S_{\textrm{obs}}^{95}$ ) and the number of signal events ($S_{\textrm{exp}}^{95}$)
given the expected number of background events (and $\pm 1\sigma$ excursions of the expectation).}
\label{tab:p0_UL}
\end{table}
 
 
\clearpage
 
\begin{figure}
\begin{center}
\subfigure[]{\includegraphics[width=0.49\textwidth]{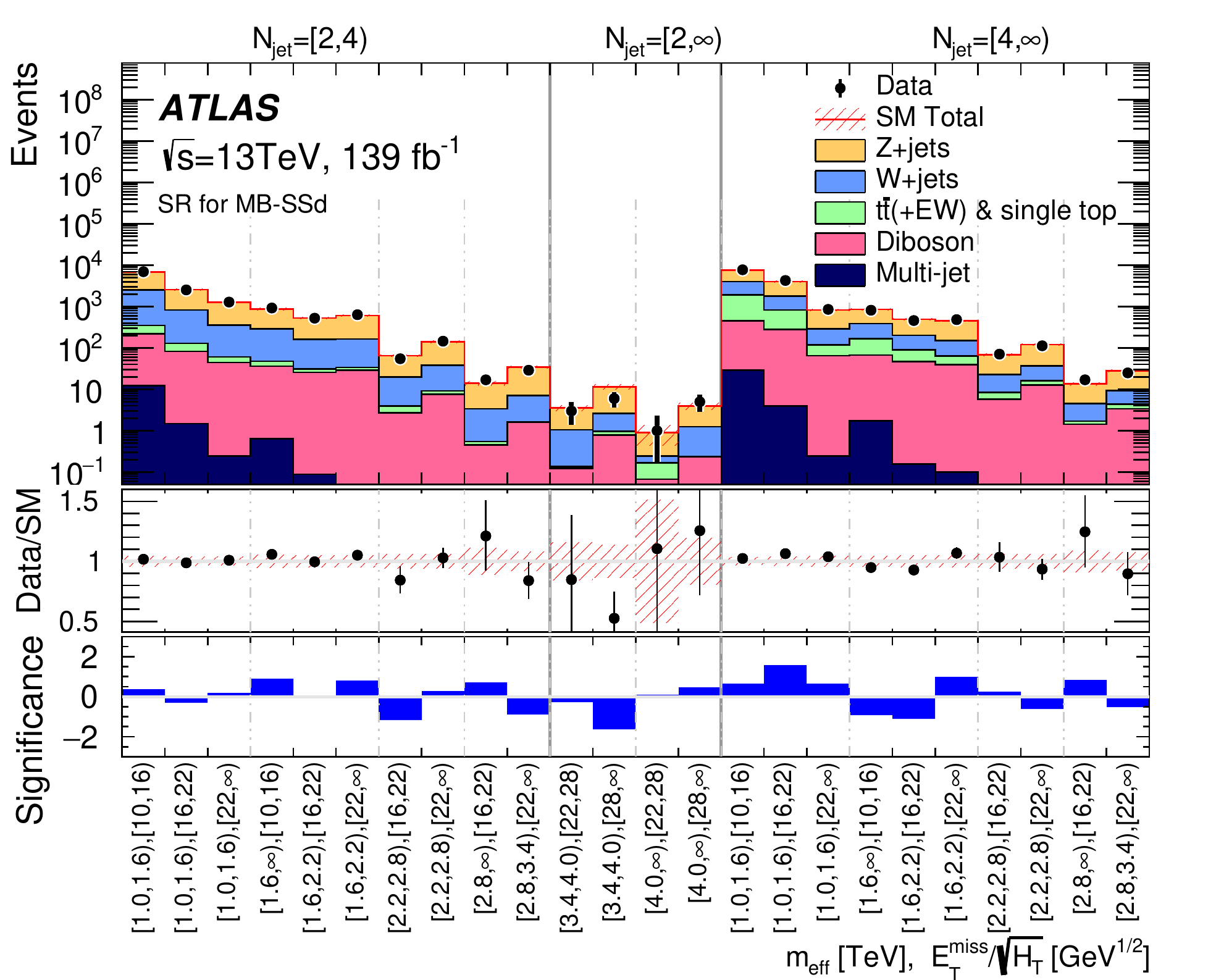}}
\subfigure[]{\includegraphics[width=0.49\textwidth]{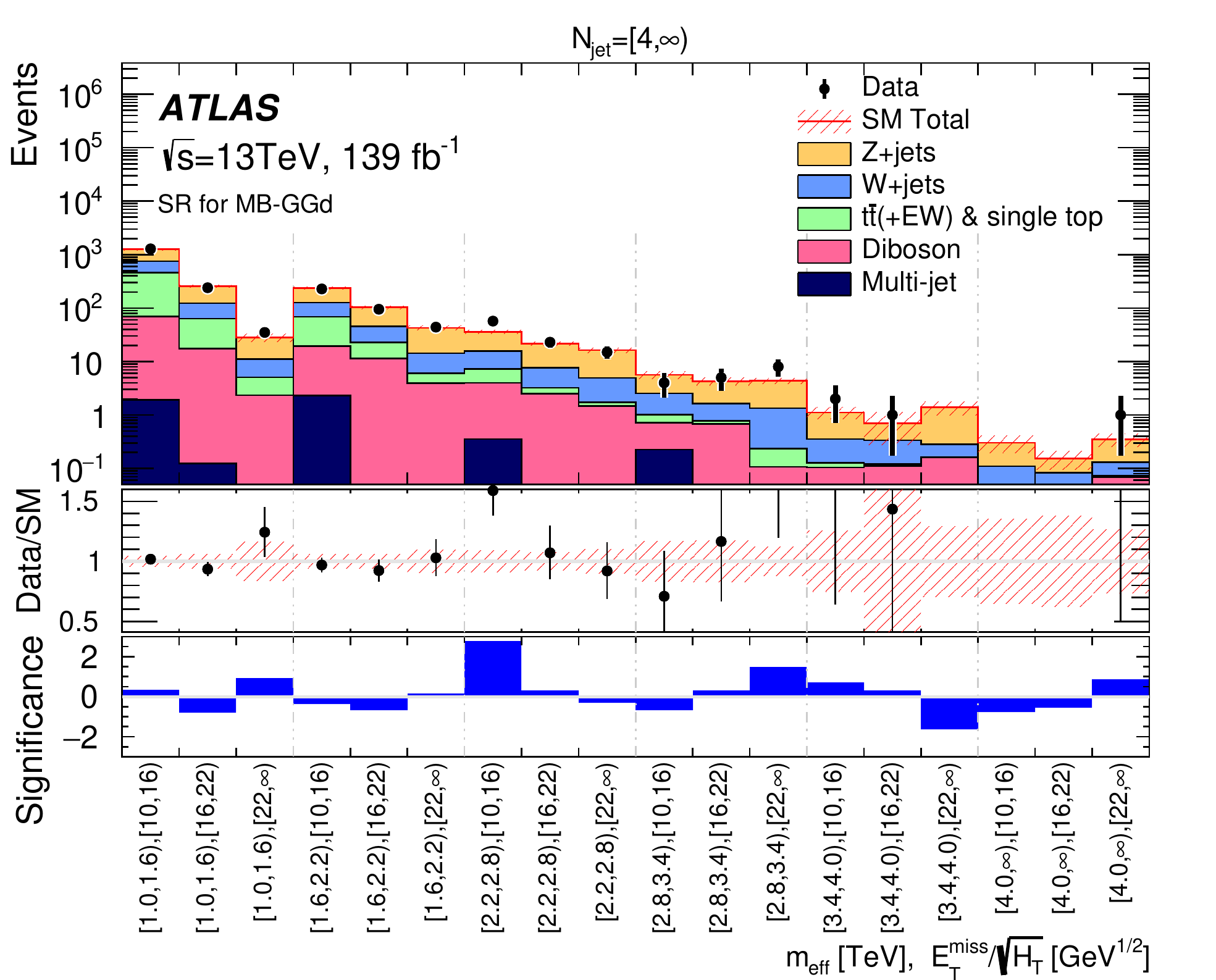}}\\
\subfigure[]{\includegraphics[width=0.49\textwidth]{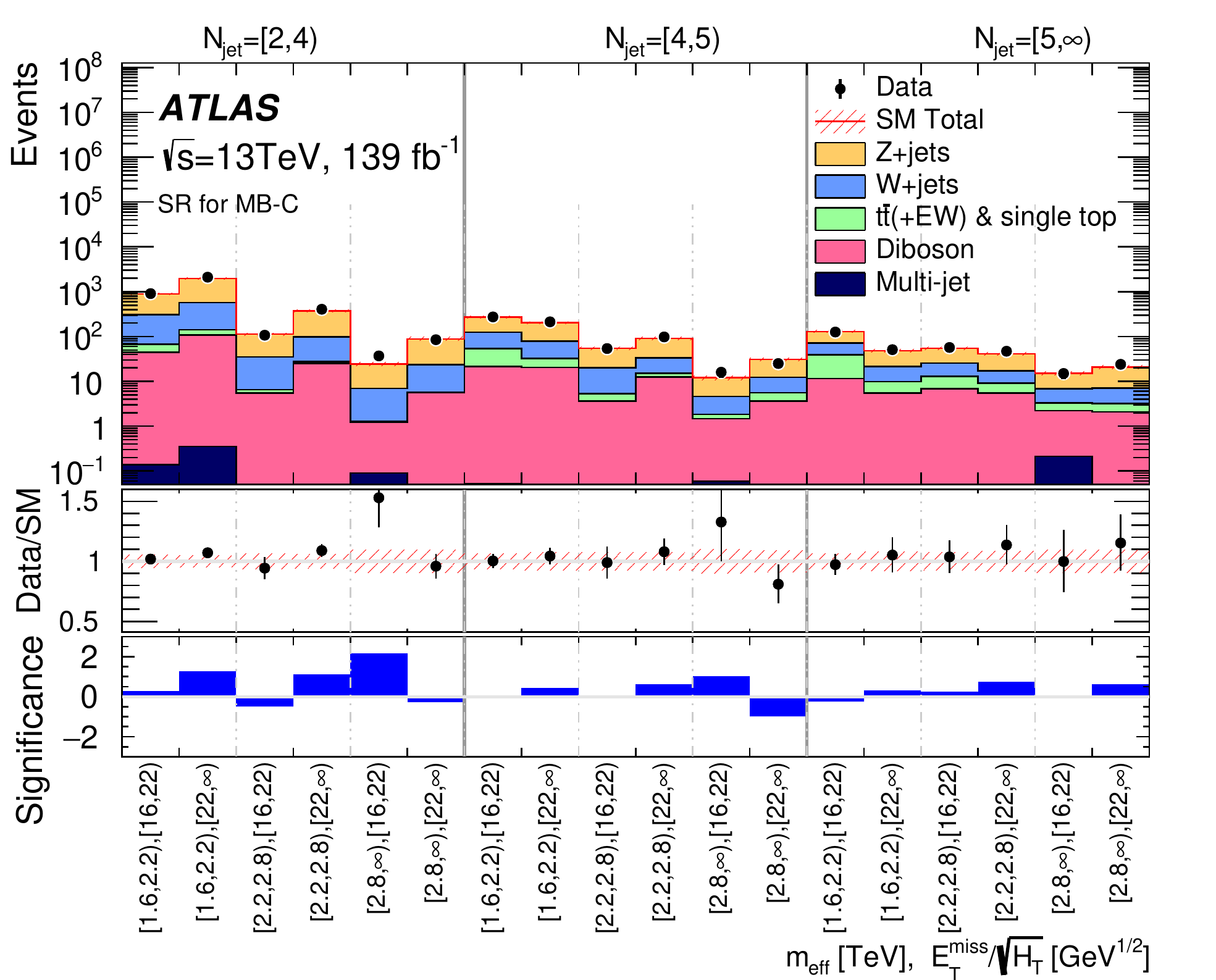}}
\subfigure[]{\includegraphics[width=0.49\textwidth]{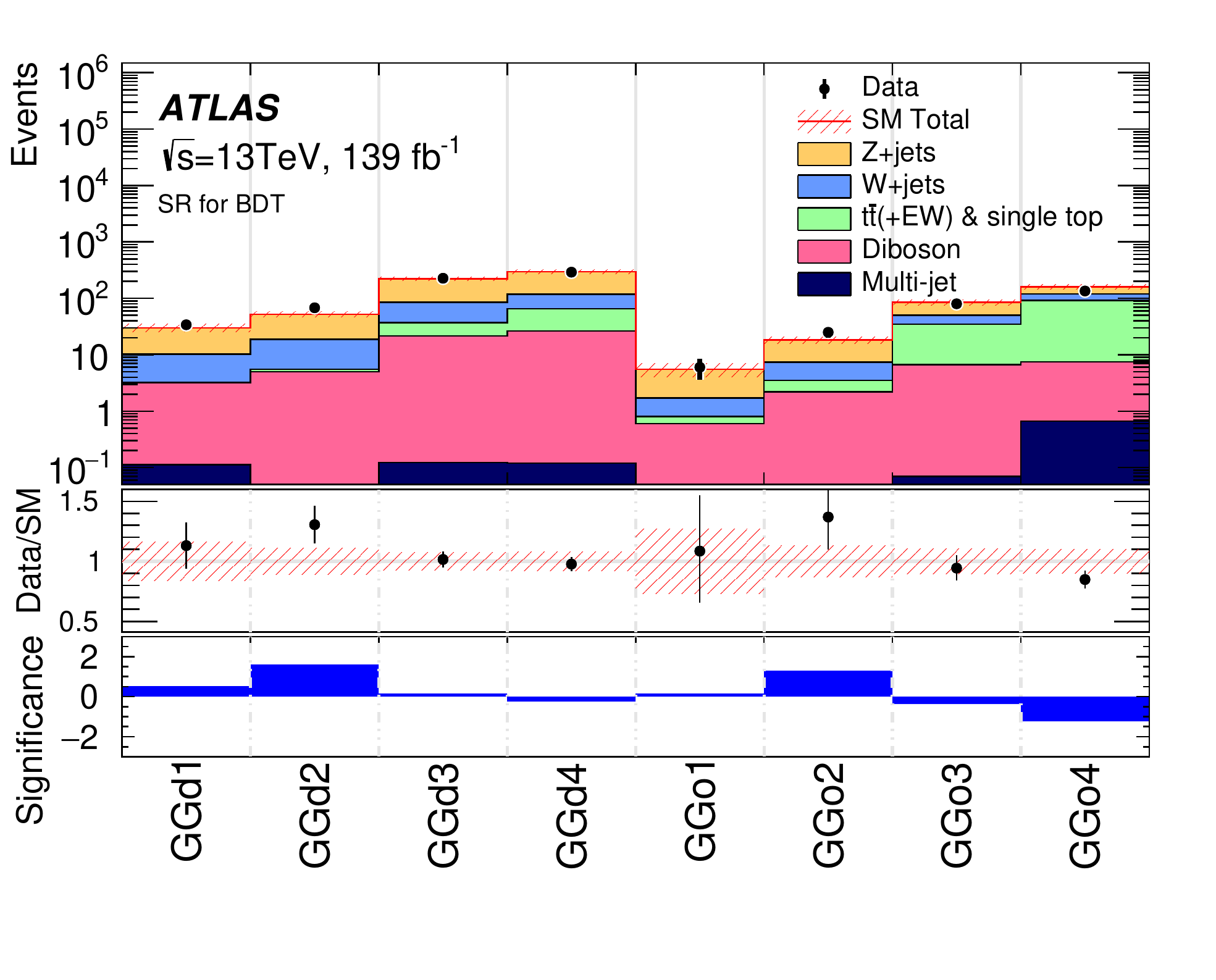}}
\end{center}
\caption{Comparison of the observed and expected event yields as a function of signal region in the (a) MB-SSd, (b) MB-GGd, (c) MB-C regions from the multi-bin search, and (d) regions from the BDT search. The background predictions are those obtained from the background-only fits, as discussed in the text. The lower panels in each case show the ratio of observed data yields to the total predicted background and the observed significance of the data relative to the background-only hypothesis. The significance is computed following the profile likelihood method of Ref.~\cite{Cousins:2007bmb} in the case where the observed yield exceeds the prediction, and using the same expression with an overall minus sign if the yield is below the prediction. The hatched (red) error bands indicate the combined post-fit experimental, theoretical and MC statistical uncertainties.}
\label{fig:PlotSR}
\end{figure}
 
\clearpage
 
\begin{figure}
\begin{center}
\includegraphics[width=0.49\textwidth]{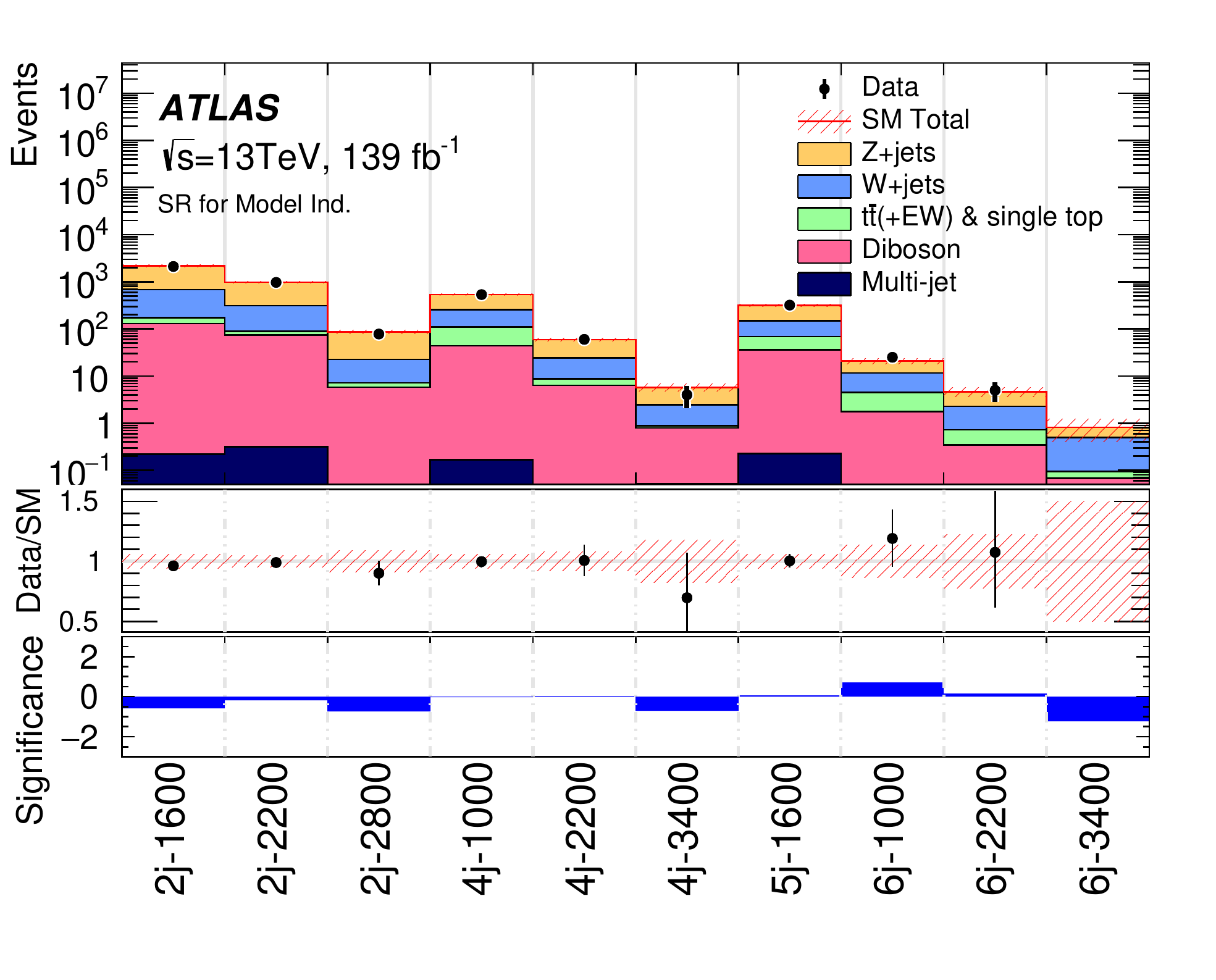}
\end{center}
\caption{Comparison of the observed and expected event yields as a function of signal region in the model-independent search. The background predictions are those obtained from the background-only fits, as discussed in the text. The lower panels show the ratio of observed data yields to the total predicted background and the observed significance of the data relative to the background-only hypothesis. The significance is computed following the profile likelihood method of Ref.~\cite{Cousins:2007bmb} in the case where the observed yield exceeds the prediction, and using the same expression with an overall minus sign if the yield is below the prediction. The hatched (red) error band indicates the combined post-fit experimental, theoretical and MC statistical uncertainties.}
\label{fig:PlotSRb}
\end{figure}
 
Model-dependent fits \cite{HFpaper} in all the SRs are used to set limits on
specific classes of SUSY models, using asymptotic formulae~\cite{statforumlimits} except in cases where the limit corresponds to a signal yield of fewer than three events. Such a fit proceeds in the same way as the model-independent fit, except that both the signal yield in the signal region and the signal contamination in the CRs are taken into account. Correlations between signal and background systematic uncertainties are taken into account where appropriate. Systematic uncertainties in the assumed signal yields due to detector effects and the theoretical uncertainties in the signal acceptance are included in the fit. The results of the three search strategies, multi-bin, BDT and model-independent, presented in this paper are all considered when constructing the final observed and expected 95\% CL exclusion limits. For each considered physics model the observed and expected exclusion limits obtained from the signal region with the best expected CL$_{\textrm s}$ value are used. The limits are driven for most models by the multi-bin search, which additionally exploits the shapes of the expected signal distributions. The BDT search is most powerful for models characterised by complex topologies with large jet multiplicities, such as one-step gluino decay models with significant mass splitting between SUSY states. All the fits for the various model points and parameter spaces considered yield fitted SUSY signal cross-sections consistent with zero within uncertainties.
 
Figure~\ref{fig:limit-sqsq_direct} shows the exclusion limits in simplified models with squark pair production and subsequent direct squark decays into a quark and the lightest neutralino.  The expected and observed exclusion limits shown in the figure are obtained by using the signal region from the three search strategies with the best expected sensitivity at each point.
These regions are usually those from the multi-bin search, although all signal regions are considered in the optimisation. Limits are shown both for a hypothesis of eight mass-degenerate light-flavour squarks and for a hypothesis of a single non-mass-degenerate light-flavour squark.
From the observed limits in the former case, neutralino masses below about 800~\GeV\ can be excluded for squark masses of
1300~\GeV, while squark masses below 1850~\GeV\ are excluded for a massless neutralino, using the optimised signal regions from the multi-bin search.
 
\begin{figure}[h]
\begin{center}
\includegraphics[width=0.9\textwidth]{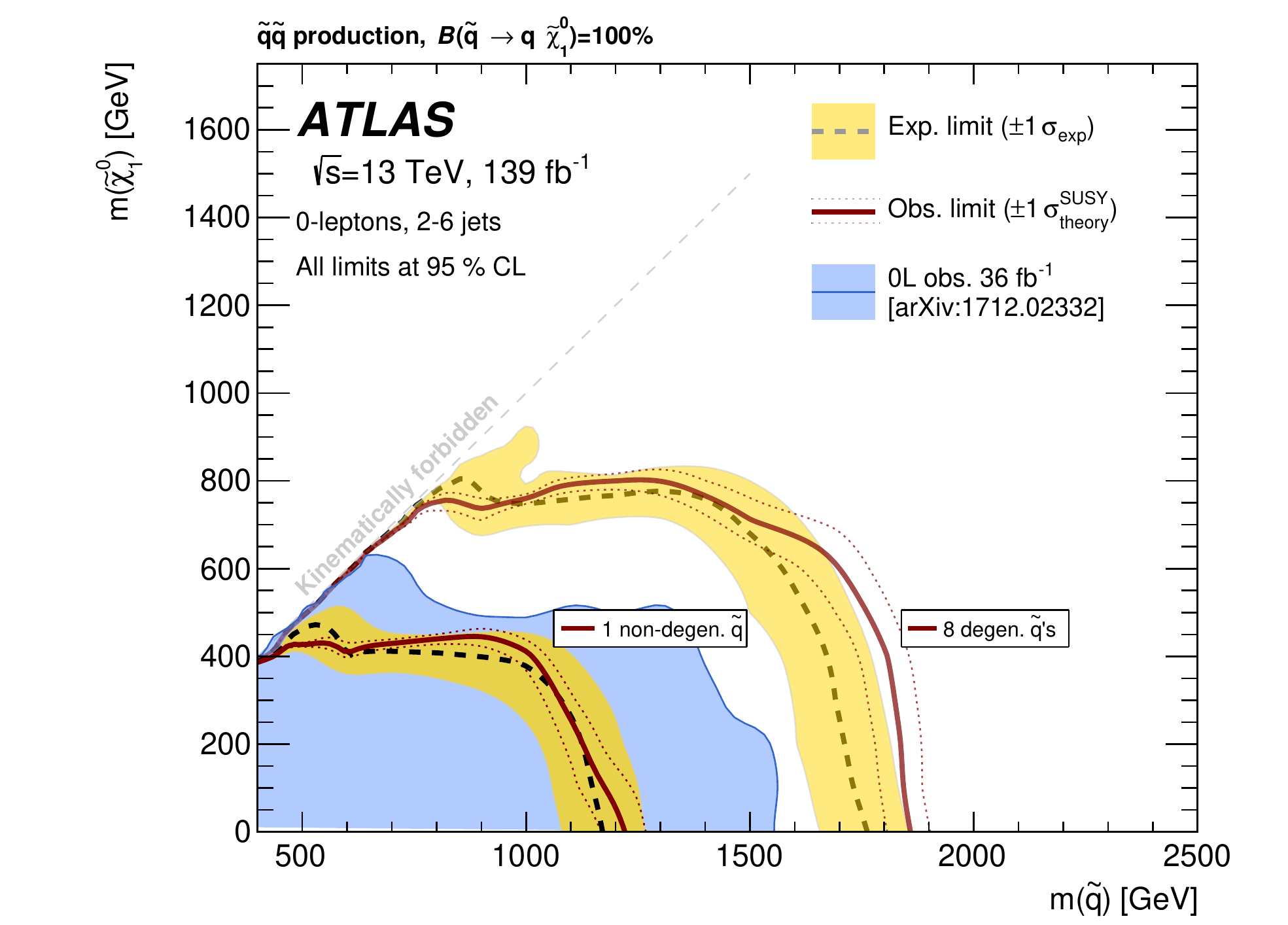}
\end{center}
\caption{Exclusion limits in the mass plane of the lightest neutralino and first- and second-generation squarks  assuming squark pair production and direct decays $\squark \to q \ninoone$ obtained by using the signal region with the best expected sensitivity at each point. Observed limits are indicated by the medium dark (maroon) curve where the solid contour represents the nominal limit, and the dotted lines are obtained by varying the signal cross-section by the renormalisation and factorisation scale and PDF uncertainties. The expected limits are indicated with a dark dashed curve, with the light (yellow) band indicating the $1\sigma$ excursions due to experimental and background-only theoretical uncertainties.
Limits are shown both for a hypothesis of eight mass-degenerate light-flavour squarks and for a hypothesis of a single non-mass-degenerate light-flavour squark.
Results are compared with the observed limits for the hypothesis of eight mass-degenerate light-flavour squarks obtained by the previous ATLAS search with jets, missing transverse momentum, and no leptons~\cite{0Lpaper2017}. }
\label{fig:limit-sqsq_direct}
\end{figure}

Another example of a direct decay is shown in Figure~\ref{fig:limit-glgl_direct}, where gluino pair production with the subsequent decay
$\tilde{g}\rightarrow q\bar{q}\ninoone$ is considered.
Due to the higher production cross-sections compared to squark pair production, higher mass limits can be obtained. For gluino masses up to about 1000~\GeV, neutralino masses can be excluded up to 950~\GeV, close to the kinematic limit near the diagonal. These limits are driven by the multi-bin signal regions dedicated to models with small mass differences. For small neutralino masses the observed lower limit on the gluino mass is as large as 2300~\GeV. For gluino masses up to about 1700~\GeV\ the best sensitivity is obtained with the optimised BDT regions, excluding neutralino masses below about 1160~\GeV.
 
\begin{figure}[h]
\begin{center}
\includegraphics[width=0.9\textwidth]{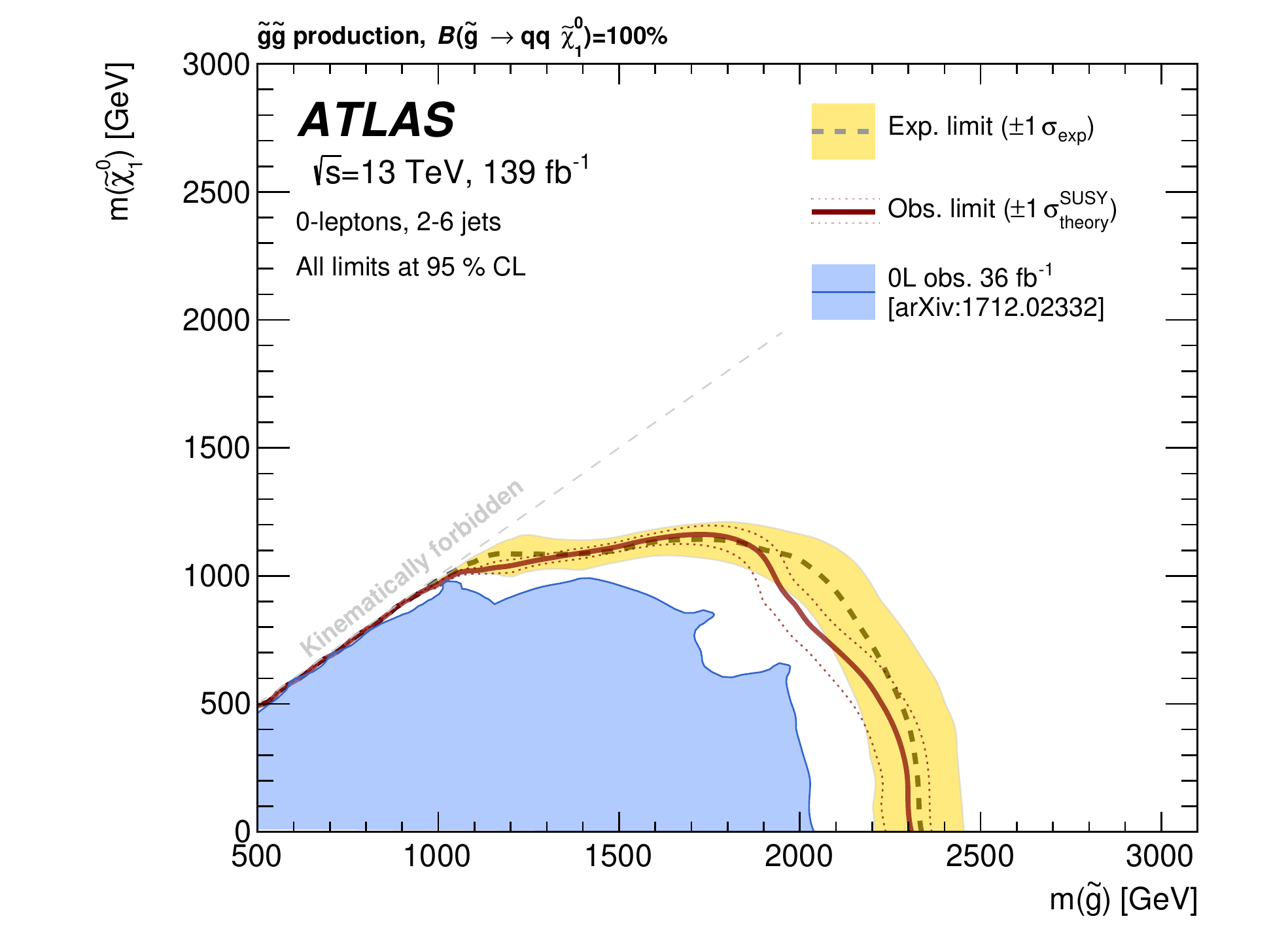}
\end{center}
\caption{Exclusion limits in the mass plane of the lightest neutralino and the gluino for gluino pair production with direct decays $\tilde{g}\rightarrow q\bar{q}\ninoone$ obtained by using the signal region with the best expected sensitivity at each point. Observed limits are indicated by the medium dark (maroon) curve where the solid contour represents the nominal limit, and the dotted lines are obtained by varying the signal cross-section by the renormalisation and factorisation scale and PDF uncertainties. The expected limits are indicated with a dark dashed curve, with the light (yellow) band indicating the $1\sigma$ excursions due to experimental and background-only theoretical uncertainties.
Results are compared with the observed limits obtained by the previous ATLAS searches with jets, missing transverse momentum, and no leptons~\cite{0Lpaper2017}.}
\label{fig:limit-glgl_direct}
\end{figure}
 
\clearpage
 
Figure~\ref{fig:limit-sqsq_onestep} shows the exclusion limits for squark pair production where the squark decays via an intermediate chargino (one-step) into a quark, $W$ boson and neutralino.  For the model presented in Figure~\ref{fig:limit-sqsq_onestep}(a) the chargino mass is fixed at $m(\chinoonepm)=(m(\tilde{q})+m(\ninoone))/2$ and the result is shown in the ($m(\squark), m(\ninoone)$) plane.
In the region close to the kinematic limit near the diagonal, neutralino and squark masses up to 600~\GeV\ are excluded, again driven by the multi-bin signal regions dedicated to models with small mass differences.
For massless neutralinos, squark masses are excluded below 1310~\GeV.  Figure~\ref{fig:limit-sqsq_onestep}(b) shows the exclusion limits in the ($m(\squark), X$) plane, for $X=\Delta m(\chinoonepm, \ninoone)/ \Delta m ({\tilde{q}}, \ninoone)$, in models with the neutralino mass fixed to 60~\GeV. Squark masses are excluded up to 1350~\GeV\ for the most favourable $X$ values. For low values of $m(\ninoone)$, the observed exclusion limits are less stringent than those expected, due to a small excess of events in one bin of the MB-GGd SR with $N_\mathrm{j}\ge 4$, $\meff=[2200,2800)$~\GeV\ and $\met / \sqrt{\HT}=[10,16)$~\GeV$^{1/2}$ (see Figure~\ref{fig:PlotSR}(b)). While the MB-GGd event selection criteria are optimised for sensitivity to gluino pair production with direct decays, they also provide sensitivity to these one-step squark pair production models due to their increased jet multiplicity.

\begin{figure}[h]
\begin{center}
\subfigure[]{\includegraphics[width=0.49\textwidth]{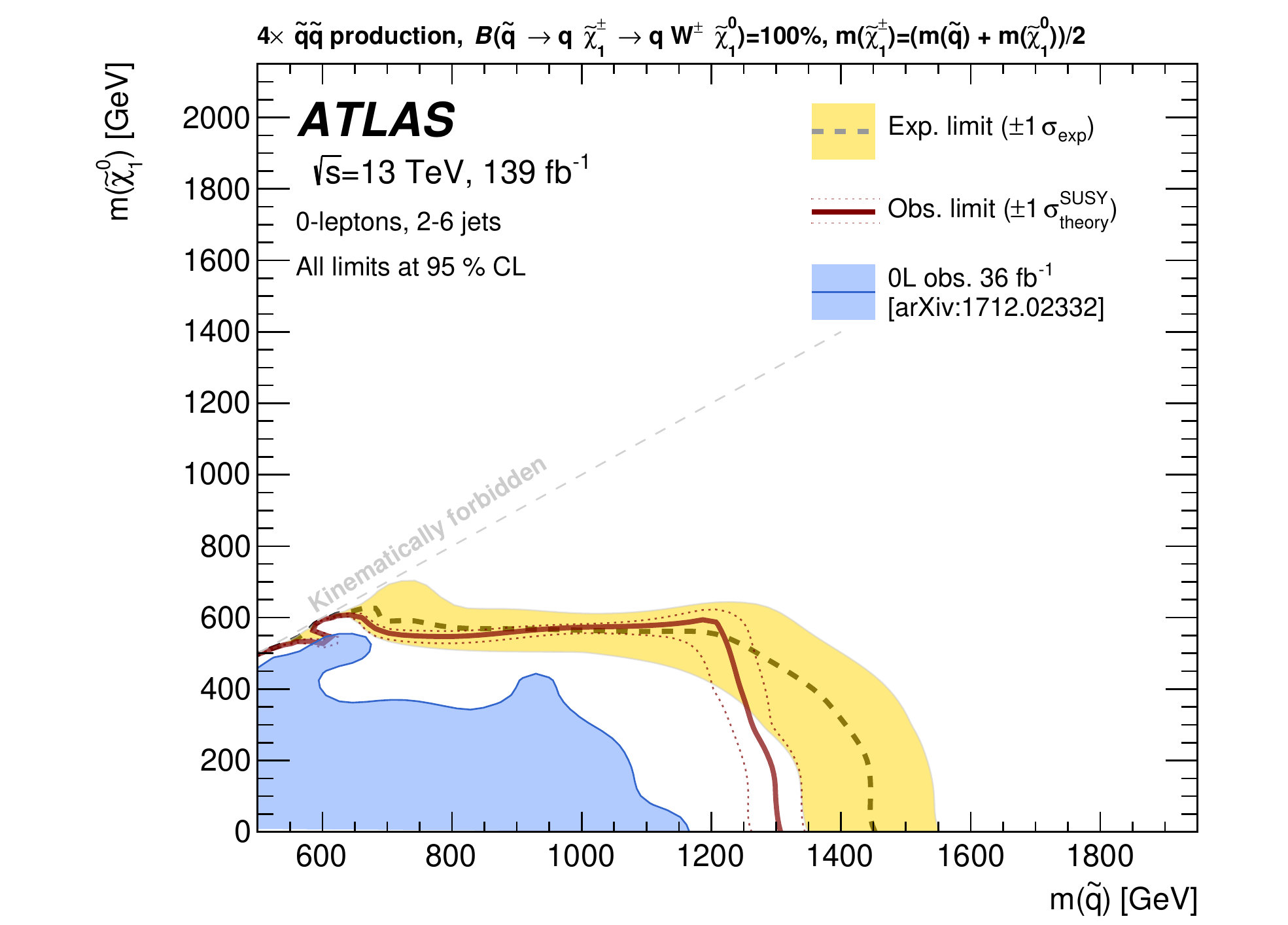}}
\subfigure[]{\includegraphics[width=0.49\textwidth]{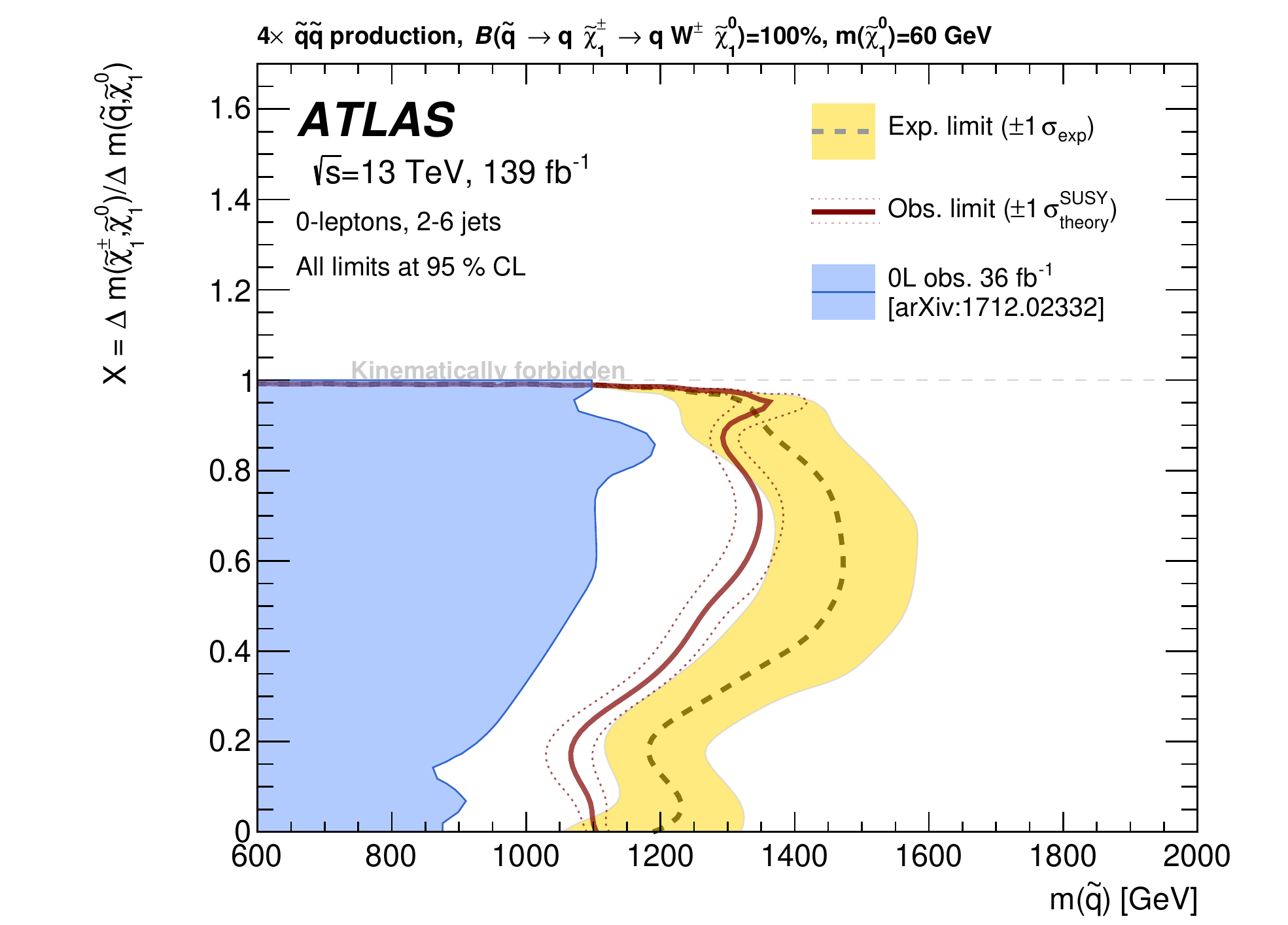}}
\end{center}
\caption{Exclusion limits for squark pair production with a one-step decay via an intermediate chargino into $qW\ninoone$.
Figure (a)  shows the limits in the ($m(\squark), m(\ninoone)$) plane for a chargino mass fixed
at $m(\chinoonepm)=(m(\tilde{q})+m(\ninoone))/2$.  Alternatively in Figure (b), the neutralino mass is fixed at 60~\GeV\ and exclusion limits are given for $X=\Delta m(\chinoonepm, \ninoone)/ \Delta m ({\tilde{q}}, \ninoone)$, as a function of the squark mass.   Exclusion limits are obtained by using the signal region with the best expected sensitivity at each point. Observed limits are indicated by the medium dark (maroon) curves where the solid contour represents the nominal limit, and the dotted lines are obtained by varying the signal cross-section by the renormalisation and factorisation scale and PDF uncertainties. The expected limits are indicated with dark dashed curves, with the light (yellow) bands indicating the $1\sigma$ excursions due to experimental and background-only theoretical uncertainties.
Results are compared with the observed limits obtained by the previous ATLAS searches with jets, missing transverse momentum, and no leptons~\cite{0Lpaper2017}.}
\label{fig:limit-sqsq_onestep}
\end{figure}
 
The results of the search for gluino pair production with a one-step decay via an intermediate chargino into $q\bar{q}'W\ninoone$ are shown in
Figure~\ref{fig:limit-glgl_onestep}.
Figure~\ref{fig:limit-glgl_onestep}(a) shows the limit for a chargino mass chosen such that $m(\chinoonepm)=(m(\tilde{g})+m(\ninoone))/2$.
In the region close to the kinematic limit near the diagonal, neutralino and gluino masses up to 900~\GeV\  are excluded, driven by the multi-bin signal regions dedicated to models with small mass differences.  For massless neutralinos, gluino masses are excluded below 2220~\GeV.
Figure~\ref{fig:limit-glgl_onestep}(b) shows limits on $X=\Delta m(\chinoonepm, \ninoone)/ \Delta m ({\tilde{g}}, \ninoone)$, for a neutralino mass of 60~\GeV. Gluino masses are excluded up to 2210~\GeV\ for the most favourable values of $X$. The narrow corridor of decreased sensitivity to the gluino mass at $X\sim 0.06$ corresponds to models for which $\Delta m(\chinoonepm, \ninoone)\sim m(W)$ and hence the chargino decay products are produced at rest in the chargino rest frame, leading to reduced signal acceptance.
 
\begin{figure}[h]
\begin{center}
\subfigure[]{\includegraphics[width=0.49\textwidth]{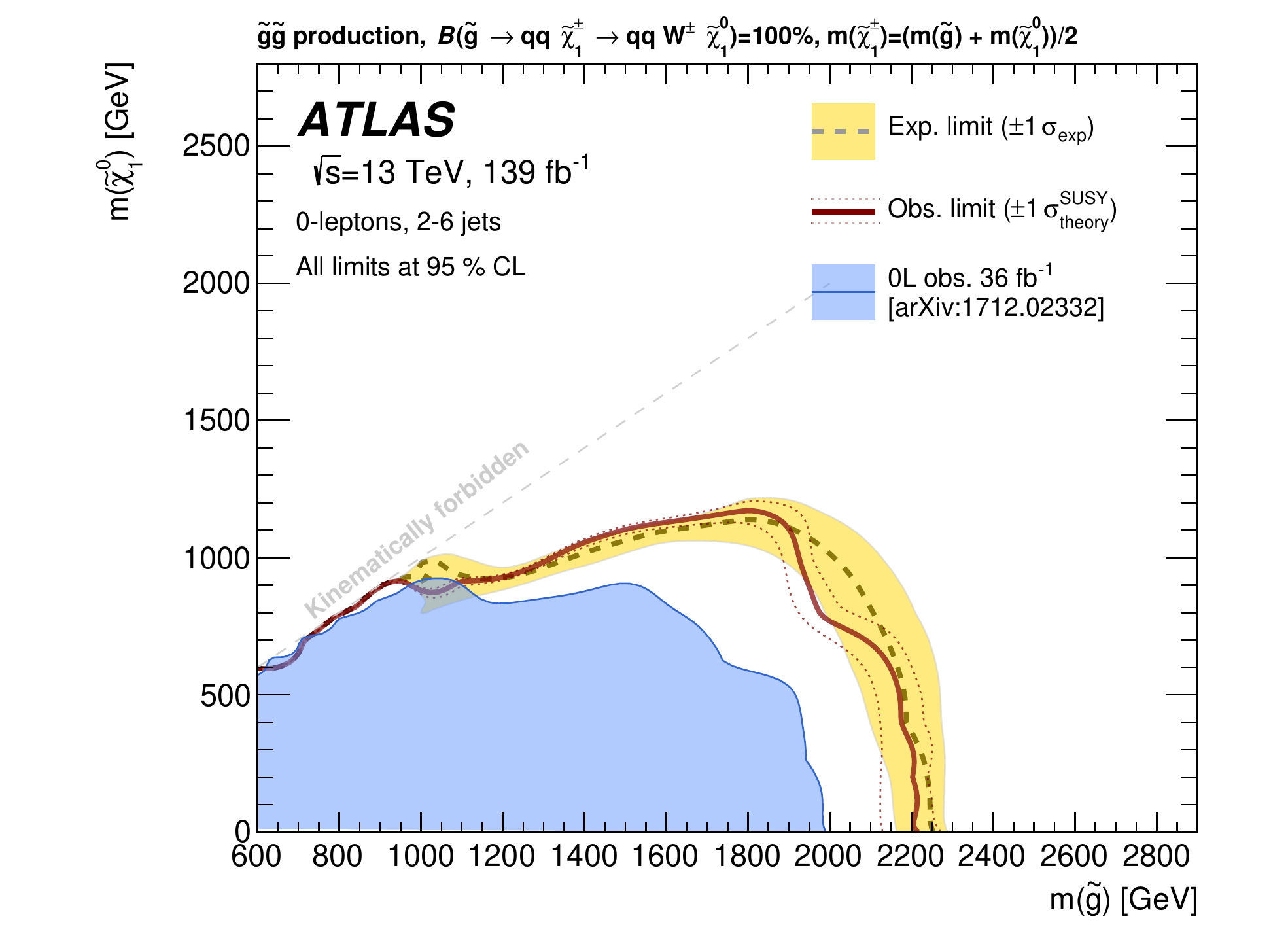}}
\subfigure[]{\includegraphics[width=0.49\textwidth]{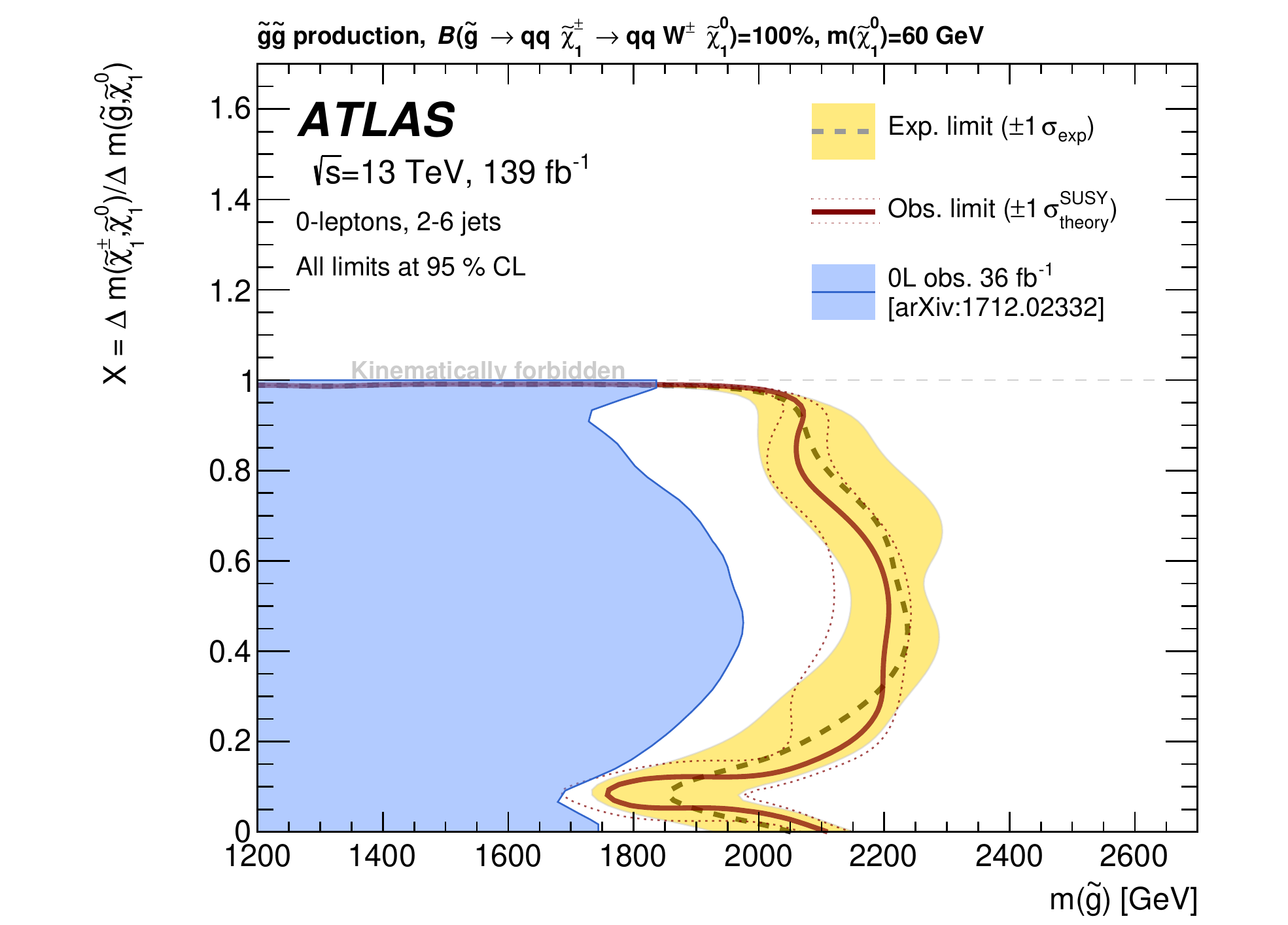}}
\end{center}
\caption{Exclusion limits for gluino pair production with a one-step decay via an intermediate chargino into $q\bar{q}'W\ninoone$.
Figure (a) shows the limits in the ($m(\gluino), m(\ninoone)$) plane for a chargino mass fixed
at $m(\chinoonepm)=(m(\tilde{g})+m(\ninoone))/2$.  Alternatively in Figure (b), the neutralino mass is fixed at 60~\GeV\ and exclusion limits are given for $X=\Delta m(\chinoonepm, \ninoone)/ \Delta m ({\tilde{g}}, \ninoone)$, as a function of the gluino mass.  Exclusion limits are obtained by using the signal region with the best expected sensitivity at each point. Observed limits are indicated by the medium dark (maroon) curves where the solid contour represents the nominal limit, and the dotted lines are obtained by varying the signal cross-section by the renormalisation and factorisation scale and PDF uncertainties. The expected limits are indicated with dark dashed curves, with the light (yellow) bands indicating the $1\sigma$ excursions due to experimental and background-only theoretical uncertainties.
Results are compared with the observed limits obtained by the previous ATLAS searches with jets, missing transverse momentum, and no leptons~\cite{0Lpaper2017}.}
\label{fig:limit-glgl_onestep}
\end{figure}

Figure~\ref{fig:limit-pmssm} expresses the mass limits in the ($m(\tilde{g}), m(\tilde{q})$) plane in the model with combined production of squark pairs, gluino pairs, and squark--gluino pairs, for different assumptions about the neutralino mass: $m(\ninoone)=$ 0~\GeV, 995~\GeV\ or 1495~\GeV, motivated by the assumptions used in Ref.~\cite{0Lpaper2017}. Depending on the mass hierarchy, the $\gluino \to \squark q$ and $\squark \to \gluino q $ one-step decays are taken into account.
A lower limit of 3000~\GeV\ for equal squark and gluino masses is found for the scenario with a massless $\ninoone$.
The squark production cross-section, which in the considered models is strongly dominated by $t$- and $u$-channel diagrams, decreases with increasing gluino mass, leading to weaker limits in regions of the mass plane where gluino masses are high. In regions where the gluino mass becomes greater than 8~\TeV, the kinematics is expected to stay the same, and the change of the production cross-section is
expected to provide a smooth transition of the exclusion limits between a gluino mass of 8.5~\TeV\ and the decoupled gluino scenario.
In scenarios with $m(\ninoone)=$ 995~\GeV, the search becomes less sensitive to models with very small mass difference between the particles, as seen in models with gluino masses around 6~\TeV\ and squark masses around 1~\TeV. In similar compressed regions, with the squark (gluino) mass close to the mass of the LSP and the gluino (squark) mass as high as 4~\TeV, the search still has sensitivity to such models due to $\squark\gluino$ production processes that provide sufficient acceptance.

\begin{figure}[h]
\begin{center}
\subfigure[]{\includegraphics[width=0.49\textwidth]{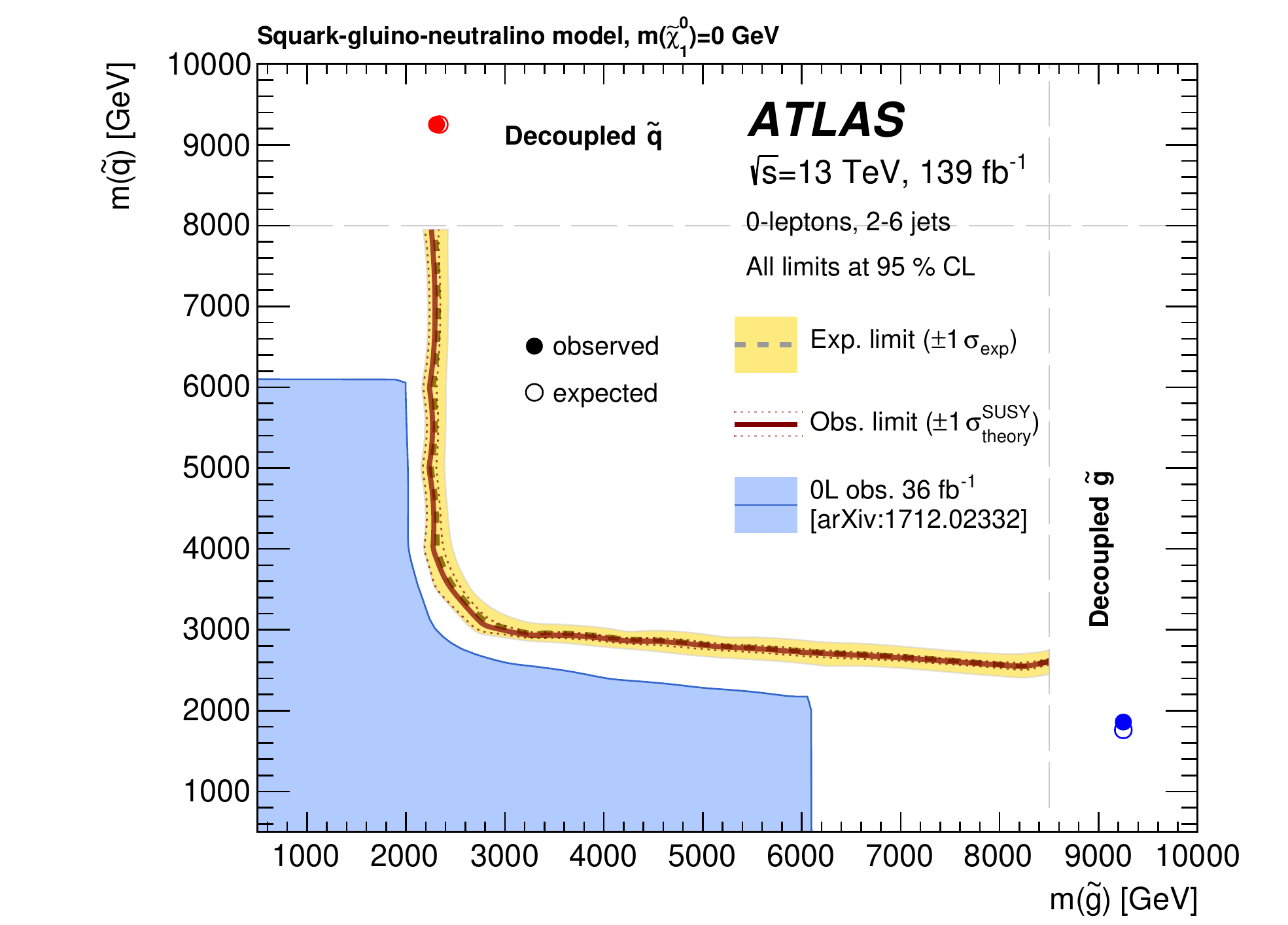}}
\subfigure[]{\includegraphics[width=0.49\textwidth]{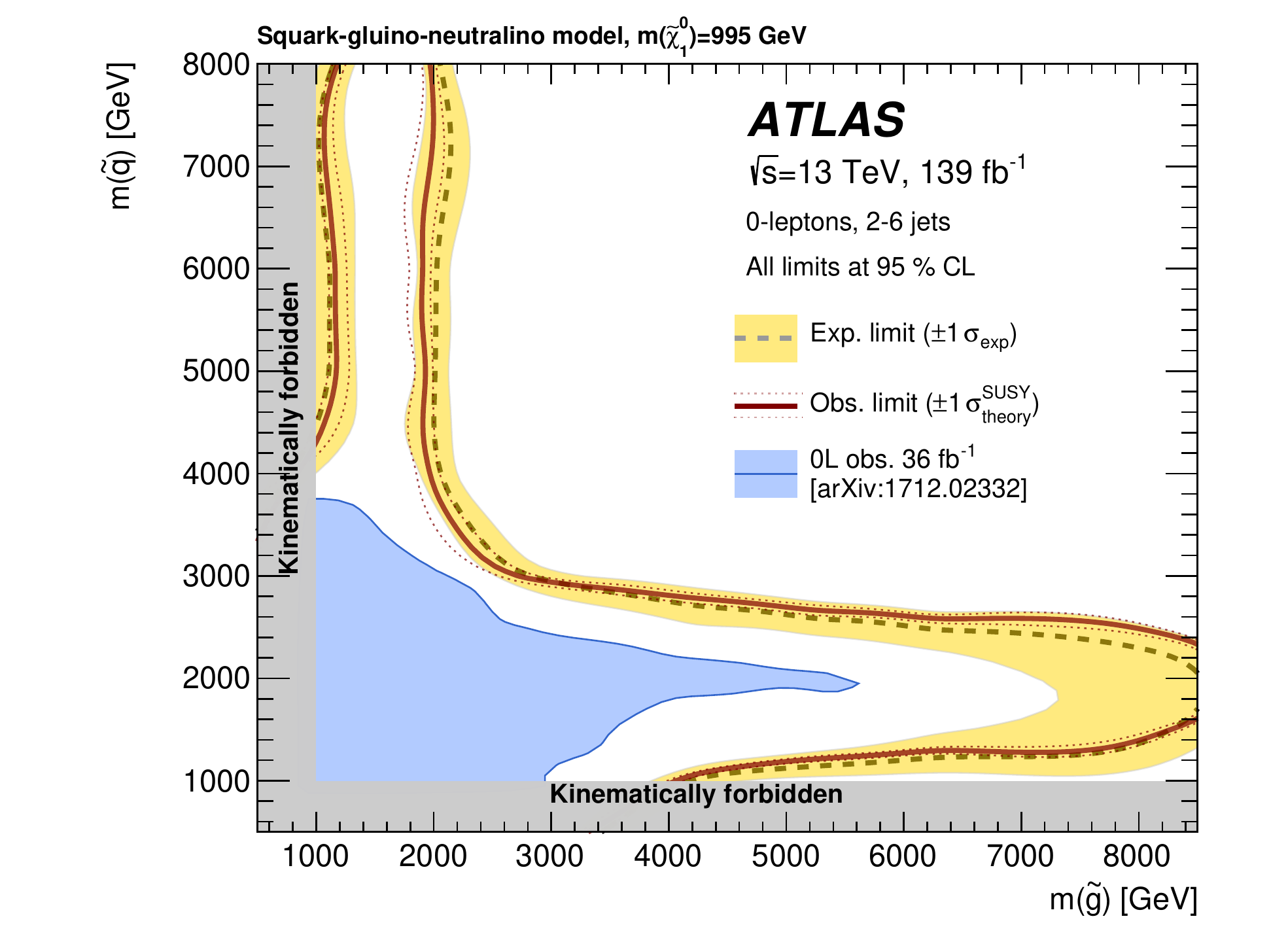}} \\
\subfigure[]{\includegraphics[width=0.49\textwidth]{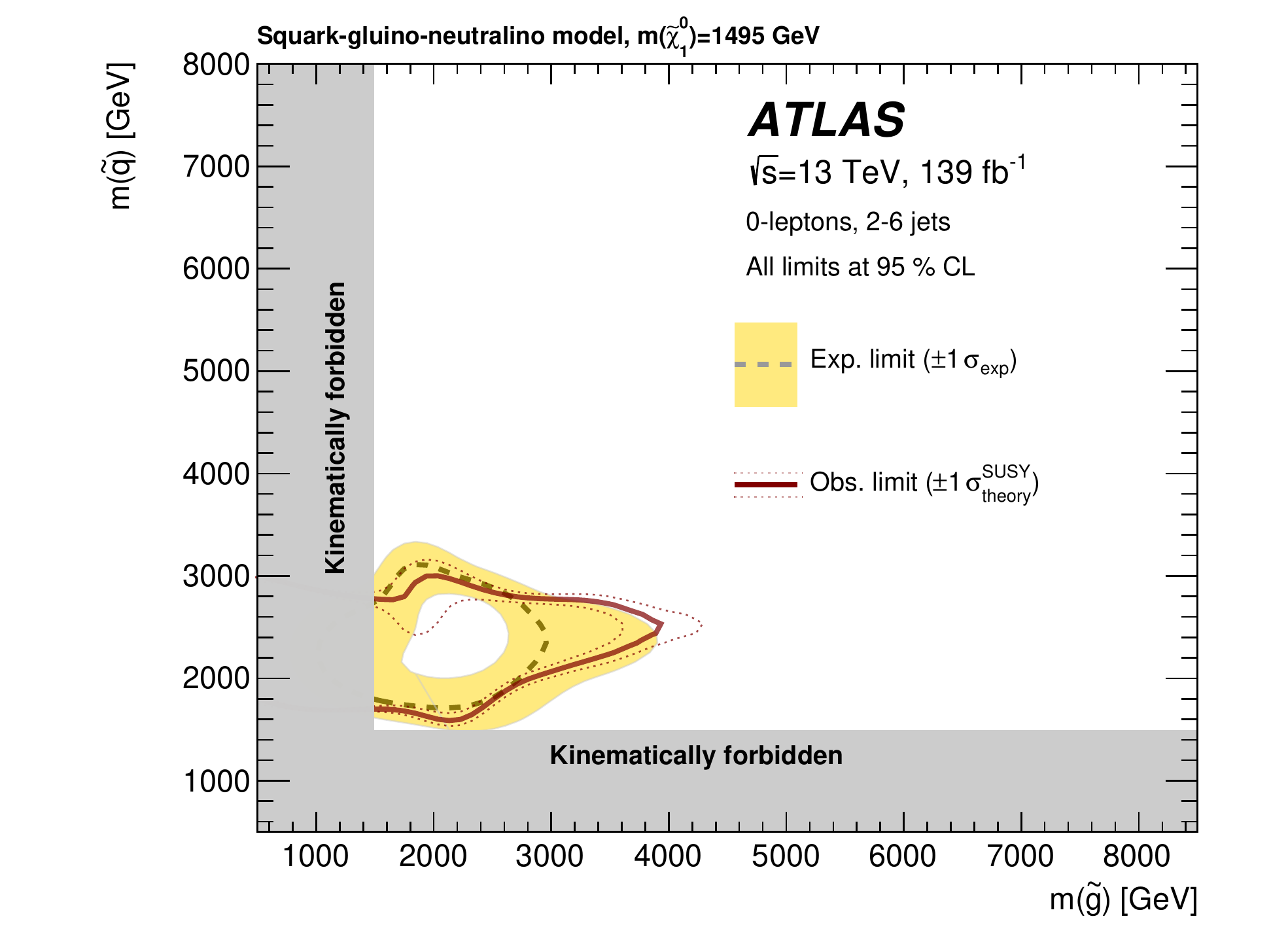}}
\end{center}
\caption{Exclusion limits for the model with combined production of squark pairs, gluino pairs, and squark--gluino pairs, for different assumptions about the neutralino mass: (a) $m(\tilde{\chi}^{0}_{1})=0~\GeV$, (b) $m(\ninoone)=995~\GeV$ and (c) $m(\tilde{\chi}^{0}_{1})=1495~\GeV$, varying values of $m(\gluino)$ and $m(\squark)$ and assuming a purely bino $\ninoone$. Exclusion limits are obtained by using the signal region with the best expected sensitivity at each point. Observed limits are indicated by the medium dark (maroon) curves where the solid contour represents the nominal limit, and the dotted lines are obtained by varying the signal cross-section by the renormalisation and factorisation scale and PDF uncertainties. The expected limits are indicated with dark dashed curves, with the light (yellow) bands indicating the $1\sigma$ excursions due to experimental and background-only theoretical uncertainties. In Figure (a) observed and expected limits on squark (gluino) masses are also shown, assuming gluino (squark) masses are decoupled as in simplified models presented in Figure~\ref{fig:limit-sqsq_direct} (\ref{fig:limit-glgl_direct}).
Results (a) and (b) are compared with the observed limits obtained by the previous ATLAS searches with no leptons, jets and missing transverse momentum~\cite{0Lpaper2017}.
}
\label{fig:limit-pmssm}
\end{figure}

\FloatBarrier
\newpage
\section{Conclusions}
\label{sec:conclusions}
 
This paper presents the results of three search strategies for squarks and gluinos in final states containing
high-\pT{} jets, large missing transverse momentum but no electrons or
muons, based on a 139~\ifb\ dataset of $\sqrt{s}=13~\TeV$
proton--proton collisions recorded by the ATLAS experiment at the LHC. No significant deviation from the background expectation is found.
 
Results are interpreted in terms of simplified models
with only first- and second-generation squarks, or gluinos, together with a neutralino
LSP, with the masses of all the other SUSY particles set such that the
particles are effectively decoupled.
For a massless lightest neutralino, gluino masses below 2.30~\TeV\ are excluded at the 95\% confidence level in a simplified model with only gluinos and the lightest neutralino.
For a simplified model involving the strong production of squarks of the first and second generations, with decays to a massless lightest neutralino, squark masses below 1.85~\TeV\ are excluded, assuming mass-degenerate squarks of the first two generations.
In simplified models with pair-produced squarks and gluinos, each decaying via an intermediate $\chinoonepm$ into one quark or two quarks, a $W$ boson and a $\ninoone$,
squark masses below $1.31~\TeV$ and gluino masses below 2.22~\TeV\ are excluded for massless $\ninoone$.
In models with combined production of squark pairs, gluino pairs, and squark--gluino pairs, a lower limit of 3000~\GeV\ for equal squark and gluino masses is found for the scenario with a massless $\ninoone$.
 
These results extend the region of supersymmetric parameter space excluded by ATLAS searches substantially beyond that obtained previously.


\section*{Acknowledgements}
 

We thank CERN for the very successful operation of the LHC, as well as the
support staff from our institutions without whom ATLAS could not be
operated efficiently.
 
We acknowledge the support of ANPCyT, Argentina; YerPhI, Armenia; ARC, Australia; BMWFW and FWF, Austria; ANAS, Azerbaijan; SSTC, Belarus; CNPq and FAPESP, Brazil; NSERC, NRC and CFI, Canada; CERN; ANID, Chile; CAS, MOST and NSFC, China; COLCIENCIAS, Colombia; MSMT CR, MPO CR and VSC CR, Czech Republic; DNRF and DNSRC, Denmark; IN2P3-CNRS and CEA-DRF/IRFU, France; SRNSFG, Georgia; BMBF, HGF and MPG, Germany; GSRT, Greece; RGC and Hong Kong SAR, China; ISF and Benoziyo Center, Israel; INFN, Italy; MEXT and JSPS, Japan; CNRST, Morocco; NWO, Netherlands; RCN, Norway; MNiSW and NCN, Poland; FCT, Portugal; MNE/IFA, Romania; JINR; MES of Russia and NRC KI, Russian Federation; MESTD, Serbia; MSSR, Slovakia; ARRS and MIZ\v{S}, Slovenia; DST/NRF, South Africa; MICINN, Spain; SRC and Wallenberg Foundation, Sweden; SERI, SNSF and Cantons of Bern and Geneva, Switzerland; MOST, Taiwan; TAEK, Turkey; STFC, United Kingdom; DOE and NSF, United States of America. In addition, individual groups and members have received support from BCKDF, CANARIE, Compute Canada, CRC and IVADO, Canada; Beijing Municipal Science \& Technology Commission, China; COST, ERC, ERDF, Horizon 2020 and Marie Sk{\l}odowska-Curie Actions, European Union; Investissements d'Avenir Labex, Investissements d'Avenir Idex and ANR, France; DFG and AvH Foundation, Germany; Herakleitos, Thales and Aristeia programmes co-financed by EU-ESF and the Greek NSRF, Greece; BSF-NSF and GIF, Israel; La Caixa Banking Foundation, CERCA Programme Generalitat de Catalunya and PROMETEO and GenT Programmes Generalitat Valenciana, Spain; G\"{o}ran Gustafssons Stiftelse, Sweden; The Royal Society and Leverhulme Trust, United Kingdom.
 
The crucial computing support from all WLCG partners is acknowledged gratefully, in particular from CERN, the ATLAS Tier-1 facilities at TRIUMF (Canada), NDGF (Denmark, Norway, Sweden), CC-IN2P3 (France), KIT/GridKA (Germany), INFN-CNAF (Italy), NL-T1 (Netherlands), PIC (Spain), ASGC (Taiwan), RAL (UK) and BNL (USA), the Tier-2 facilities worldwide and large non-WLCG resource providers. Major contributors of computing resources are listed in Ref.~\cite{ATL-SOFT-PUB-2020-001}.
 

\printbibliography
\newpage

\clearpage 
 
\begin{flushleft}
\hypersetup{urlcolor=black}
{\Large The ATLAS Collaboration}

\bigskip

\AtlasOrcid[0000-0002-6665-4934]{G.~Aad}$^\textrm{\scriptsize 102}$,    
\AtlasOrcid[0000-0002-5888-2734]{B.~Abbott}$^\textrm{\scriptsize 128}$,    
\AtlasOrcid{D.C.~Abbott}$^\textrm{\scriptsize 103}$,    
\AtlasOrcid[0000-0002-2788-3822]{A.~Abed~Abud}$^\textrm{\scriptsize 36}$,    
\AtlasOrcid[0000-0002-1002-1652]{K.~Abeling}$^\textrm{\scriptsize 53}$,    
\AtlasOrcid[0000-0002-2987-4006]{D.K.~Abhayasinghe}$^\textrm{\scriptsize 94}$,    
\AtlasOrcid[0000-0002-8496-9294]{S.H.~Abidi}$^\textrm{\scriptsize 167}$,    
\AtlasOrcid[0000-0002-8279-9324]{O.S.~AbouZeid}$^\textrm{\scriptsize 40}$,    
\AtlasOrcid{N.L.~Abraham}$^\textrm{\scriptsize 156}$,    
\AtlasOrcid[0000-0001-5329-6640]{H.~Abramowicz}$^\textrm{\scriptsize 161}$
% ...
\bigskip
\\

% ...
$^{36}$CERN, Geneva; Switzerland.\\
% ...
$^{40}$Niels Bohr Institute, University of Copenhagen, Copenhagen; Denmark.\\
% ...
$^{53}$II. Physikalisches Institut, Georg-August-Universit\"{a}t G\"ottingen, G\"ottingen; Germany.\\
% ...
$^{94}$Department of Physics, Royal Holloway University of London, Egham; United Kingdom.\\
% ...
$^{102}$CPPM, Aix-Marseille Universit\'e, CNRS/IN2P3, Marseille; France.\\
$^{103}$Department of Physics, University of Massachusetts, Amherst MA; United States of America.\\
% ...
$^{128}$Homer L. Dodge Department of Physics and Astronomy, University of Oklahoma, Norman OK; United States of America.\\
% ...
$^{156}$Department of Physics and Astronomy, University of Sussex, Brighton; United Kingdom.\\
% ...
$^{161}$Raymond and Beverly Sackler School of Physics and Astronomy, Tel Aviv University, Tel Aviv; Israel.\\
% ...
$^{167}$Department of Physics, University of Toronto, Toronto ON; Canada.\\
% ...

\end{flushleft}


\end{document}